\documentclass[aps,prd,twocolumn,groupedaddress,nofootinbib]{revtex4-1}

\usepackage{graphicx,color,amsmath,amssymb}
\usepackage{slashed}

\newcommand{\nocontentsline}[3]{}
\newcommand{\toclesslab}[3]{\bgroup\let\addcontentsline=\nocontentsline#1{#2\label{#3}}\egroup}
\newcommand{\tocless}[2]{\bgroup\let\addcontentsline=\nocontentsline#1{#2}\egroup}

\usepackage[export]{adjustbox}
\usepackage{tikz}
\usepackage{tkz-euclide}
\usetikzlibrary{decorations.pathmorphing}	
\tikzset{
    v/.style={decorate, decoration={snake, segment length=3mm, amplitude=0.75mm}, draw},
    f/.style={draw=black, postaction={decorate},
        decoration={markings,mark=at position .6 with {\arrow[very thick]{latex}}}},
    fb/.style={draw=black, postaction={decorate},
        decoration={markings,mark=at position .4 with {\arrowreversed[very thick]{latex}}}},
    fnar/.style={draw=black},
    g/.style={decorate, draw=black,
        decoration={coil,amplitude=3pt, segment length=3.5pt}},
    s/.style={dashed,draw=black, postaction={decorate},
        decoration={markings,mark=at position .55 with {\arrow[very thick]{latex}}}},
    sb/.style={dashed,draw=black, postaction={decorate},
        decoration={markings,mark=at position .55 with {\arrowreversed[draw=black,very thick]{latex}}}},
    snar/.style={dashed,draw=black,line width =1.25pt},
}

\usepackage{hyperref} 
\hypersetup{
    colorlinks=true,
    citecolor=blue,
    linkcolor=magenta,
    filecolor=magenta,
    urlcolor=blue,
}

\newcommand{\cm}{{\, {\rm cm}}}

\newcommand{\eV}{{\, {\rm eV}}}
\newcommand{\keV}{{\, {\rm keV}}}
\newcommand{\MeV}{{\, {\rm MeV}}}
\newcommand{\GeV}{{\, {\rm GeV}}}

\newcommand{\GHz}{{\, {\rm GHz}}}

\newcommand{\nm}{{\, {\rm nm}}}

\newcommand{\kms}{{\, {\rm km \, s^{-1}}}}
\newcommand{\meV}{{\, {\rm meV}}}

\newcommand{\LL}{{\mathcal{L}}}
\newcommand{\OO}{{\mathcal{O}}}

\DeclareMathOperator{\tr}{tr}
\DeclareMathOperator{\Imag}{Im}
\DeclareMathOperator{\Real}{Re}

\definecolor{mypurple}{RGB}{164,64,214}

\begin{document}

\title{DM-electron scattering in materials: sum rules and heterostructures}

\author{Robert Lasenby}
\email{rlasenby@stanford.edu}
\affiliation{Stanford Institute for Theoretical Physics, Stanford University, Stanford, CA 94305, USA}
\author{Anirudh Prabhu}
\email{aniprabhu@stanford.edu}
\affiliation{Stanford Institute for Theoretical Physics, Stanford University, Stanford, CA 94305, USA}

\date{\today}

\begin{abstract}
 In recent years, a growing experimental program
 has begun to search for sub-GeV dark matter
 through its scattering with electrons. An
 associated theoretical challenge is to compute
 the dark matter scattering rate in experimental
 targets, and to find materials with large
 scattering rates. In this paper we point out
 that, if dark matter scatters through a mediator that
 couples to EM charge, then electromagnetic sum
 rules place limits on the achievable scattering
 rates. These limits serve as a useful
	sanity check for calculations,
	as well as setting a theoretical target
	for proposed detection methods.
	Motivated by this analysis,
 we explore how conductor-dielectric
 heterostructures can result in enhanced
 scattering rates compared to bulk conductors,
 for dark matter masses $\lesssim \MeV$. These effects
 could be especially important in computing the
 scattering rates from thin-film targets, e.g.\
 superconducting detectors such as SNSPDs, TESs
 or MKIDs, for which the scattering rate could be
 enhanced by orders of magnitude at low
	enough dark matter masses,
	as well as introducing or enhancing directional
	dependence.
\end{abstract}

\maketitle

{\hypersetup{linkcolor=blue}
\tableofcontents
}


\section{Introduction}

There is very strong evidence that some 
form of non-relativistic, non-Standard-Model
matter makes up most of the universe's matter
density. 
While it is possible that this `dark matter' (DM)
only interacts gravitationally, in many theories
it possesses other interactions with Standard Model (SM) particles,
which may allow its detection in laboratory experiments.

For fermionic dark matter candidates, or
those for which some symmetry prevents absorption,
the leading interaction with an SM target
is usually via scattering. An extensive
experimental program searching for the scattering
of heavy ($\gg$ nucleon mass) DM particles
has been in progress for decades, with the latest
detectors operating at the multi-ton scale~\cite{1903.03026}.
So far, no convincing DM signals have been seen,
which --- along with other observations --- has ruled out some of the most natural models
for electroweak-scale DM.

Recently, there have been efforts to extend
searches for DM scattering to smaller masses
($m_{\rm DM} \ll \GeV$). While such DM particles
would be produced in too large an abundance
via weak-scale thermal freeze-out~\cite{10.1103/PhysRevLett.39.165},
other early-universe production mechanisms are possible,
including freeze-out via lighter mediators~\cite{10.1016/j.nuclphysb.2004.01.015},
or thermal freeze-in~\cite{1911.03389,10.1103/PhysRevD.99.115009,10.1007/JHEP03(2010)080}.
The small energy depositions arising from such scatterings
mean that they would not be detectable
in standard WIMP direct detection experiments. 
Consequently, new experiments with lower energy
thresholds are required, and there has been an extensive theoretical
effort to identify suitable target materials
and detection strategies~\cite{Essig:2011nj,Essig:2012yx,Essig:2015cda,Hochberg:2015pha,Hochberg:2015fth,Hochberg:2016ntt,1708.08929,Derenzo:2016fse,Kurinsky:2019pgb,10.1103/PhysRevD.101.055004,Blanco:2019lrf,Trickle:2019nya,1910.02091,10.1103/PhysRevLett.123.151802,1909.09170,Griffin:2020lgd}.

An important set of models are those in which the DM scatters
through a mediator that couples to EM charge. This includes
models with a `dark photon' mediator, which are some
of the best-motivated and least-constrained possibilities
for light DM~\cite{10.1103/PhysRevD.96.115021}. In addition,
for models in which the mediator is not nucleophilic, it is often
the case that electrons dominate the target response,
so the scattering is very similar to that for a
mediator which couples to charge.
Recently, it has been emphasised~\cite{2101.08263,2101.08275} that,
for these models,
the scattering
rate of non-relativistic DM particles
in a material is controlled by the material's 
`energy loss function', $\Imag(-\epsilon_L^{-1})$, where
$\epsilon_L$ is the longitudinal dielectric permittivity~\cite{10.1103/PhysRev.113.1254}.
Electromagnetism constrains the properties of this energy
loss function; for example, it must satisfy `sum rules'
imposed by causality~\cite{10.1007/978-1-4757-5714-9,Dressel2002}, including
\begin{equation}
	\int_0^\infty \frac{d\omega}{\omega} \Imag\left(\frac{-1}{\epsilon_L(\omega,k)}\right)
	= \frac{\pi}{2} \left(1 - \frac{1}{\epsilon_L(0,k)}\right)
	\label{eqsr1}
\end{equation}
for any wavenumber $k$.
We point out that this sum rule imposes non-trivial 
constraints on the maximum DM scattering rate;
parametrically, it shows
that $\overline{\Gamma} \lesssim g_{\rm DM}^2 g_e^2
m_{\rm DM} v_{\rm DM}$, where
$g_{\rm DM}$ is the DM-mediator coupling,
$g_e$ is the electron-mediator coupling, and
$v_{\rm DM}$ is the typical DM velocity.
We derive precise bounds in Section~\ref{secsrate}.

In addition to serving as a sanity check,
the sum rule constraint 
sets an obvious target --- can we find materials
which come close to saturating the achievable scattering rates?
While this can be achieved with 
theoretically simple dielectric functions --- e.g.\
a plasmon pole at a frequency close to the DM kinetic
energy scale  ---
finding practical materials with the appropriate properties
can be difficult.

We discuss how conductor-dielectric heterostructures
could enable more optimized response functions,
compared to bulk materials, for DM masses $\lesssim \MeV$.
As well as analysing toy examples of periodic bulk heterostructures,
we analyse the very simple system of a single conductive layer.
This is the physical form taken by low-energy-threshold
detectors such as transition
edges sensors (TESs)~\cite{Dreyling-Eschweiler:2014mxa,Dreyling-Eschweiler:2015pja,Cabrera1998,Karasik:2012rb,Lita:08,Bastidon:2015aha},
microwave kinetic inductance detectors (MKIDs)~\cite{Mazin,DayLeduc,GaoMazin},
and superconducting nanowires (SNSPDs)~\cite{Rosfjord:06, Reddy20,
verma2020singlephoton, Wollman17}, and we illustrate
how the scattering rates of low-mass DM in such devices
may be orders of magnitude larger than a naive prediction
based on the bulk material properties may suggest.
While existing detectors usually 
have energy thresholds that are too high
for such effects to be significant (e.g.\ the
results reported in~\cite{yonit}),
they will become important for future devices.

In addition to modifying the overall scattering
rate, conductor-dielectric heterostructures also introduce
preferred directions, even for isotropic constituent materials,
resulting in directional dependence of the DM scattering
rate. Since the DM velocity distribution at
Earth is expected to be anisotropic,
this leads to modulation of the DM scattering
rate as the Earth rotates over the course of the day.
By introducing (or, for anisotropic materials,
potentially enhancing) this modulation, 
heterostructures could help to distinguish DM
signals from other backgrounds.

We also comment on how, when the dynamics of nuclei
are important, the DM scattering rate
for mediators which do not couple to EM charge can exceed
the sum rule bounds. This is true even for mediators
which only couple to electrons.
We illustrate how scattering
into acoustic phonons may have significantly higher
rates than into optical phonons, for mediators
with couplings not precisely those of a dark photon.


\section{Dark matter scattering}

Suppose that a DM state $\chi$ with mass $m_\chi$ couples to a (scalar or vector)
mediator of mass $m$, with coupling strength $g_\chi$. 
If the mediator couples to EM charges, with coupling strength
$g_e$ (i.e. $\LL \supset g_e X_\mu (\bar e \gamma^\mu
e - \bar p \gamma^\mu p)$ for its couplings to electrons
and protons), then as discussed in~\cite{2101.08263,2101.08275}
the scattering rate of sufficiently light,
non-relativistic DM in a material will be given by
\begin{equation}
	\Gamma 
	\simeq \frac{2 g_\chi^2 g_e^2}{e^2}
	\int \frac{d^3 k}{(2\pi)^3} \frac{k^2}{(k^2 + m^2)^2}
	{\rm Im}\left(\frac{-1}{\epsilon_L(\omega_k,k)}\right)
	\label{eqsrate1}
\end{equation}
where $\epsilon_L(\omega,k)$ is the material's longitudinal
permittivity in response to charge density perturbations
with frequency $\omega$ and wavevector $k$, and
$\omega_k = k \cdot v - k^2  /(2 m_\chi)$ 
is the energy loss corresponding to
momentum transfer $k$ from a DM particle with velocity $v$.
The integral is over momentum transfers $k$ such that
$\omega_k \ge 0$ (we are neglecting the temperature of 
the medium, so up-scattering does not occur).
We re-derive this result using in-medium effective propagators
in Appendix~\ref{apprates},
reviewing the approximations made.
Even if the mediator does not couple to SM charge,
as long as the electrons dominate the material
response, Eq.~\eqref{eqsrate1} will be a good
approximation (we discuss this further in Section~\ref{secother}).
The rate $\Gamma$ in Eq.~\eqref{eqsrate1}
corresponds to the scattering rate for a single
DM particle passing through the medium ---
in a volume $V$, the total scattering rate
will be given by $\Gamma_{\rm tot} = \Gamma n_\chi V$,
where $n_\chi$ is the DM number density.

\subsection{EM sum rules}
\label{secsrules}

To make Eq.~\eqref{eqsrate1} more precise,
we need to define $\epsilon_L$ more carefully.
We will suppose that we have some periodic structure, 
and will consider its response to a small longitudinal
free charge density perturbation ($\rho_f = \rho_0 e^{-i(
\omega t - k \cdot x)}$, with associated
current perturbation $J_f = J_0 \hat k e^{-i (\omega t - k \cdot x)}$,
where $k J_0 = \omega \rho_0$).\footnote{as usual,
complex quantities of this kind are used
as shorthand for their real parts.}
The `displacement' field is defined as 
$D = \hat k \rho_f/i k$, and the 
effective (inverse) permittivity is defined as the (position-dependent) linear response
function for the electric field, $E_i = \epsilon^{-1}_{ij} D_j$.
Then, we define the effective longitudinal dielectric
function as $\epsilon_L^{-1}(\omega,k) \equiv 
\overline{\hat k_i \hat k_j \epsilon^{-1}_{ij}}$, 
where the overline denotes spatial averaging.

At high enough frequencies, faster than the response
times of system's matter, $\epsilon_L^{-1}(\omega,k) \rightarrow 1$.
Consequently, via the Kramers-Kronig relations, we have
\begin{equation}
	1 - \epsilon_L^{-1}(0,k) 
	= \frac{2}{\pi} \int_0^\infty \frac{d\omega}{\omega} {\rm Im}
	\left(-\epsilon_L^{-1}(\omega,k)\right)
\end{equation}
($\epsilon_L^{-1}$ is real at $\omega = 0$, since
its imaginary part is an odd function of $\omega$).
There are also other sum rules~\cite{10.1007/978-1-4757-5714-9,10.1007/978-1-4757-5714-9}, as reviewed in Appendix~\ref{appsum},
but this one will be most useful for our purposes.
For a physical system in its ground state,
we should have $\Imag(-\epsilon_L^{-1}(\omega)) \ge 0$
for all frequencies,
corresponding to the system always absorbing (rather than emitting)
energy in response to a perturbation.\footnote{this
condition will not necessarily apply to a system
in a metastable state, such as the `magnetic
bubble chamber' proposal of~\cite{1701.06566}.} So,
integrating over any range of positive frequencies,
we should have
\begin{equation}
	\int \frac{d\omega}{\omega} {\rm Im}
	\left(-\epsilon_L^{-1}(\omega,k)\right)
	\le \frac{\pi}{2} \left(1- \epsilon_L^{-1}(0,k)\right)
	\label{eqsrule1}
\end{equation}
This lets us bound the integral over any range
of $\omega$ in terms of the (inverse)
static dielectric function
$\epsilon_L^{-1}(0,k)$ at the appropriate $k$.
Since $\epsilon_L^{-1}(0,k=0)$ must be non-negative
for a stable system~\cite{10.1103/RevModPhys.53.81}, it
follows
by continuity that $-\epsilon_L^{-1}(0,k)$ should be small
for $k$ much less than relevant momentum scales
in the system.
While it is possible for $-\epsilon_L^{-1}(0,k)$
to be positive for non-zero $k$ --- indeed, 
this is probably the case for some metals, such as aluminium~\cite{10.1103/RevModPhys.53.81}
--- it only becomes large and positive for systems
close to a point of instability, corresponding
to diverging response to a charge density 
perturbation~\cite{10.1103/RevModPhys.53.81}. As a result, for most
materials, the RHS of Eq.~\ref{eqsrule1}
will be $\OO(1)$.

\subsection{Scattering rates}
\label{secsrate}

We can use the sum-rule bound from Eq.~\eqref{eqsrule1}
to bound the DM scattering rate in our target system.
Starting from the scattering rate in Eq.~\eqref{eqsrate1},
we want to average over DM velocities, to obtain the average
scattering rate
\begin{align}
	\overline \Gamma
	&\simeq \frac{2 g_\chi^2 g_e^2}{e^2}
	\int \frac{d^3 k}{(2\pi)^3} \frac{k^2}{(k^2 + m^2)^2}
	\times \nonumber \\
	&\int d^3v \, p(v)
	{\rm Im}\left(\frac{-1}{\epsilon_L(\omega_{k,v},k)}\right)
	\label{eqscatt1}
\end{align}
where $p(v)$ is the probability distribution for DM velocities.

Properly, we should consider mounting our target in a particular
lab-frame orientation, and then changing this orientation relative
to the Galactic frame according to the Earth's rotation.
Instead, to simplify our initial calculations, we will average over
\emph{all} detector orientations relative to the Galactic
frame, which can equivalently be viewed as specifying
an appropriate \emph{isotropic} $p(v)$ in Eq.~\eqref{eqscatt1}
(we discuss anisotropic velocity distributions in
Section~\ref{sec_anisotropic}).
For isotropic materials, this gives the correct rate directly;
for other materials, it still provides the expected
rate for a randomly-chosen orientation.
In general, if we are allowed to tune the medium
properties and the initial DM velocity, we can
obtain arbitrarily large scattering rates,
via matching the on-shell momentum transfers
possible for the DM particle to the dispersion of
weakly-damped excitations in the medium (so that
we obtain resonant scattering at all momentum
transfers). However, if we are interested
in the scattering rate averaged over different directions,
such tuning is no longer possible, and as we will see, 
it is possible to set general limits on the scattering rate.

For a given $k$, the frequency $\omega_{k,v} 
= k \cdot v - k^2/(2 m_\chi)$ only depends on $|k|$ and $k \cdot v$,
so it only depends on the component $v_k$ of the velocity 
in the $\hat k$ direction.
\color{black}Thus, if we write $p_1(v_k)$ as the probability distribution
for the projection of the DM velocity onto a particular axis
(this is independent of the axis, since we are assuming that
$p(v)$ is isotropic), then
\begin{align}
	\overline \Gamma
	&\simeq \frac{2 g_\chi^2 g_e^2}{e^2}
	\int \frac{d^3 k}{(2\pi)^3} \frac{k^2}{(k^2 + m^2)^2} \nonumber \\
	&\times
	\int_{k/(2 m_\chi)}^\infty dv_k p_1(v_k)
	{\rm Im}\left(\frac{-1}{\epsilon_L(\omega_{k,v_k},k)}\right)
\end{align}
Since $d\omega_{k,v} = k dv_k$, this is equal to
\begin{align}
	\overline \Gamma &= \frac{2 g_\chi^2 g_e^2}{e^2}
	\int \frac{d^3 k}{(2\pi)^3} \frac{k}{(k^2 + m^2)^2} \nonumber \\
	&\times \int_0^\infty d\omega p_1(v_k(\omega))
	{\rm Im}\left(\frac{-1}{\epsilon_L(\omega,k)}\right).
	\label{eqp1}
\end{align}
Now, we can use the sum-rule bound from Eq.~\eqref{eqsrule1},
which implies that
\begin{align}
	&\int_0^\infty \frac{d\omega}{\omega} \omega p_1(v_k(\omega)) 
{\rm Im}\left(\frac{-1}{\epsilon_L(\omega,k)}\right) 
\nonumber \\
	&\le \frac{\pi}{2}
	\left(1 - \frac{1}{\epsilon_L(0,k)}\right)\max_\omega \left( \omega p_1 (v_k(\omega)) \right)
\end{align}
Consequently, if we write $g_0(k)\equiv 1 -\epsilon_L^{-1}(0,k)$, then
\begin{align}
	\overline \Gamma &\le
	\frac{\pi g_\chi^2 g_e^2}{e^2}
	\int \frac{d^3 k}{(2\pi)^3} \frac{k}{(k^2 + m^2)^2}
	g_0(k)
	\max_\omega \left( \omega p_1 (v_k(\omega)) \right)
	\nonumber \\
	&=
	\frac{g_\chi^2 g_e^2}{2\pi e^2}
	\int dk \frac{k^3}{(k^2 + m^2)^2}
	g_0(k)
	\max_\omega \left( \omega p_1 (v_k(\omega)) \right)
	\label{eqgmaxw}
\end{align}
(where the $g_0(k)$ in the second line is angle-averaged).
We can use an explicit form for $p_1$ to evaluate
this expression. 
For 
an isotropic velocity distribution
at a single speed $v_\chi$, i.e.\
$p(v) \propto \delta(|v| - v_\chi)$, we have
$p_1(v_k) = \frac{1}{2v_\chi} \mathbf{1}_{|v_k| \le v_\chi}$,
and so
$\max_\omega(\omega p_1(v_k(\omega)))
= \frac{k}{2} - \frac{k^2}{4 m_\chi v_\chi}$
for $k \le 2 m_\chi v_\chi$. Consequently, 
if we have an upper bound $g_0$ for $g_0(k)$,
then for a massless mediator ($m=0$),
\begin{equation}
	\overline \Gamma \le
	\frac{g_\chi^2 g_e^2 g_0}{2\pi e^2}
	\int_0^{2 m_\chi v_\chi} \frac{dk}{k}
	\left(\frac{k}{2} - \frac{k^2}{4 m_\chi v_\chi}\right)
	= \frac{g_\chi^2 g_e^2 g_0}{4 \pi e^2} m_\chi v_\chi
	\label{eq_gbm0}
\end{equation}
(if we have an explicit form for $g_0(k)$,
we can use this instead).
In the opposite limit, for a heavy mediator,
$m \gg 2 m_\chi v_\chi$, 
\begin{align}
	\overline \Gamma &\le
	\frac{g_\chi^2 g_e^2 g_0}{2\pi e^2}
	\int_0^{2 m_\chi v_\chi} dk \frac{k^3}{m^4}
	\left(\frac{k}{2} - \frac{k^2}{4 m_\chi v_\chi}\right)
	\nonumber \\
	&= \frac{g_\chi^2 g_e^2 g_0}{4 \pi e^2} \frac{16}{15} m_\chi v_\chi \left(\frac{m_\chi v_\chi}{m}\right)^4
\end{align}

These results for a single-speed distribution can be 
directly applied to scattering inside
a deep gravitational well, e.g. inside stars~\cite{wdr}, where the DM velocity
is close to the escape velocity.
Bounds for more complicated velocity distributions can
most simply be obtained by averaging over the single-speed
bounds. However, this is not necessarily optimal;
by using Eq.~\eqref{eqgmaxw} with a specific $p_1$ directly, we can generally
obtain tighter bounds.
For example, using 
the truncated Maxwell-Boltzmann velocity distribution
from Appendix~\ref{appdmvel} gives
\begin{equation}
	\overline\Gamma \le {\overline \Gamma}_{\rm opt} \equiv 0.68 \times \frac{g_\chi^2 g_e^2
	g_0}{4 \pi e^2} m_\chi v_0
	\label{eq_glim}
\end{equation}
for a massless mediator, where $v_0 \simeq 230 \kms$ is the
characteristic halo velocity scale
(the escape velocity $v_{\rm esc} \simeq 600 \kms$
affects the 0.68 coefficient, though only weakly),
and
\begin{equation}
\overline \Gamma \le 9.1 \times \frac{g_\chi^2 g_e^2
	g_0}{4 \pi e^2} m_\chi v_0 \left(\frac{m_\chi
	v_0}{m}\right)^4
\end{equation}
for a 
heavy mediator.

\begin{figure*}[t]
\includegraphics[width=0.49\textwidth]{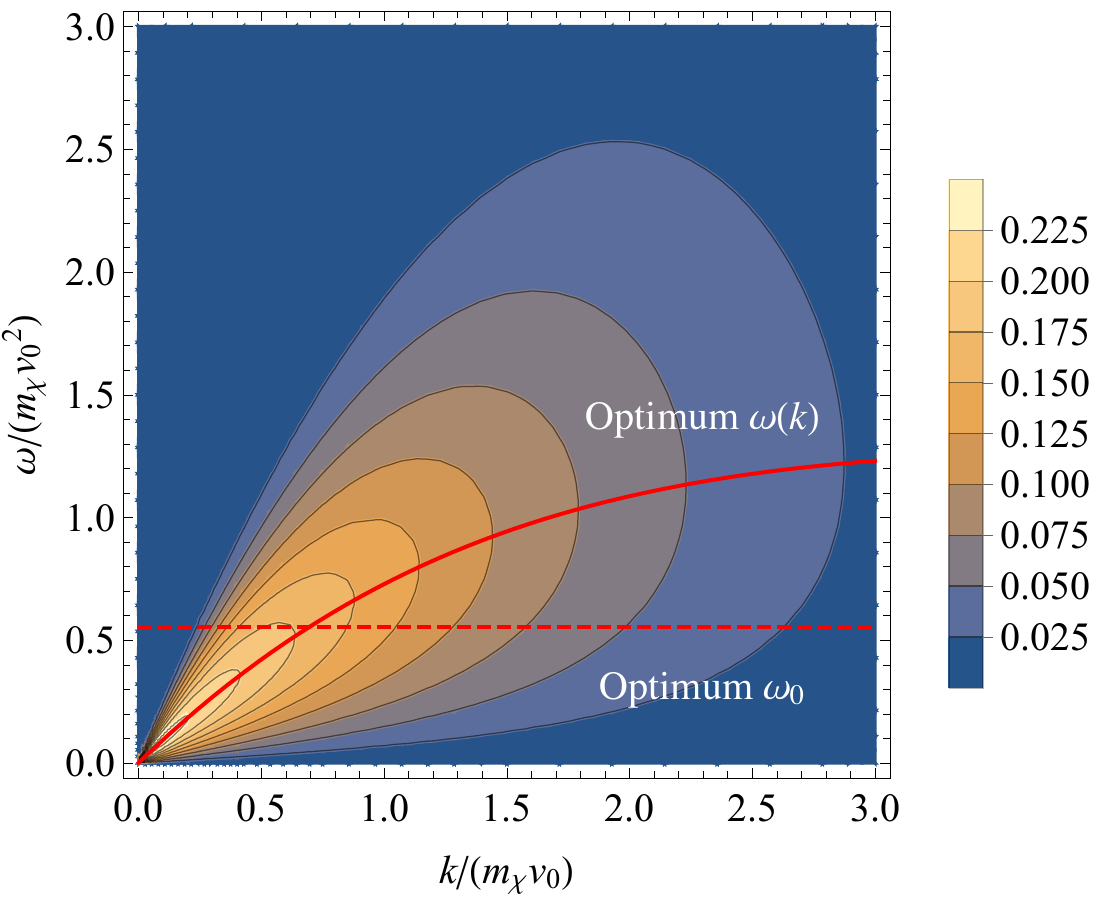}
\includegraphics[width=0.49\textwidth]{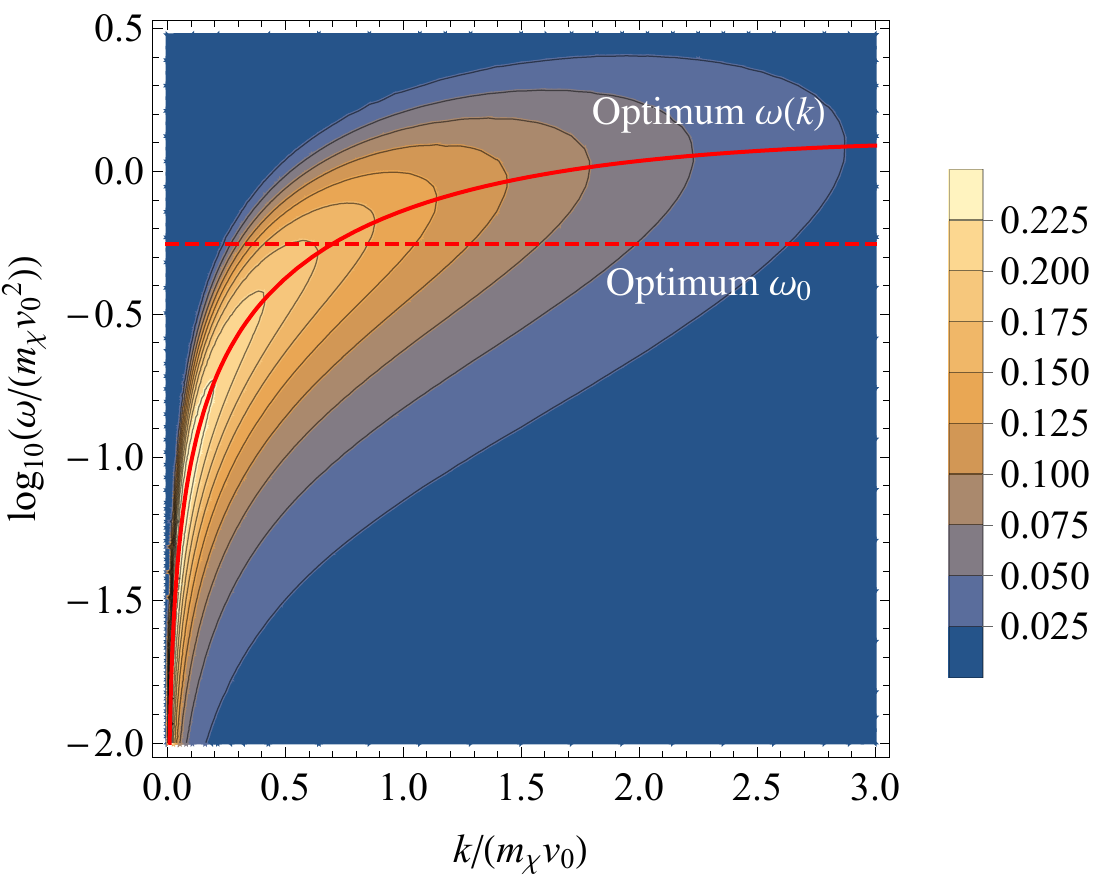}
	\caption{\emph{Left panel:} Integrand $I(\omega,k) = \frac{\omega}{k} p_1\left(\frac{\omega}{k} + \frac{k}{2 m_\chi}\right)$ in Eq.~\eqref{eq_integrand} for
	the DM scattering rate (with a massless mediator),
	taking $p_1$ for the truncated Maxwell-Boltzmann velocity distribution
	from Appendix~\ref{appdmvel}. Here, $m_\chi$ corresponds
	to the DM mass, and $v_0 = 230 \kms$ to the characteristic
	velocity dispersion in the DM halo. The solid red line
	shows the location of the integrand's maximum for fixed $k$,
	while the dashed red line corresponds
	to the $\omega_0$ such that the integral
	with $\omega(k) = \omega_0$ constant is largest.
	\emph{Right panel:} as per left panel, but with logarithmic
	$\omega$ axis; the integral in Eq.~\eqref{eq_integrand}
	is with respect to $\frac{d\omega}{\omega}$,
	for the given integrand.}
    \label{fig_integrand}
\end{figure*}

We can gain some more insight into the expression
in Eq.~\eqref{eqp1} by separating it into
integrals over momentum and solid angle,
\begin{align}
	\overline \Gamma &\simeq \frac{2 g_\chi^2 g_e^2}{e^2}
	\int dk \int \frac{d\omega}{\omega}  \overbrace{\left[\frac{k^3 \omega}{(k^2 + m^2)^2}
	p_1(v_k(\omega))\right]}^{I(\omega,k)}\nonumber \\
	&\times \int \frac{d\Omega_k}{(2\pi)^3}
	{\rm Im}\left(\frac{-1}{\epsilon_L(\omega,k)}\right)
	\label{eq_integrand}
\end{align}
Figure~\ref{fig_integrand} plots the integrand term $I(\omega,k)$
for a massless mediator ($m=0$),
and $p_1$ corresponding
to the truncated Maxwell-Boltzmann distribution from
Appendix~\ref{appdmvel}
(this will be our default $p_1$ going forwards).
For each $k$, the integral is maximized
by taking ${\rm Im}(-\epsilon_L^{-1})$ 
to be a delta function at the $\omega$
which maximizes $I(\omega,k)$; as expected, this 
is $\omega \simeq k v_0$ for $k \lesssim m_\chi v_0$.

One important feature is that, even for a massless
mediator, obtaining $\overline \Gamma$
of order the limit in Eq.~\eqref{eq_glim}
requires that most of the contribution
to Eq.~\eqref{eq_integrand}
comes from $k \sim m_\chi v_0$, $\omega \sim m_\chi v_0^2$
(assuming that $g_0(k)$ is order-1 throughout).
Conversely, if we scatter entirely into excitations
with energy below some threshold $\omega_s$,
where 
$\omega_s \ll m_\chi v_0^2$, then 
from Eq.~\eqref{eqgmaxw}, $\frac{\overline \Gamma}
{\overline \Gamma_{\rm opt}} \lesssim \frac{\omega_s}{m_\chi v_0^2}$.\footnote{This might naively seem surprising, since e.g.\
the Coulomb scattering rate in a plasma in usually dominated
by soft scatterings. However, if we are in this regime,
then we can increase the scattering rate by increasing the 
electron density --- this starts having diminishing returns
once the screening scale becomes small enough,
which corresponds to when soft scatterings
stop dominating the rate.}
This illustrates that, while schemes with very low
energy detection thresholds (such as~\cite{2104.05737,2108.05283})
are important for detecting low-velocity particles,
they do not offer a volumetric enhancement for detecting
virialized DM.

Another feature to note is that, while attaining
these bounds
requires that $\Imag(-\epsilon_L^{-1})$ is concentrated
at an optimum $\omega$ for each $\vec k$, 
as illustrated in Figure~\ref{fig_integrand},
taking it to be concentrated at a $k$-independent value
$\omega_0$ can be $\OO(1)$ optimal.
Numerically, we maximize the rate (for constant $g_0$)
by taking $\omega_0 \simeq 0.55 m_\chi v_0^2$,
which gives $\overline \Gamma \simeq 0.48 \times \frac{g_\chi^2 g_e^2 g_0}
{4\pi e^2} m_\chi v_0$, compared
to the bound in Eq.~\eqref{eq_glim}
which allowed the $\omega$ value to change with $k$.

\section{Material projections}

A wide range of papers have investigated the 
DM-electron scattering rates in different 
materials. By comparing these to the bounds
derived in the previous section, we can
sanity-check such calculations, as well as identifying
where significant improvements might be possible.

\begin{figure}[h]
\includegraphics[width=\columnwidth]{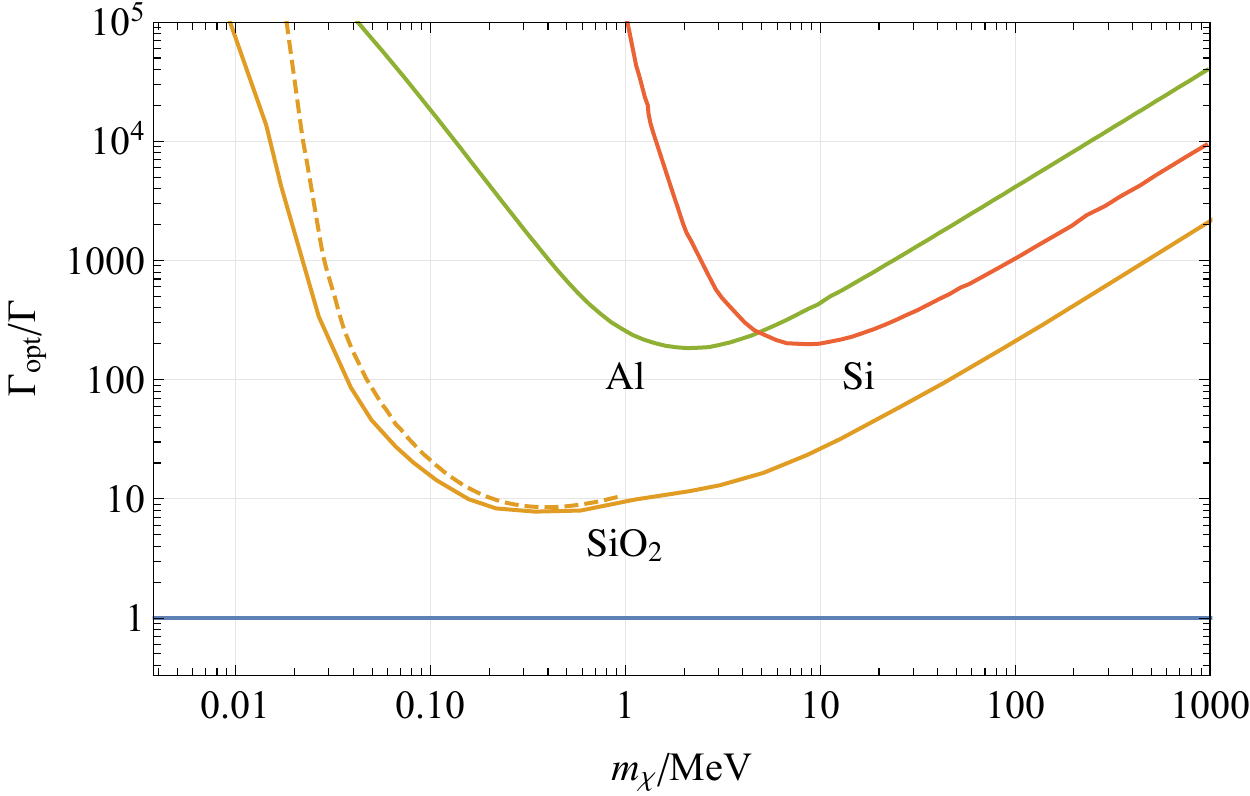}
    \caption{Comparison of projections
	for DM scattering rates (via a light dark 
	photon mediator) to the theoretical bound
	from Eq.~\eqref{eq_glim} (taking $g_0 = 1$).
	The Al and Si curves correspond to the
	projections for electronic excitations
	in aluminium
	and silicon
	from~\cite{2101.08275} (see also~\cite{2101.08263}),
	using approximations to the energy loss function
	(the Al curve corresponds to an energy
	threshold $\omega_{\rm min} = 10 \meV$,
	while the Silicon curve corresponds to excitations
	above the bandgap).
	These illustrate 
	that the scattering rates are significantly
	below the theoretical optimum,
	especially for DM masses
	$\ll \MeV$ or $\gg \MeV$.
	The solid SiO$_2$ curve corresponds to an 
	updated~\cite{kzhang} density functional theory projection
	from~\cite{10.1103/PhysRevD.101.055004}
	for phonons in quartz,
	and the dashed curve to an energy loss function
	calculation from~\cite{2104.12786}, illustrating that
	scattering into optical phonons can approach the
	sum rule limit more closely
	at suitable DM masses.}
    \label{figrel1}
\end{figure}

Figure~\ref{figrel1} compares a number 
of projections for different materials,
in the case of a light mediator coupling to EM charge,\footnote{this can be either a light dark photon mediator, or simply 
the SM photon itself (in the case of millicharged DM).}
to the bound on the per-volume scattering rate from
Eq.~\eqref{eq_glim}.
As discussed in Section~\ref{secsrate}, for sensitivity to
DM masses $\sim m_0$, a material with response
function concentrated around frequencies
$\sim m_0 v_0^2$ is almost as good as one with an
optimal response function.
At frequencies $\sim 10 - 100 \meV$, polar materials
can support optical phonon excitations, which
can have $\int \frac{d\omega}{\omega} \Imag(-\epsilon_L^{-1})$
up to $\sim 0.3$ over the relevant frequency range~\cite{10.1063/1674-0068/29/cjcp1605110} (at small $k$). 
Approximate energy-loss-function-based calculations
of the DM scattering rate in such materials~\cite{2104.12786},
as well as density functional theory calculations~\cite{10.1103/PhysRevD.101.055004,2102.09567},
indicate that they
are promising candidates for DM detection~\cite{1712.06598,1807.10291,10.1103/PhysRevD.101.055004,2102.09567}.
The SiO$_2$ curves in Figure~\ref{figrel1}
illustrate that, at DM masses in the $\sim 0.1 - 1\MeV$ range,
the scattering rate into optical phonons can be within
an order of magnitude of the sum rule bound.\footnote{The SiO$_2$ projection
in the current version (arXiv v2) of~\cite{10.1103/PhysRevD.101.055004}
corresponds to rates higher than the sum rule bound for
$m_\chi \sim {\rm few} \times 10^{-2} \MeV$, due to
a bug in the density functional theory calculation~\cite{kzhang},
illustrating the usefulness of the sum rule
bounds as a sanity check. The projections
in \cite{2102.09567} and Figure~\ref{figrel1}~\cite{kzhang} have
been updated to correct this.}

Other materials proposed for DM scattering
experiments, such as aluminium or semiconductors,
have response functions with most of their
support at energies $\gtrsim 10 \eV$~\cite{2101.08263,2101.08275,2104.12786}, for low $k$.
This means that their scattering rates are some way 
from the theoretical optimum, at all DM masses.
For DM masses $\lesssim 20 \MeV$, the response function is
concentrated at overly high frequencies,
while for higher DM masses,
the associated momentum transfers are $\gtrsim 10 \keV$,
which is large enough that $\Imag \epsilon_L^{-1}(\omega,k)$ is significantly reduced. This is illustrated in Figure~\ref{figrel1},
which shows that the scattering rates in aluminium and 
silicon 
are always at least two orders of magnitude
smaller than the volumetric optimum.\footnote{Some earlier
projections for superconducting materials, such
as those in~\cite{10.1103/PhysRevLett.123.151802},
did not take into account `screening' effects
--- effectively, the $1/|\epsilon|^2$ term in 
$\Imag(-1/\epsilon) = -\Imag(\epsilon)/|\epsilon|^2$
--- resulting in rates exceeding 
the sum rule bounds.
What was not widely appreciated until recently~\cite{2006.13909}
is that this `screening' suppression also
applies for a scalar mediator, as well
as vector mediators. An advantage of the energy
loss function formalism~\cite{2101.08263,2101.08275,10.1103/PhysRev.113.1254} is that it makes this physics transparent.}

The sum-rule rate in Figure~\ref{figrel1}
was obtained by setting $g_0 = 1$ in 
Eq.~\eqref{eq_glim}. This may not be
precisely correct, since as discussed below
Eq.~\eqref{eqsrule1}, $g_0(k) \equiv 1 - \epsilon_L^{-1}(0,k)$
may be $> 1$ for large enough $k$. 
However, while we do not have full $\epsilon_L^{-1}(0,k)$
calculations or measurements for these materials,
it does not seem likely that $\epsilon_L^{-1}(0,k)$
becomes large and negative --- for example, the 
values for aluminium presented in~\cite{10.1103/RevModPhys.53.81}
reach a minimum value of $\epsilon_L^{-1}(0,k) \simeq -0.2$
at $k$ around half of the reciprocal lattice vector.
Also, as mentioned above, $\epsilon_L^{-1}(0,k)$
should be non-negative for small enough $k$,
so for $k$ smaller than inverse lattice scales,
$g_0 \simeq 1$ should be a good approximation.
Overall, given that we are using Eq.~\eqref{eq_glim}
as a parametric bound, we do not expect
taking $g_0 = 1$ to be a problem.\footnote{An interesting
question is whether there are practical materials
for which $\epsilon_L^{-1}(0,k)$ is large and negative
at relevant $k$, so that $g_0(k) \gg 1$,
and the DM scattering rate is enhanced.
\cite{10.1103/RevModPhys.53.81} gives the example of molten
salt, which is predicted to have
$\epsilon_L^{-1}(0,k) \simeq -20$ for $k a \sim {\rm few}$,
where $a$ is the inter-atomic distance~\cite{10.1016/0375-9601(78)90013-0} (though such high-temperature systems
are unlikely to be useful for DM detection).\label{fn5}}

From Figure~\ref{fig_integrand}, we can see that,
to have sensitivity to a wide range of DM masses, 
a material's response function should be concentrated 
around $\omega \simeq k v_0$. While this brings
to mind the linear dispersion relations that can 
realised in e.g.\ Dirac materials (which have been 
proposed as targets for DM scattering~\cite{1708.08929,1806.06040,1909.09170,1910.02091}), 
explicit models for the permittivity in these
materials, such as those given in 
\cite{1708.08929,2101.08263}, do not have
$\Imag(-\epsilon_L(\omega,k)^{-1})$ concentrated in this way.\footnote{The rate
projections for the zero-gap model in \cite{1708.08929} seem 
to be unphysically high.}

Materials with good response function support in the $\sim \eV$
range may be useful for probing $\sim \MeV$ mass DM.
Possible examples include transparent conducting oxides~\cite{10.1116/1.580384},
or non-elemental superconductors~\cite{10.1103/PhysRevB.42.1969}. We are not aware
of proper measurements of the frequency- and momentum-dependent
loss function for such materials, so cannot
make reliable projections. However, low-momentum measurements
suggest that they may have good scattering rates.
Whether excitations deposited in such materials
can be reliably detected is, of course, 
a separate but important question.

Bulk materials with good response function support at 
very low frequencies, $\lesssim 50 \meV$,
are hard to achieve. However, heterostructures ---
structured combinations of different materials ---
can have different behaviour. Taking an extreme case,
conducting cavities at $\sim$ metre scales allow the low-$k$ response function
to be concentrated at $\sim \GHz$ frequencies.
For DM scattering, we are interested in the 
response function for $k \sim 10^3 \omega$
(as illustrated in Figure~\ref{fig_integrand}), 
so we need spatial structure at or below
the scale $k^{-1}$.
As we demonstrate below, straightforward combinations
of conductors and insulators could allow for tailored
response functions, concentrated at frequencies
well below those for the bulk materials themselves.

\section{Conducting layers}
\label{seccl}

\begin{figure}[t!]
\includegraphics[width=0.6\columnwidth]{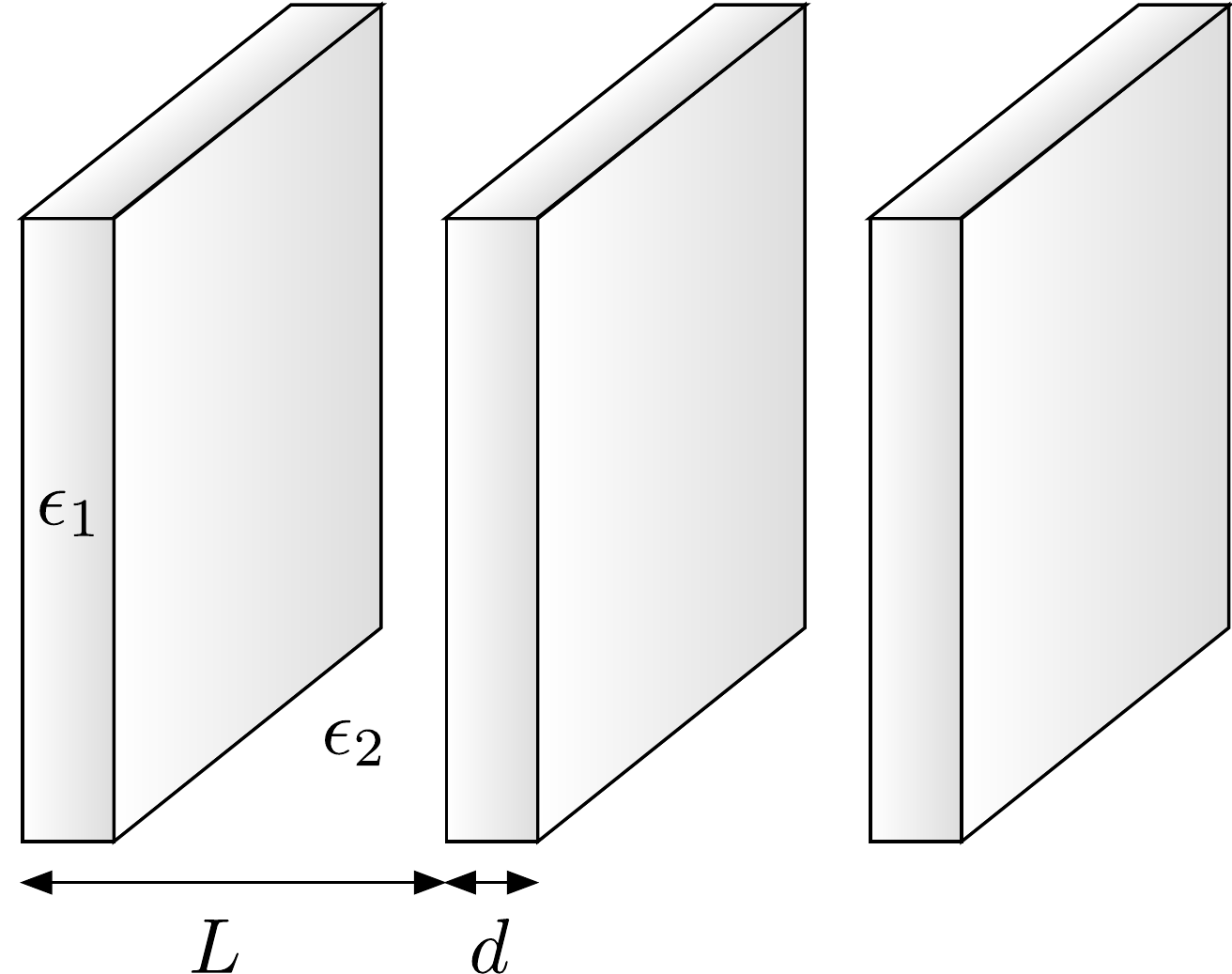}
    \caption{Diagram of a layered heterostructure,
	consisting of conductive layers of thickness $d$
	and permittivity $\epsilon_1$, placed with spacing
	$L$ in a dielectric medium of permittivity $\epsilon_2$.}
    \label{fig_layers1}
\end{figure}

One of the simplest examples of a metal-dielectric
heterostructure is a alternating array of planar layers, 
as illustrated in Figure~\ref{fig_layers1}. 
In response to a charge density perturbation with
long wavelength (compared to the layer separation), 
and wavevector parallel to the planes, the effective carrier
density should roughly be given by the metal's carrier
density, multiplied by the volumetric filling fraction of the metal
layers. Consequently, the effective plasma frequency
should be decreased from its bulk value, according
to the (square root of the) filling factor. Thus, even if the metal's bulk 
plasma frequency is significantly larger
than the DM kinetic energy scale, it may be possible
to increase the scattering rate by choosing the layer
thicknesses and spacings appropriately.

To analyse the response quantitatively,
we will assume that we are interested in non-relativistic
scatterings with $k \gg \omega$,
so that magnetic fields are unimportant,
and the dynamics are effectively electrostatic.
For simplicity, we will take the dielectric
function in each uniform medium to be isotropic
and $k$-independent, so we want to solve
$\nabla^2 \phi = - \rho_f / \epsilon_i(\omega)$,
where $\rho_f = \rho_0 e^{-i(\omega t - k \cdot x)}$
as in Section~\ref{secsrules}, and $\epsilon_i$
is the dielectric function for the medium.
At the medium boundaries, we need $\epsilon_1 \hat n \cdot \nabla
\phi_1 = \epsilon_2 \hat n \cdot \nabla \phi_2$, where
$\hat n$ is the normal to the boundary, and
$\phi_{1,2}$ are the solutions on each side.

Figure~\ref{fig_layers1} illustrates the geometry of
our setup.
Taking the layers to be normal to the $x$ direction,
we can, without loss of generality, write $k = k_x \hat x + k_z \hat z$.
Writing $\phi(x,z,t) = \psi(x) e^{-i(\omega t - k_x x - k_z z)}$,
we want to solve for $\psi$. Once we have this,
we can use it to compute
the electric field, from which we can derive
the effective longitudinal response function,
\begin{equation}
	\overline{\hat{k} \cdot E}
	= \epsilon^{-1}_{{\rm eff},kk} \frac{\rho_f}{i k}
	= \epsilon^{-1}_L \frac{\rho_f}{i k}
\end{equation}
The general expression for $\epsilon_L^{-1}$ is
rather complicated. However, in the $d/L \ll 1$
limit, where $d$ is the width
of the $\epsilon_1$ layers and $L-d$
the width of the $\epsilon_2$ layers, it has the simple form
\begin{equation}
	\epsilon^{-1}_L(\omega,k)	\simeq 
	\frac{(1-\frac{d}{L})\epsilon_1 + \frac{d}{L} \epsilon_2}
	{\epsilon_1 \epsilon_2 + (1-\frac{d}{L}) \frac{d}{L} (\epsilon_1 - \epsilon_2)^2 \frac{k_z^2}{k^2}}
	\label{eqel1}
\end{equation}
(where $\epsilon_1$ and $\epsilon_2$ are in general functions of $\omega$ and $k$). For $k_z = 0$, this is simply the volume-weighted sum
of $\epsilon_1^{-1}$ and $\epsilon_2^{-1}$, as we would expect.
However, for $k_z \neq 0$, the behaviour can be
significantly different (if
$k_x = 0$, we have $\epsilon_L^{-1} = \left(
\frac{d}{L}\epsilon_1 + \left(1 - \frac{d}{L}\right)\epsilon_2\right)^{-1}$). In particular, the response
poles will be at different frequencies.

If we take the limit $d/L \ll 1$, then
the denominator of Eq.~\eqref{eqel1} vanishes when $\epsilon_1 \simeq
-\epsilon_2 \frac{k_z^2}{k^2} \frac{d}{L}$,
or $\epsilon_1 \simeq -\epsilon_2\frac{k^2}{k_z^2} \frac{L}{d}$.
For example, if we take a simple Drude model,
$\epsilon_1 \simeq 1 - \omega_p^2/\omega^2$,
then the latter equality occurs for $\omega^2 \simeq \frac{1}{\epsilon_2} \frac{k_z^2}{k^2} \frac{d}{L} \omega_p^2$.
This corresponds to the expected result that the effective plasma
frequency (squared) is suppressed by the filling fraction of the metal.

\begin{figure}[t]
\includegraphics[width=\columnwidth]{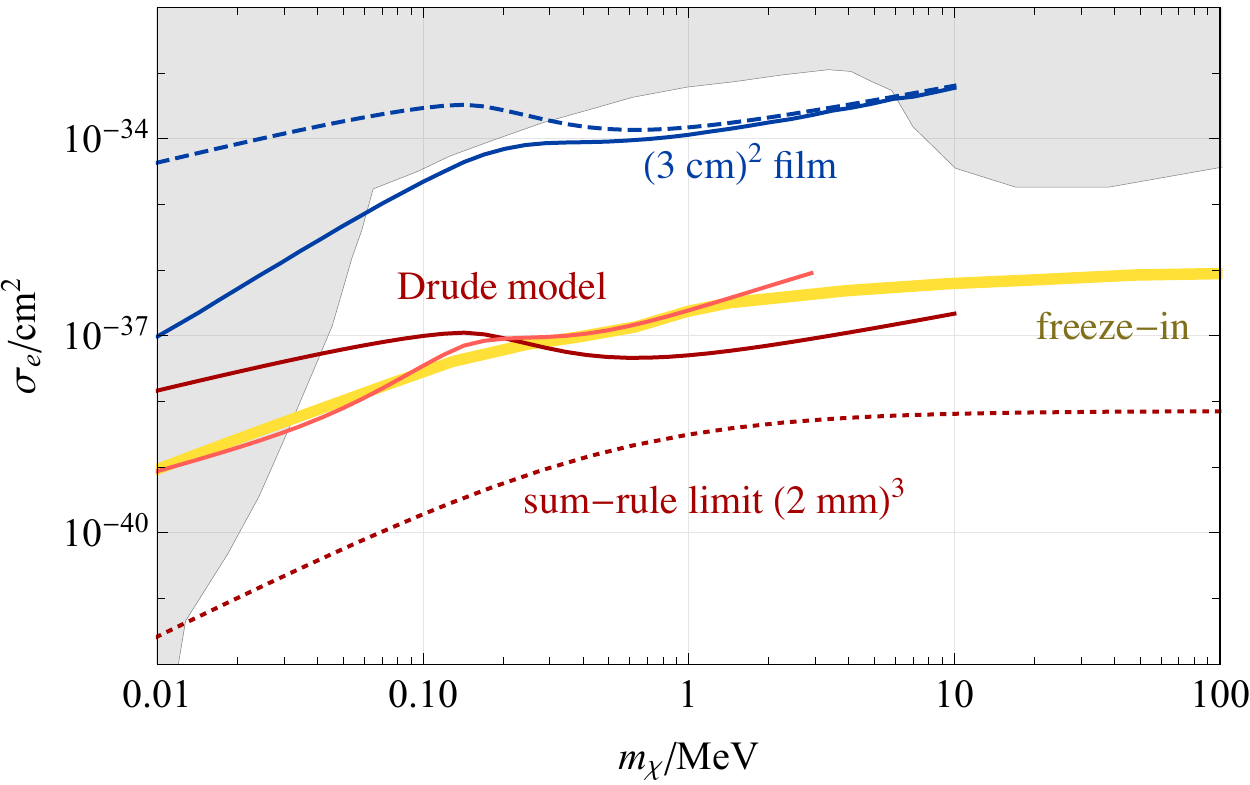}
    \caption{Plot of DM-electron scattering cross section
	sensitivity
	versus DM mass, assuming the DM couples through a 
	low-mass dark photon mediator.
	The gray shaded area shows the existing 
	constraints~\cite{10.1088/1475-7516/2014/02/029,10.1007/JHEP09(2018)051,10.1103/PhysRevLett.109.021301,10.1103/PhysRevD.96.043017}, while
	the yellow band shows the parameters for which 
	early-universe freeze-in~\cite{10.1103/PhysRevD.99.115009} 
	gives the correct DM abundance, assuming no pre-existing hidden
	sector population.
	The red curves correspond to the background-free sensitivity
	reach for a 1-year exposure with
	a $(2 {\rm \, mm})^3$ target volume
	(the sensitivity reach is taken to be the cross
	section that would result in 3 expected events
	during the exposure).
	The dotted red curve corresponds to the theoretical 
	section limit from Eq.~\eqref{eq_glim}.
	The dark red curve corresponds to a bulk material
	target, with Drude model permittivity as described in Section~\ref{seccl}.
	The lighter red curve corresponds to a
	layered material, with $d = 1 {\rm \, nm}$ thick layers of
	this material, alternating with $L - d = 4 {\rm \, nm}$ thick
	dielectric ($\epsilon = 1$) layers. 
	As the plot shows, this has worse sensitivity at larger DM
	masses, but better sensitivity at smaller masses.
	The blue curve corresponds to the sensitivity reach
	(for a background-free 1-year exposure)
	from a $3 \nm$ layer of material with area $(3 {\rm \, cm})^2$,
	for the same Drude model permittivity
	(taking an energy threshold $\omega_{\rm min} = 1 \meV$).
	The dashed blue curve shows the sensitivity reach from an 
	equivalent bulk volume of the same material,
	showing how, at low DM masses, taking into account
	the geometrical effects of the thin layer is very important.}
    \label{fig_sigma1}
\end{figure}

Consequently, compared to a bulk conductor, a
heterostructure of conductor-dielectric layers will have
its energy loss function concentrated at lower frequencies,
so can have better sensitivity to low-mass dark matter.
In Figure~\ref{fig_sigma1}, we illustrate this with a toy
example, taking the bulk conductor
to have a Drude-model permittivity, 
$\epsilon(\omega) = \epsilon_\infty \left(1 - 
\frac{\omega_p^2}{\omega(\omega + i \gamma)}\right)$,
where we take $\epsilon_\infty = 10$, 
$\omega_p = 0.5 \eV$, and $\gamma = 0.1 \omega_p$
(this is in rough analogy to 
the optical response function for NbN~\cite{nbn}).
Taking the DM to couple through a 
massless (or sufficiently low-mass) dark photon mediator,
the dark red curve shows the background-free
sensitivity reach for a $(2 {\rm \, mm})^3$ bulk volume
of this material with a one-year exposure. The lighter
red curve shows the sensitivity reach for the same volume
of a conductor-dielectric heterostructure,
where we take the conductive layer thicknesses to be
$d = 1 \nm$, and the dielectric ($\epsilon = 1$)
layer thicknesses to be $L - d = 4 \nm$. 
As expected, the layered material has better sensitivity at small DM
masses, compared to an equivalent volume of the bulk conductor,
and worse sensitivity at larger DM masses.

Especially at high DM masses, our toy model calculation
will not be realistic. For $m_\chi \gtrsim \MeV$,
the characteristic momentum scale is $m_\chi v_0 \gtrsim 
770 \eV = \frac{2\pi}{1.6 \nm}$, which is close enough to 
atomic lattice scales that the 
dielectric function will have
non-negligible momentum dependence~\cite{2101.08275,2104.12786}. However,
for $m_\chi \lesssim \MeV$, our calculations illustrate
the kind of behaviour expected.

Layered structures represent only one possible kind
of heterostructure. Other examples include
conductor-dielectric mixtures with random structures~\cite{Cai2010},
or granular inclusions (such as granular aluminium~\cite{Maleeva2018},
which is superconducting for small enough grain separations).
We leave investigation of such possibilities
to future work.

\section{Thin conducting films} \label{sec:thinfilm}

\begin{figure*}[t]
	\includegraphics[width=0.4\textwidth]{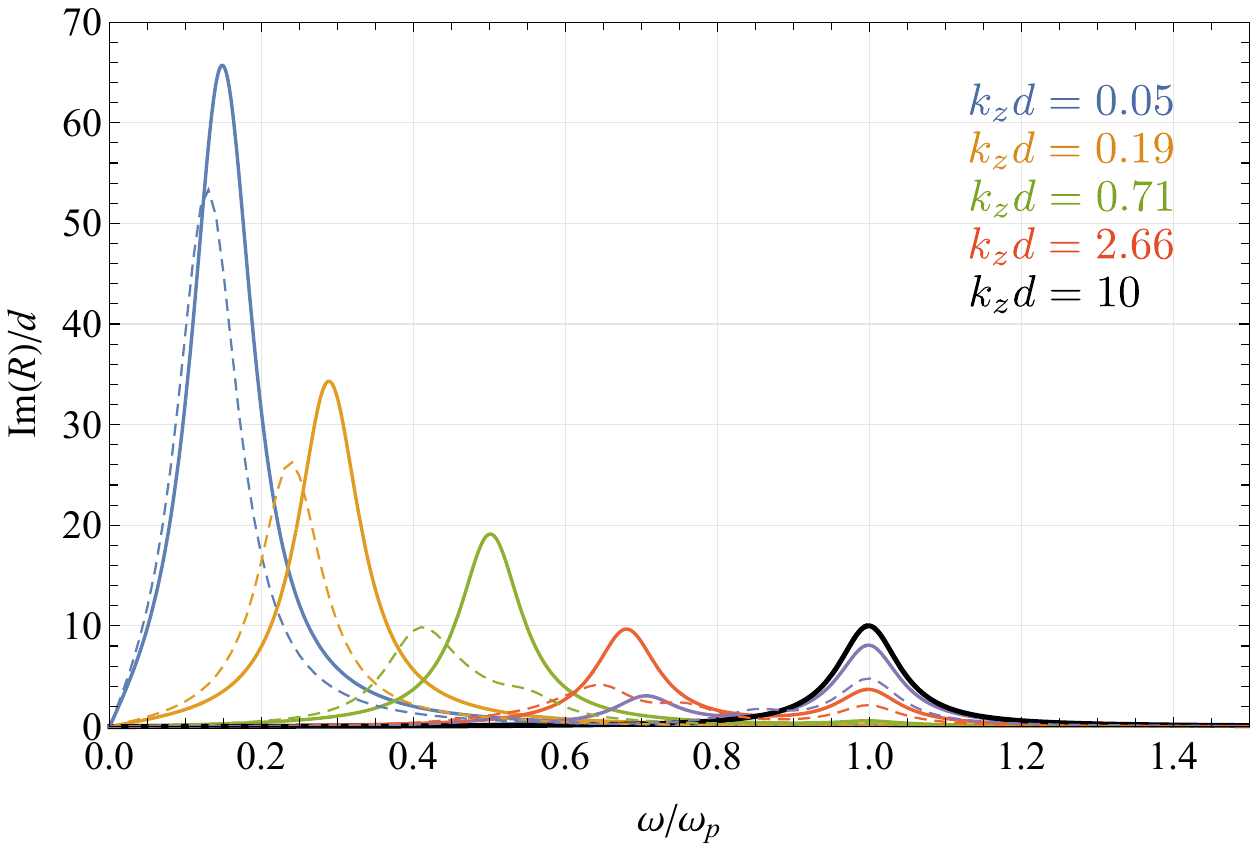}
	\includegraphics[width=0.4\textwidth]{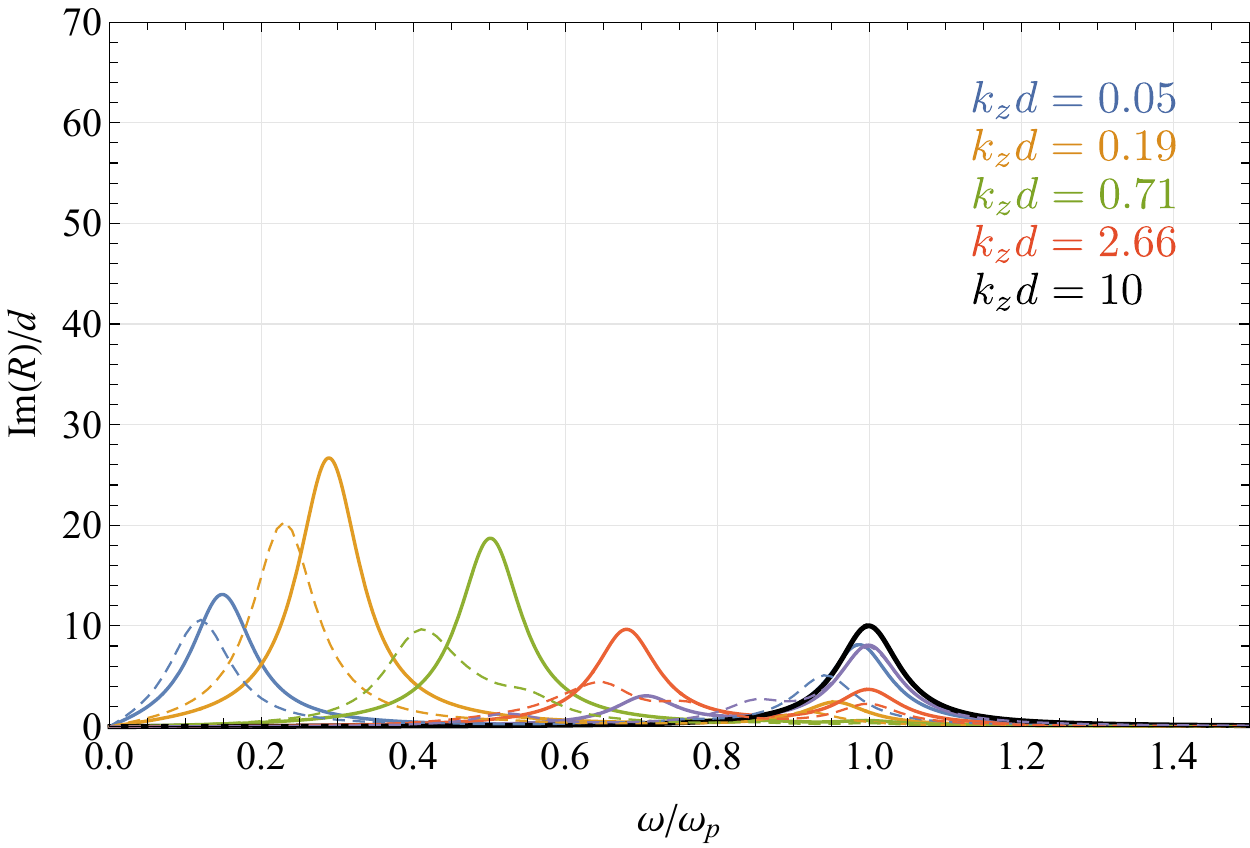}
	\caption{\emph{Left panel:} the solid curves show
	the thin-layer response function
	$R$ from Eq.~\eqref{eqr1}, taking a Drude model
	dielectric function
	$\epsilon = 1 - \frac{\omega_p^2}{
		\omega(\omega + 0.1 i \omega_p)}$
	and $k_x = 0$, for $k_z d$ values from 
	0.05 to 10. The black curve corresponds
	to the response function for an equivalent 
	volume of bulk material.
	The dashed curves corresponds to the 
	response for the rectangular wire geometry
	diagrammed in Figure~\ref{fig:snspdside}, 
	with $w = 6d$ and $h = 10d$.
	\emph{Right panel:} as per left panel,
	but taking $k_x d = 0.1$.}
	\label{fig_wr}
\end{figure*}

Volume-filling heterostructures, such as the layered materials
considered 
in the previous subsection, may be an interesting
option for DM scattering experiments. However, whether such 
materials can be constructed, and whether excitations
deposited in them can be detected,
are topics for future research. Nevertheless,
similar calculations apply to a more concrete
prospect --- detection of DM scattering in superconducting
thin-film detectors themselves.
 
Detectors for low-energy-threshold excitations,
such as TESs, MKIDs, and SNSPDs, often taken the
form of thin, superconducting layers.
In~\cite{10.1103/PhysRevLett.123.151802}, DM scattering in SNSPDs was considered, but their scattering
rate was based on bulk scattering rate
in the conductive material. Here, we point out that, 
for momentum transfers smaller than the inverse
thickness scale of the film,
the geometrical structure of the substrate-film-air
system needs to be taken into account.
Since typical film thickness are a few nm, these
effects are important for $m_\chi \lesssim \MeV$.\footnote{For absorption
of light bosonic DM, as opposed to DM scattering,
the typical momentum transfer is $\sim m_{\rm DM} v_0$,
which is much less than the energy transfer $\sim m_{\rm DM}$.
Consequently, for comparable energy depositions,
geometrical effects will be more important for DM
absorption; for example,
the bulk-material-based calculations in~\cite{10.1103/PhysRevLett.123.151802}
will be modified, as mentioned in~\cite{yonit}.\label{fn_abs}}

For simplicity, we first analyse the case
of a single, infinite layer with thickness $d$
and permittivity $\epsilon$, surrounded by a medium
of unit permittivity.
Since the thickness is finite, instead of
the per-volume scattering rate 
being set by $\Imag(-\epsilon_L^{-1})$,
the per-area scattering rate 
is set by $\Imag(R)$, where $R$ is
the appropriate response function with dimensions of length.
The time-averaged power absorbed
from a longitudinal free
charge perturbation $J_f$ is
$\langle P \rangle = \langle \int dV E \cdot J_f \rangle$, where $E$
is the electric field response, and angle brackets
denote time averaging. Assuming that the response
is effectively electrostatic, and writing
$E = - \nabla \phi = ((-\psi' + i k_x \psi)\hat x
+ i k_z \psi \hat z) e^{-i (\omega t - k \cdot x)}$
as above (taking the layer to be normal to the $x$ direction), we have
\begin{align}
	\langle P \rangle 
	&= - \frac{1}{2} 
		\int dV \frac{\omega}{k^2} 
		\Real \left[\rho_0^* (i k^2 \psi + k_x \psi')\right]
		\\ 
		&\equiv \frac{1}{2} A \frac{\omega}{k^2}|\rho_0|^2 \Imag(R)
\end{align}
where $A$ is the area we are considering.
For comparison, in a bulk material, we have
$P = \frac{1}{2} V \frac{\omega}{k^2} |\rho_0|^2
\Imag(-\epsilon_L^{-1})$.

The general expression for $R$ is somewhat complicated,
but if we consider an excitation with $k \parallel \hat z$, then
\begin{equation}
 R = \frac{1}{k \epsilon}\frac{2 - k d \coth(k d/2) + \epsilon (2 \epsilon - 4 - k d)}{\coth(k d/2) + \epsilon}
	\label{eqr1}
\end{equation}
The denominator vanishes when $\epsilon = - \coth(k d/2)$.
For a simple Drude model, $\epsilon(\omega) = 1-\omega_p^2/\omega^2$,
so for $k d \ll 1$, this corresponds to a resonant
frequency of $\omega^2 \simeq \frac{k d}{2} \omega_p^2$. Intuitively,
the relevant filling fraction is the ratio of the layer
thickness to the scattering wavelength.
There is also a divergence at $\epsilon = 0$, corresponding
to the bulk material resonance, but the contribution
of this is suppressed for $k d \ll 1$, since
$\frac{2 - k d \coth(k d/2)}{\coth(k d/2)}
= -\frac{(k d)^3}{12} + \OO((k d)^5)$.
When $k d \gg 1$, we have $R \simeq -d/\epsilon$,
so $\Imag(R) = d \, \Imag(-\epsilon^{-1})$, as expected.

As well as moving the response to lower frequencies,
the $k d\ll 1$ regime can also increase the frequency-integrated
response. For example, suppose that we work
in an approximation where $\epsilon \rightarrow \epsilon_\infty
> 1$ as $\omega \rightarrow \infty$.\footnote{This will not be true in a strict physical sense,
but can be a good approximation if e.g.\ there are some
effectively-decoupled, higher-frequency dynamics
which contribute a background permittivity $\epsilon_\infty$.
For example, the optical energy loss function for
SiO$_2$ has features below $\sim 200 \meV$, corresponding
to phonon dynamics, but then most of the $\int \frac{d\omega}{\omega}
\Imag (-\epsilon_L^{-1})$
integral comes from electronic excitations at $\omega\gtrsim 10 \eV$~\cite{10.1063/1674-0068/29/cjcp1605110}.} Then, for a bulk material,
we have $\int \frac{d\omega}{\omega} \Imag(-\epsilon_L^{-1}) \le 
\frac{\pi}{2} \epsilon_\infty^{-1} (1 - \epsilon_L^{-1}(0,k))$,
so the frequency-averaged absorption is suppressed
by $\epsilon_\infty^{-1}$.
However, for the thin layer, we have
\begin{equation}
	\int_0^\infty \frac{d\omega}{\omega}
	\Imag R(\omega) \simeq \frac{\pi}{2} k^{-1}
\end{equation}
for $d k \ll 1$. As well as being enhanced over 
the equivalent volume of an ideal bulk material
by $1/(d k)$, this is not suppressed by $\epsilon_\infty^{-1}$
(intuitively, this occurs because the response is spread across a
full wavelength around the layer, most of which
is in vacuum).

These features are illustrated in the left panel
of Figure~\ref{fig_wr}, which plots $\Imag R(\omega)$
for a simple Drude-model dielectric function, at
different $k$ values. At large $k d$, the response
is almost the same as for a bulk material, while
for small $k d$, it is moved to lower frequencies and
enhanced. 

The above formulae applied to the $k_x = 0$ case. If we
take the opposite limit, $k \parallel \hat x$, then we just 
have the usual bulk material response, $R = - d /\epsilon$.
For intermediate directions, we interpolate between
these two extremes, as illustrated in the
right-hand panel of Figure~\ref{fig_wr}.

The geometric effects discussed above can have important
consequences for the scattering rate of low-mass
DM.
As illustrated in Figure~\ref{fig_wr},
the shift of the $\Imag R(\omega)$ distribution
to lower frequencies means that a thin layer
can have a larger total scattering rate for low-mass
dark matter than a thicker layer, even if the latter
has larger volume.
In Figure~\ref{fig_sigma1}, the blue curve corresponds
the sensitivity reach for scattering from 
a $3 \nm$ thick film with area $(3 \cm)^2$,
assuming a background-free exposure of 1 year. We take the Drude-model permittivity 
from the previous subsection, and for extra realism, assume that the layer is mounted
on a dielectric substrate with permittivity $\epsilon = 11$
(corresponding to that of silica). The dotted blue curve corresponds to the sensitivity 
for the equivalent bulk volume of conductor.
For large DM masses, $\gtrsim \MeV$, the geometric effects
are only $\OO(1)$. However, for smaller masses,
they can increase the scattering rate by orders
of magnitude. 

It should be emphasised that the calculations presented here
apply to toy models. To calculate limits or sensitivity projections
for actual devices, more realistic models of the materials'
dielectric functions would be required --- ideally,
derived from actual measurements of such devices.\footnote{This is especially
important since, for such thin layers, one might expect
the response function to differ quite significantly from
that of a bulk material, due to 
surface effects~\cite{10.1364/OE.25.025574}.} Similarly,
the detectability of excitations absorbed in this way would
need to be quantified.
\cite{yonit}, which appeared on arXiv simultaneously with
this paper, uses 
techniques from this paper, and data from the 
tungsten silicide SNSPD used
in the LAMPOST dark photon DM detection experiment~\cite{lampost} to estimate limits on dark matter scattering with
electrons, as well as making projections for
future SNSPD experiments.
They find that the energy threshold for
this SNSPD is too high for geometrical effects
to be important in DM scattering, but that these should be significant
for future generations of SNSPDs
(for dark photon absorption within the SNSPD,
which~\cite{yonit} also estimates, geometric
effects will be more important,
as per Footnote~\ref{fn_abs}).

\subsection{Lossy dielectrics}

If a thin film is not surrounded by vacuum
(e.g.\ it is mounted on a substrate), then
the surrounding dielectric will have some
imaginary part to its permittivity. If we naively
integrate over the entire spatial volume, this may
result in the absorbed power being dominated by
the bulk absorption in the dielectric.

Because of how thin-film detectors
such as SNSPDs operate,
we are interested in
the rate of scatterings which deposit enough
energy \emph{into the conductor}, quickly enough, to register
as an excitation~\cite{10.3390/nano10061198}. Depending on the transport properties
of energy deposited in the dielectric material,
this rate may actually be dominated by bulk absorption
in the dielectric. For example, this is the design
principle behind detectors based on exciting optical
phonons in polar crystals~\cite{Knapen_2018,Griffin2018} --- the idea is that
such excitations decay into non-thermal
quasi-particles,
which then propagate until they are absorbed by
a superconducting detector.

To be conservative, we can restrict ourselves
to excitations where the energy is directly
deposited in the conductor itself. To do so, we
can calculate the electric potential response
$\psi$, as per above, and calculate how much
energy is dissipated inside the conductor given
this response.  If the dielectric
within $\sim k^{-1}$ of the conductor is not
very lossy, then this will be dominated by
the conductor, giving a result analogous the
lossless-dielectric case considered above. 
These considerations will also apply to
the absorption of light bosonic DM,
mentioned in footnote 6.

\subsection{Non-uniform geometries} \label{sec:npg}

\begin{figure}
    \centering
    \includegraphics[width=\columnwidth]{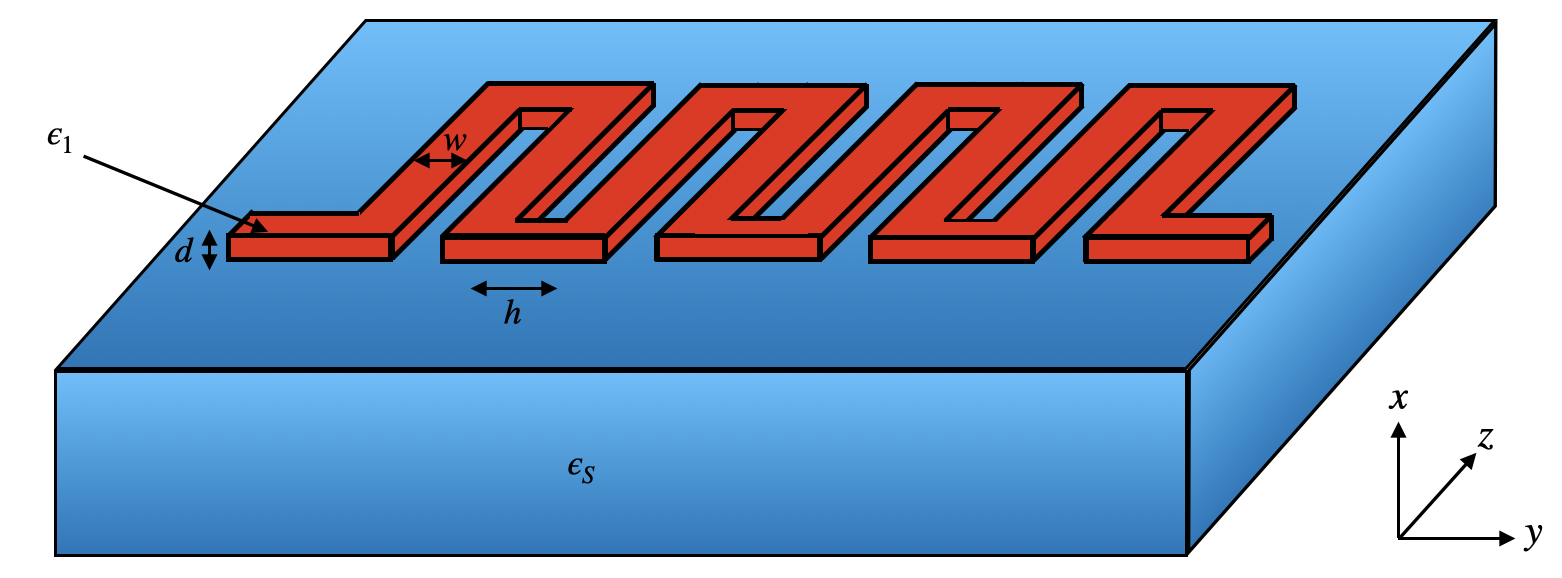}
    \includegraphics[width=\columnwidth]{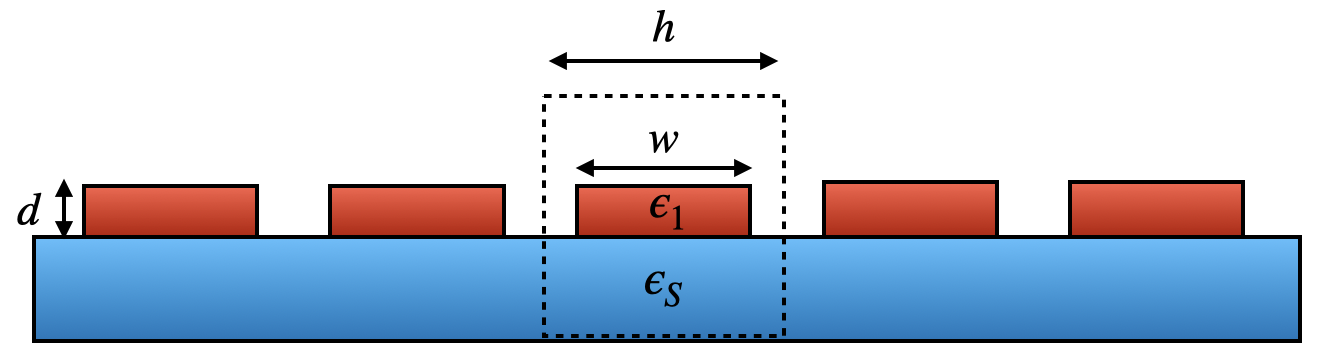}
    \caption{
		\emph{Top}: schematic of an superconducting nanowire single-photon detector (SNSPD) consisting of a superconducting wire meander (red) deposited on an electrically insulating substrate (blue).
		\emph{Bottom}: two-dimensional cross section (fixed $z$) of the SNSPD.}
    \label{fig:snspdside}
\end{figure}

The calculations above assumed an
infinite, uniform conductive plane. This can
be a good approximation when the inverse
momentum transfer is much smaller than
geometric features other than the thickness
of the film. However, some types of thin film detectors
have transverse structure on small scales.
For examples, SNSPDs~\cite{Rosfjord06,Reddy20,verma2020} use a wire meander with small width (10s to 100s of nm),
as illustrated in Figure~\ref{fig:snspdside}, so for momentum transfers
$\lesssim 0.1 \keV$, we might expect this
structure to have some effects on scattering 
rates. 

To estimate these effects, we can solve
for the 2D electrostatic response across
the wire's cross-section (the meander length is generally
long enough that end effects are unimportant).
Similarly to the 1D case, we want to solve
the Poisson equation, $\nabla \cdot (\epsilon
\nabla\phi)= -\rho$,
with $\phi(x,y,z,t) = \psi(x,y) e^{-i(\omega t - k \cdot x)}$.
Since doing this analytically is somewhat difficult
for general geometries, we can instead discretise
it on a 2D grid, and solve the resulting 
system of equations numerically to obtain
$\psi$. 

\begin{figure*}
\includegraphics[width=1\textwidth]{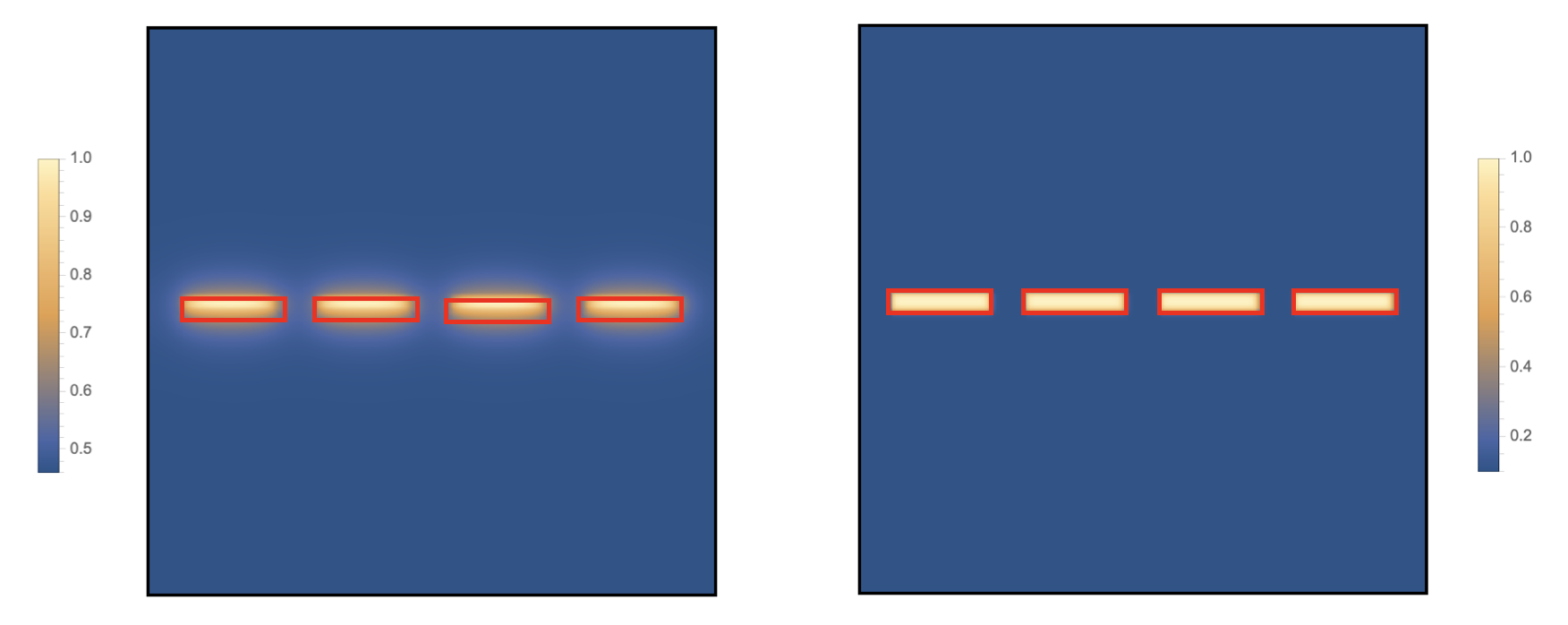}
	\caption{Numerical solution for the response function $\psi(x,y)$, discussed in Sec.~\ref{sec:npg} for a periodic array of rectangular cross-section wires (depicted by red rectangles) and $k_z d = 1$ (\emph{left panel}), $k_z d = 10$ (\emph{right panel}), where $d$ is the wire thickness. Orange
	corresponds to higher magnitudes for $\psi$, and blue to smaller magnitudes (color bars are in arbitrary units).}
    \label{fig_phi_integrand}
\end{figure*}

The simplest way to do this
is to impose periodic boundary conditions,
which means that we are effectively 
solving for the response of
a series of equally-spaced, 
infinitely-long wires. The corresponding
two-dimensional cross-section
is shown in the lower panel of Figure~\ref{fig:snspdside},
and some example numerical solutions
for $\psi(x,y)$ are shown in Figure~\ref{fig_phi_integrand}.
The latter illustrate that, for 
$k \gtrsim d^{-1}$, the response is dominantly
contained within the conductor,
while for $k \ll d^{-1}$, the response 
extends over a range $\sim 1/k$,
and approximates that from a uniform
layer. The dashed curves
in Figure~\ref{fig_wr} compare
the numerical scattering rates derived
from these $\psi$ solutions to the analytic
rates for an equivalent uniform film, illustrating
that these match well at $k \ll d^{-1}$,
while being volumetrically suppressed at larger $k$.

\section{Anisotropic velocity distributions}
\label{sec_anisotropic}

\begin{figure}[t]
	\includegraphics[width=\columnwidth]{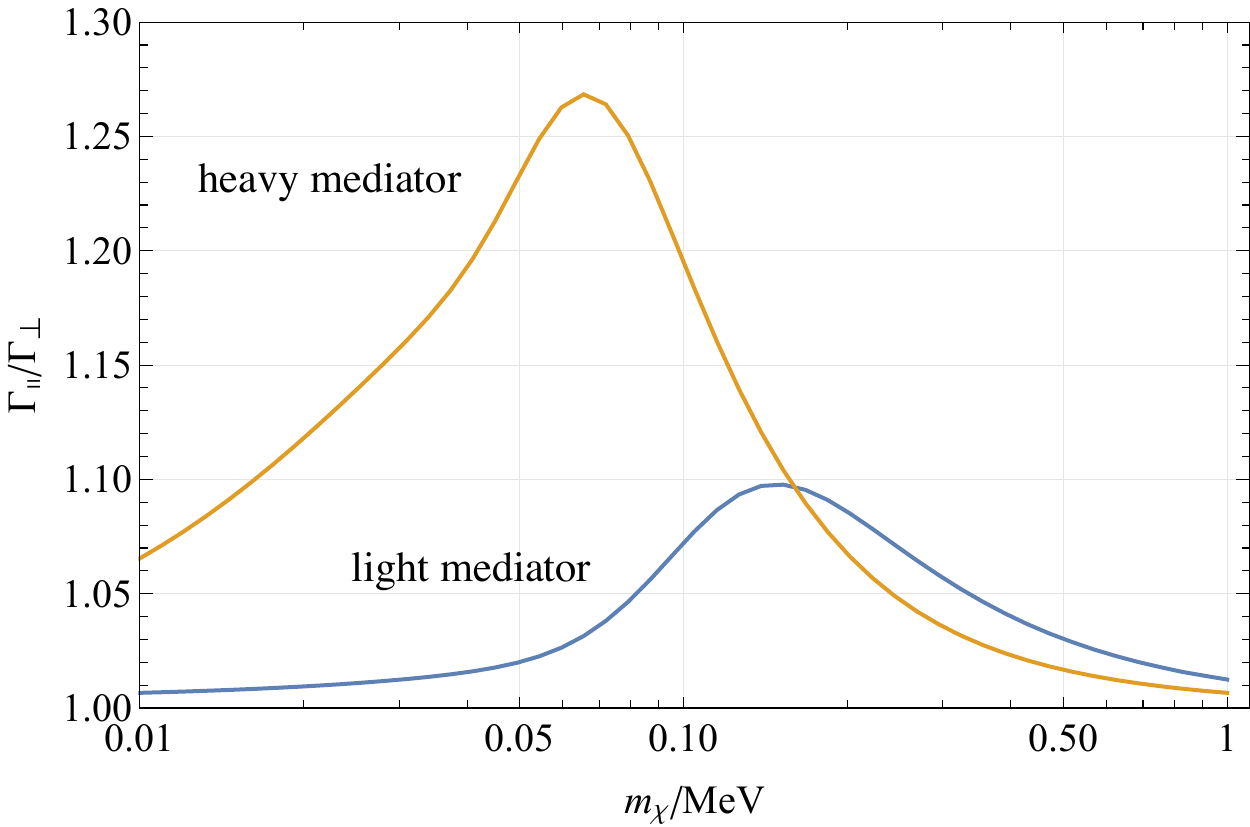}
	\caption{Ratio of velocity-averaged scattering
	rates for a thin film (with properties as in 
	Figure~\ref{fig_sigma1}),
	given a truncated Maxwell-Boltzmann DM
	velocity distribution (Appendix~\ref{appdmvel}),
	where $\Gamma_{\parallel}$ corresponds 
	to the average DM velocity being parallel
	to the film, versus
	perpendicular to the film for $\Gamma_{\perp}$
	(we take an energy threshold
	$\omega_{\rm min} = 1\meV$).
	The blue curve assumes a light mediator,
	with mass much smaller than relevant momentum
	transfer scales,
	while the orange curve assumes a 
	heavy mediator, with mass much larger
	than relevant momentum transfer scales.
	}
	\label{fig_vmod}
\end{figure}

In our calculations so far, we have adopted
the approximation of an isotropic DM 
velocity distribution.
However, it is expected that, due to the velocity
of the Earth with respect to the Galactic
frame, the DM velocity distribution
in the laboratory will
be significantly anisotropic (c.f.\ Appendix~\ref{appdmvel}).
 
Since the direction of this anisotropy
in the lab frame
will vary over each day as the Earth rotates,
a detector for which the scattering rate depends on
the direction of the incoming DM will
see a daily modulation in scattering rate.
The conductor-dielectric heterostructures
we have been considering do have 
anisotropic structures, so 
even in the approximation where the individual
materials have isotropic response functions,
the overall scattering rate will still
depend on the DM direction.

For the truncated Maxwell-Boltzmann
distribution described in Appendix~\ref{appdmvel},
and considering a thin-film detector
with the parameters given in Section~\ref{sec:thinfilm},
Figure~\ref{fig_vmod} shows the ratio of the DM
scattering rates for the extreme cases
of the velocity offset being parallel and perpendicular
to the film (other directions give intermediate
rates). For a light mediator, the effect on the overall
scattering rate is $\lesssim 10\%$.
Roughly speaking, this is because the phase space
volume of mostly-parallel momentum transfers
(which maximize collective effects)
is larger for parallel DM velocities,
but sits at smaller $k$ for perpendicular DM
velocities. Since small-$k$ scatterings are enhanced
for a light mediator, these effects
partially cancel out, reducing the difference
between perpendicular and parallel DM velocities.
For heavy mediators,
the ratio can be $\gtrsim 25\%$.

More complicated geometries, which
modify the $k$-dependence of the structure's
response, can also enhance the ratio
between scattering rates for different DM
directions. For example,
in the `SNSPD' geometry considered
in Section~\ref{sec:npg}, the wire width
$w$ provides an additional scale.
Numerical calculations, of the kind illustrated
in Figure~\ref{fig_phi_integrand}, indicate that
this could significantly increase
the directional dependence (integrating
over the full velocity distribution to obtain
the analogue of Figure~\ref{fig_vmod} would
be possible, but computationally expensive
--- we leave detailed investigations to future work).

It should be emphasised that the specific calculations
described above assume
that the conductor and dielectric materials both
have isotropic and $k$-independent $\epsilon$.
For large enough frequencies and momentum transfers,
this will be a poor approximation.
In addition, for thin enough layers, edge effects
may become important, even for materials
with fairly isotropic bulk permittivities, complicating
matters still further. As a result, 
Figure~\ref{fig_vmod} should not be taken as
a realistic prediction of daily modulation
amplitudes. However, it does illustrate that the geometrical properties of 
heterostructures can lead to significant
directional dependence, even in situations
where the bulk material properties would not
do so. This could help to distinguish a dark matter
signal from laboratory backgrounds.

As well as daily modulation due
to the rotation of the Earth, there is also
an annual effect caused by the Earth's changing 
velocity around the Sun. This leads
to the Earth's velocity relative to the Galactic frame
changing by $\sim 60 \kms$ over the course of the year,
with the RMS DM speed varying by 
$\sim 3\%$. For a light mediator,
this generally leads to small (percent-level) 
differences in the scattering rate, with
larger ($\OO(10\%)$) differences for
a heavy mediator.

\section{Other mediators}
\label{secother}

\begin{figure}[t!]
\includegraphics[width=\columnwidth]{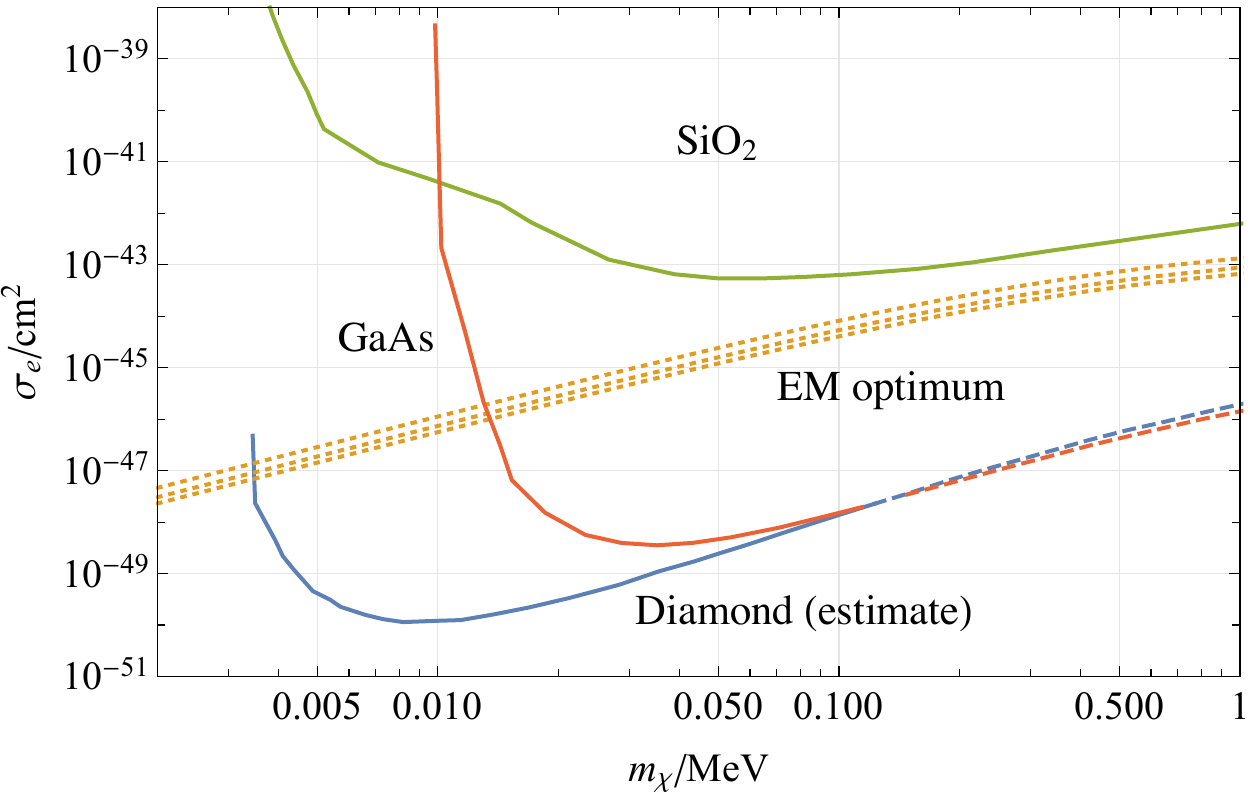}
    \caption{Plot of sensitivity estimates for a
	kg-year background-free exposure (i.e.\
	the cross section corresponding to 3 expected events), for
	different mediator couplings
	(taking an energy threshold
	$\omega_{\rm min} = 1 \meV$).
	The red curve shows the calculation
	for GaAs with a leptophilic scalar mediator
	from~\cite{Trickle:2019nya} 
	(the higher-DM-mass part of this curve
	is shown dashed, since neglected effects
	such as screening should be least important at small
	DM masses, where acoustic phonons dominate the rate).
	The blue curve shows the estimated
	rate for scattering in diamond
	via a leptophilic mediator, based on the 
	nucleophilic mediator result
	from~\cite{10.1103/PhysRevD.101.055004}.
	The green curve shows an 
	updated~\cite{kzhang} density functional theory projection
	from~\cite{10.1103/PhysRevD.101.055004}
	for phonons in SiO$_2$,
 assuming a light mediator coupling to EM charge.
	The orange dotted lines shows the optimum sensitivity
	for a mediator coupling to charge (the upper line
	corresponds to the density of GaAs,
	the middle line to the density of diamond, and the lower
	line to the density of SiO$_2$).}
    \label{fig_diamond}
\end{figure}

As mentioned above, the $\Imag(-\epsilon_L^{-1})$
prescription, and the associated sum rules,
apply in the case of a mediator that couples
to EM charge. For mediators with different SM
couplings, we need to consider the in-medium self-energy
of that mediator, rather than the SM photon,
as outlined in Appendix~\ref{apprates}.

In many circumstances, for light DM with a 
mediator that is not nucleophilic,
the material's response is dominated
by the more mobile electrons, and the $\Imag(-\epsilon_L^{-1})$
formulae give approximately the right results. This
is true for most excitations at frequencies $\gtrsim \eV$. 
However, for excitations in which nuclei play
a significant part, such as phonons, this will no longer
be the case. In particular, scattering into such excitations
can violate the sum rule bounds, and allow larger
rates than those for a dark photon mediator.

As an example, we can consider DM scattering via
a scalar mediator which couples to electrons,
but not to nucleons. 
For momentum transfers small enough compared
to the material's inverse lattice scale, 
the mediator's effect will correspond to 
a coherent forcing, and we can excite acoustic phonons,
rather than just optical phonons. 
If the material's sound speed is large
enough, then the enhancement due to the coherent coupling to
acoustic phonons can be greater than
the suppression due to the velocity mismatch between acoustic
phonons and typical DM velocities.
Consequently, acoustic phonons can dominate
the scattering rate. 

This case was analysed in~\cite{Griffin2018,Trickle:2019nya};
in particular,
\cite{Trickle:2019nya} performed
a density functional theory calculation for the scattering
rate in GaAs, plotted in Figure~\ref{fig_diamond}.
While this calculation did not take into account
screening, the scattering rate was dominated
by acoustic
phonons at small DM masses, for which screening should
not be an important effect. As 
Figure~\ref{fig_diamond} shows, the
scattering rate is orders of magnitude below
the sum rule limit for a mediator coupling to charge,
illustrating how these limits do not apply for
other types of coupling.
To obtain other examples, we can translate the scattering
rates for a nucleophilic scalar mediator 
calculated in \cite{10.1103/PhysRevD.101.055004};
at small enough momentum transfers, the coupling
of a leptophilic mediator to acoustic phonons
can be related to that of a nucleophilic mediator by
comparing the nucleon density to the electron density.
Figure~\ref{fig_diamond} shows this translation
for the diamond calculation from~\cite{10.1103/PhysRevD.101.055004}, illustrating how diamond's faster sound speed
results in larger scattering rates for small DM masses.

Even for a scalar mediator with equal and opposite
couplings to electrons and protons, the different
velocities of electrons and protons in materials
will mean that it has some non-zero coupling to neutral
bulk matter. Consequently, it can couple coherently
to acoustic phonons. Given that un-suppressed couplings
to acoustic phonons can result in very large
scattering rates (c.f.\ Figure~\ref{fig_diamond}),
we might wonder whether corrections suppressed by the SM
fermion velocities could dominate the scattering
rate at low DM masses, even for a scalar mediator coupling
to charge. To estimate this, 
we can note that $\bar f \gamma^0 f 
\simeq (1 + v^2/2) \bar f f$ for a non-relativistic fermion.
Typical inner-shell electron velocites
are $\sim Z \alpha$, while proton velocities
in nuclei are $\OO(0.1)$.
As a result, we expect typical deviations from
bulk neutrality at the $\OO(10^{-2})$ level,
with the consequence that scattering into acoustic phonons
may well be important for low-mass DM.
Of course,
it is difficult for DM models with a
non-nucleophilic scalar mediator to account
for all of dark matter without running into other
constraints~\cite{10.1103/PhysRevD.96.115021},
and for a small enough dark matter sub-component,
even the scattering rates possible with
a leptophilic mediator are somewhat hard
to probe experimentally~\cite{Trickle:2019nya},
given existing bounds. Consequently, models
where the scattering rate is further suppressed,
such as a scalar mediator coupling to charge,
would be even harder to see.

\section{Discussion}

In this paper, we have discussed two main 
topics; how electromagnetic sum rules
place bounds on the DM-electron scattering rate in 
materials, and how conductor-dielectric heterostructures
can increase the scattering rate of low-mass
DM, relative to bulk conductors.

To detect DM, there must be a high enough
DM-target scattering rate, and we must be able
to detect scatterings that occur. In most of this
paper, we have focussed on the first requirement,
but the second is also crucial.
The very simplest way to ensure detection
is for scatterings to deposit energy in the
detector material itself. 
We have pointed out that, for thin-film superconducting
detectors, which are one of the most promising
routes towards low energy thresholds,
geometric effects analogous to those for
periodic metal-dielectric heterostructures can have a
significant impact on the scattering rate
for low-mass DM. 

To achieve sensitivity to smaller DM couplings,
volume-filling targets will be required, and
further work would be needed to establish
whether heterostructures could be practically
useful. In particular, whether
suitable materials could be manufactured, and whether
excitations deposited in such materials
could be reliably detected, are not obvious.

As mentioned in footnote~\ref{fn5},
an interesting question is whether materials
with large and negative (inverse) static
dielectric function $\epsilon_L^{-1}(0,k)$,
which have larger frequency-integrated
energy loss functions,
could be useful for DM detection.
The most obvious examples of such systems,
such as materials near the threshold of crystallisation,
are high-temperature systems that are not suitable
for detecting small energy depositions.
We leave the investigation of possible alternatives
to future work.

Beyond applications to DM detection experiments in 
the laboratory, our sum rule analyses may also point to other
areas in which DM-SM scattering rate calculations
need revision. For example, many papers have
attempted to calculate the scattering rate for
DM passing through the dense interiors of neutron
stars or white dwarfs (see \cite{2004.14888,2010.13257,2104.14367} and references therein). 
However, while such calculations included Pauli blocking,
they did not include in-medium effects such
as screening.
For appropriate mediators, and sufficiently light
DM, these may significantly affect the scattering
rate. As an example, if we consider a heavy dark
photon mediator, then the appropriate
sum rule limit for the DM scattering rate
is significantly lower than both
the electron and nucleon scattering rates
given
in~\cite{2104.14367} for a white dwarf core,
for dark matter masses $\lesssim \MeV$.
These topics are explored in~\cite{wdr}.


\acknowledgments{We thank Karl Berggren, Yonit
Hochberg, Simon Knapen, 
Tongyan Lin, Tanner Trickle, and Zhengkang Zhang for
helpful conversations,
and Yonit Hochberg, Simon Knapen, Tongyan Lin and Zhengkang Zhang for comments 
on drafts. RL's research is supported in part by the National Science Foundation under Grant No.~PHYS-2014215, and the Gordon and Betty Moore Foundation Grant GBMF7946.
RL thanks the Caltech physics department for hospitality
during the completion of this work. AP's research is supported by National Science Foundation under Grant No.~PHYS-1720397 and the Gordon and Betty Moore Foundation Grant GBMF7946. AP acknowledges the
support of the Fletcher Jones Foundation and the National Science Foundation (NSF) Graduate Research Fellowship Program.}


\appendix

\section{DM scattering rate formulae}
\label{apprates}

In this appendix, we will give a condensed derivation
of the DM scattering rate formula in
Eq.~\eqref{eqsrate1}, explaining the approximations
being made. 

\toclesslab\subsection{Vector mediator}{appVM}

For concreteness, we will start by considering the case
of a DM fermion $\chi$, interacting with a vector mediator $X_\mu$,
with coupling $g_\chi X_\mu \bar \chi \gamma^\mu \chi$.

To evaluate the interaction rate of a DM fermion
travelling through a medium, we can compute the fermion's
in-medium self-energy. At leading order in $g_\chi$,
this is given by the imaginary part of 
the following diagram
\begin{center}
	\begin{tikzpicture}[line width=1] 
          \begin{scope}[shift={(2.4,0)}]
            \draw[f] (2*0.33,0)  node[above] {$ \chi $}
            -- (2*1.2,0);
            \node[] at (1.7,0.8) {$X$};
            \node[] at (4.3,0.8) {$X$};
            \node[] at (1.4,-0.3) {$P$};
            \node[] at (3.0,-0.3) {$P-Q$};
            \node[] at (4.5,-0.3) {$P$};
			\draw[f] (2*1,0) -- (2*2,0);
			  \draw[f] (2*2,0) -- (2*2.66,0) node[above] {$\chi$};
            \draw[v] (2*2,0) arc (0:180:2*0.5);
            \draw[pattern=north east lines,preaction={fill=white}] (2*1.5,2*0.52) circle (2*0.13);
          \end{scope}
	\end{tikzpicture}
\end{center} 
where the dashed circle represents the SM medium effects.
In the notation of~\cite{Bellac}, the cut self-energy is
\begin{equation}
	\Sigma^>(P)
	= g_\chi^2 \int \frac{d^4 Q}{(2\pi)^4} \gamma^\mu S_0^{>F}(P-Q) \gamma^\nu D_{\mu\nu}^>(Q)
\end{equation}
where $S_0^{>F}$ is the cut propagator for $\chi$
(in vacuum, since we assume that $\chi$ is weakly coupled
and its density is low), and
$D_{\mu\nu}^>$ is the in-medium cut propagator for
the mediator.
Using the $S_0^{>F}$ expression for a Dirac fermion, this is.
\begin{align}
	\Sigma^>(P)
	&= g_\chi^2 \int \frac{d^4 Q}{(2\pi)^4} 2\pi \delta((P-Q)^2 - m_\chi^2)\nonumber \\
	&\times \theta((P- Q)_0) \gamma^\mu (\slashed{P} - \slashed{Q} + m_\chi) \gamma^\nu
	D^>_{\mu\nu}(Q)
\end{align}
Writing $P = (E,p)$, the fermion's interaction rate
is given by~\cite{Bellac}
\begin{equation}
	\Gamma = \frac{1}{4 E} \tr\left[(\slashed{P} + m_\chi) \Sigma^>(P)\right]
\end{equation}
Evaluating the Dirac trace, we have
\begin{align}
	\tr \left[(\slashed{P} + m_\chi)\gamma^\mu (\slashed{P} - \slashed{Q} + m_\chi) \gamma^\nu\right]
	= \nonumber \\ 4 \left(P \cdot Q \eta^{\mu\nu} + P^\mu(P-Q)^\nu
	+ P^\nu(P-Q)^\mu\right)
	\label{eq_tr1}
\end{align}

Assuming that the mediator couples weakly to the SM
medium, the dominant contribution to $\Sigma^>$
comes from having only SM states in the shaded
circle. In this case, we
have (to leading order in the SM-mediator coupling)~\cite{Bellac}
\begin{align}
	D^>_{\mu\nu}(Q) &= D_{\mu\nu}^{>F}(Q) -  \\
	&\frac{2}{(Q^2 - m_X^2)^2}
	{\rm sgn}(q_0) (1 + f(q_0)) \Imag \Pi_{\mu\nu}(Q)
	\nonumber
\end{align}
where $Q = (q_0,q)$,
$D_{\mu\nu}^{>F}(Q)$ is the free cut propagator,
$f(E) \equiv (e^{E/T} - 1)^{-1}$ is the 
bosonic thermal occupation number for
the temperature $T$ of the medium,
and $\Pi_{\mu\nu}(Q)$ is the mediator's in-medium self-energy.
The real part of $\Pi_{\mu\nu}$ does not contribute,
since the integral for $\Sigma^>$
only receives contributions from
$Q^2 < 0$, where the mediator is always off-shell.
Going forwards, we will assume that the temperature
of our medium is negligible, so we can neglect 
the $f(q_0)$ term.

Using the fact that $Q_\mu \Pi^{\mu\nu}(Q) = 0$ (which
holds if the current we couple to is conserved),
we can write
\begin{align}
	\Gamma = \frac{2 g_\chi^2}{E}
	\int &\frac{d^3 q}{(2\pi)^3}
	\frac{1}{2 E'} \frac{1}{(Q^2 - m_X^2)^2}
	\times \nonumber \\
	&(- \Imag \Pi_{\mu\nu}(Q))
	\left(\frac{Q^2}{2} \eta^{\mu\nu} + 2 P^\mu P^\nu\right)
	\label{eq_gammap}
\end{align}
where $Q_\mu = (q_0,q)$ is such that $P - Q$ is on-shell.
Here, since we are assuming negligible medium temperature,
we integrate over $q$ such that $q_0 \ge 0$
(as upscattering cannot occur).

So far, our calculation has been fully relativistic.
If the incoming DM is non-relativistic
in the rest frame of the medium, so
$P \simeq m_\chi(1 + v_\chi^2/2, v_\chi)$,
then the only part of
the $\frac{Q^2}{2}\eta^{\mu\nu} + 2 P^\mu P^\nu$
term in Eq.~\eqref{eq_gammap} that is not suppressed for
$v_\chi \ll 1$ is the 00 component,
which is $\simeq 2 m_\chi^2$.
This picks out the longitudinal part $\Pi_L$
of $\Pi_{\mu\nu}$, since 
$\Pi_{00} = \frac{q^2}{Q^2} \Pi_L$
(note that our convention differs
from that of~\cite{Bellac},
which takes $\Pi_L = \Pi_{00}$).
As we will show below (in Section~\ref{sec_subleading}), considering only the
$\Imag \Pi_{00}$ term gives the leading contribution
for $v_\chi\ll 1$, with
\begin{equation}
	\Gamma \simeq 2 g_\chi^2 \int \frac{d^3 q}{(2\pi)^3}
	\frac{1}{(q^2 + m_X^2)^2} (-\Imag \Pi_L(q_0,q))
	\label{eq_gammal}
\end{equation}
since $q_0 \ll q$.

Specialising to a dark photon mediator with kinetic
mixing $\kappa$, we have
\begin{equation}
	\Imag \Pi_{\mu\nu}(Q) = \kappa^2 Q^4 \Imag (-i D_{\mu\nu}(Q))
\end{equation}
where $D_{\mu\nu}$ is the in-medium propagator
for the SM photon, in Lorenz gauge (for more details,
see Appendix D of \cite{wdr}).
The longitudinal dielectric function is related
to the longitudinal part $D_L$
(defined via $D_{\mu\nu} = -i D_L P^L_{\mu\nu}
+ \dots$, where $P^L_{\mu\nu}$ is the longitudinal projector) as $\epsilon_L^{-1}(Q) = - Q^2 D_L(Q)$~\cite{10.1016/0927-6505(92)90014-Q}, so
we can write the DM scattering rate as 
\begin{equation}
	\Gamma \simeq 2 g_\chi^2 \kappa^2
	\int \frac{d^3 q}{(2 \pi)^3} \frac{q^2}{(q^2 + m_X^2)^2}
	\Imag\left(\frac{-1}{\epsilon_L(\omega_q,q)}\right)
	\label{eqA9}
\end{equation}
where $\omega_q$ puts the DM particle on-shell, 
in agreement with Eq.~\eqref{eqsrate1}.

\toclesslab\subsubsection{Sub-leading contributions}{sec_subleading}

In the $\frac{Q^2}{2}\eta^{\mu\nu} + 2 P^\mu P^\nu$
expression, terms other than the 00 component
are suppressed by powers of $v_\chi \ll 1$. However,
if $\Imag \Pi_{\mu\nu}(Q)$ could be much larger
for these other components, then they could still
be important. For example, we might worry that,
since the exchange of a transverse photon
is unscreened for small $q_0$, whereas
longitudinal exchange is screened~\cite{Bellac}, 
transverse contributions might become important.
However, for the case of a dark
photon mediator, $D_{\mu\nu}$ obeys
additional sum rules which mean that the velocity-averaged
scattering rate is dominated
by the longitudinal-exchange expression 
in Eq.~\eqref{eq_gammal} above, as we show here.

If $J_\mu = J^{(0)}_\mu e^{i Q \cdot x}$
is a charge density perturbation with wavevector $Q_\mu$,
then the medium's EM field response 
is given by $A_\mu = -i D^R_{\mu\nu}(Q)J^\nu$,
where $D^R_{\mu\nu}$ is the retarded in-medium propagator
for the photon.
Consequently, writing $R_{\mu\nu} \equiv i D^R_{\mu\nu}$,
the time-averaged power extracted from 
the charge perturbation is
set by $i q_0 J_\mu^* A^\mu + {\rm h.c.} = q_0 (J^\mu)^* J^\nu \Imag R_{\mu\nu}$.
If the medium is in its ground state,
then this power should be positive for any perturbing
current --- that is, the medium should absorb energy from
the perturbation, rather than emitting energy.
Current conservation $\partial_\mu J^\mu = 0$ implies
that $Q_\mu J^\mu = 0$,
so for any vector $\epsilon^\mu$
perpendicular to $Q_\mu$, we should have
that $(\epsilon^\mu)^* \epsilon^\nu \Imag R_{\mu\nu} \ge 0$.

At high enough frequencies, corresponding to 
timescales much faster than the response
times of the system's matter, the response function
should be almost equal to that in vacuum.
If we fix the spatial vector $q$, then for
the spatial directions transverse
to $q$, the response
function at large $q_0$ will
be $R_{ij} \simeq -\frac{\delta_{ij}}{q_0^2}$.
Consequently,
the Kramers-Kronig
relations give
\begin{equation}
	R_{ij}(0,q) = \frac{2}{\pi} \int_0^\infty
	\frac{d q_0}{q_0}
	\Imag R_{ij}(q_0,q)
	\label{eq_sr1}
\end{equation}
This equation is not immediately useful as a sum
rule, since the integrand may not always be non-negative.
However, if we take $q = (0,0,q)$ (without loss of generality), then
$\Imag R_{11}$ and $\Imag R_{22}$ correspond
to $(\epsilon^\mu)^* \epsilon^\nu \Imag R_{\mu\nu}$
for a spatial vector $\epsilon_\mu$ which
is perpendicular to $Q_\mu$ for all $q_0$,
so are always positive. Consequently, we have 
\begin{equation}
	R_{11}(0,q) = 
	\frac{2}{\pi} \int_0^\infty
	\frac{d q_0}{q_0}
	\Imag R_{11}(q_0,q)
	\label{eq_transversesum}
\end{equation}
where the integrand is always positive,
and similarly for $R_{22}$, giving us sum rules
for the transverse components
of the propagator.

If we are interested in the $v_\chi$-averaged
scattering rate,
for an isotropic $v_\chi$ distribution,
then the appropriate integrand arising from Eq.~\eqref{eq_gammap}
is
\begin{align}
	\Bigg\langle
	&\frac{Q^2}{2} \eta^{\mu\nu}\Imag R_{\mu\nu}
	+ 2 P_0^2 \Imag R_{00}\nonumber \\
	& - 4 P_0 P_3 \Imag R_{03} + 2 P_3^2 \Imag R_{33}\nonumber \\
	&+ 2 P_1^2 \Imag R_{11}
	+ 2 P_2^2 \Imag R_{22}
	\Bigg\rangle
\end{align}
where the angle brackets denote averaging
over $v_\chi$.
The contributions from the $\Imag R_{11},\Imag R_{22}$ terms
can be bounded using the sum rule from Eq.~\eqref{eq_transversesum}. The $\Imag R_{03},\Imag R_{33}$ terms
can be related to $\epsilon_L^{-1}$,
so can be bounded using the sum rule
from Eq.~\eqref{eqsrule1}. 
Doing so, we find that all of the other terms
have $v_\chi$-suppressed contributions compared
to
$2 P_0^2 \Imag R_{00}$, which gives
rise to Eq.~\eqref{eqA9}.

\vspace{2cm}

If we are interested in the $v_\chi$-averaged
scattering rate,
for an isotropic $v_\chi$ distribution,
then the appropriate integrand arising from Eq.~\eqref{eq_gammap}
is
\begin{align}
	\Bigg\langle
	&\frac{Q^2}{2} \eta^{\mu\nu}\Imag D_{\mu\nu}
	+ 2 P_0^2 \Imag D_{00}\nonumber \\
	& - 4 P_0 P_3 \Imag D_{03} + 2 P_3^2 \Imag D_{33}\nonumber \\
	&+ 2 P_1^2 \Imag D_{11}
	+ 2 P_2^2 \Imag D_{22}
	\Bigg\rangle
\end{align}
where the angle brackets denote averaging
over $v_\chi$ (the use of the retarded propagator
does not make a difference here).
The contributions from the $\Imag D_{11},\Imag D_{22}$ terms
can be bounded using the sum rule from Eq.~\eqref{eq_transversesum}. The $\Imag D_{03},\Imag D_{33}$ terms
can be related to $\epsilon_L^{-1}$,
so can be bounded using the sum rule
from Eq.~\eqref{eqsrule1}. 
Doing so, we find that all of the other terms
have $v_\chi$-suppressed contributions compared
to
$2 P_0^2 \Imag D_{00}$, which gives
rise to Eq.~\eqref{eqA9}.

\toclesslab\subsection{Scalar mediator}{appSM}

For a scalar mediator $\phi$, we have
\begin{align}
	\Sigma^>(P)
	&= g_\chi^2 \int \frac{d^4 Q}{(2\pi)^4} 2\pi \delta((P-Q)^2 - m_\chi^2) \\
	&\times \theta((P- Q)_0) (\slashed{P} - \slashed{Q} + m_\chi)
	D^>(Q)
\end{align}
and 
\begin{align}
	D^>(Q) &= D^{>F}(Q) -  \\
	&\frac{2}{(Q^2 - m_\phi^2)^2}
	{\rm sgn}(q_0) (1 + f(q_0)) \Imag \Pi(Q)
	\nonumber
\end{align}
where $\Pi(Q)$ is the mediator's in-medium 
self-energy.
Evaluating the Dirac trace for the scattering rate,
\begin{align}
	\tr \left[(\slashed{P} + m_\chi)(\slashed{P} - \slashed{Q} + m_\chi) \right]
	=  4 \left(2 m_\chi^2 - P \cdot Q\right)
	\label{eq_tr2}
\end{align}
For a non-relativistic $P$, this is the same,
to leading order, as the leading $\mu=\nu=0$ component of the
vector mediator's trace, so we have
\begin{equation}
	\Gamma \simeq -2 g_\chi^2 \int \frac{d^3 q}{(2\pi)^3}
	 \frac{1}
	{(q^2 + m_\phi^2)^2}
	\Imag \Pi(q_0,q)
\end{equation}
If $\phi$ couples with opposite strength to electrons
and protons, then to leading order in the velocities
of the electrons and protons, $\Imag \Pi$ is the same
as the $\Imag \Pi_L$ expression for a vector mediator.

We would obtain similar expressions if we considered scalar
DM. 
Compared to the non-relativistic calculations
in~\cite{2101.08263,2101.08275}, which were spin-agnostic
by construction, our calculations illustrate
how to incorporate relativistic corrections, as
well as mediators with different couplings.

\section{DM velocity distribution}
\label{appdmvel}

The DM scattering rate in an experiment will 
depend on the DM velocity distribution at Earth.
While we do not have precise measurements of this 
distribution, a common model assumed in the direct
detection literature is the truncated Maxwell-Boltzmann
distribution~\cite{10.1103/PhysRevD.82.023530,1603.03797}.
Writing $v$ as the velocity relative to Earth,
the DM velocity distribution is taken to be
\begin{equation}
	f(v) = \frac{1}{N_0} e^{-(v+v_e)^2/v_0^2}
	\Theta(v_{\rm esc} - |v + v_e|)
\end{equation}
where
\begin{equation}
	N_0 = \pi^{3/2} v_0^2
	\left[
		v_0 {\rm erf}\left(\frac{v_{\rm esc}}{v_0}\right)
		- \frac{2 v_{\rm esc}}{\sqrt \pi}
		\exp\left(\frac{- v_{\rm esc}^2}{v_0^2}\right)\right]
\end{equation}
Standard values taken for these parameters are
$v_0 \simeq 230 \kms$,
$v_e \simeq 240 \kms$, $v_{\rm esc} \simeq 600 \kms$~\cite{10.1103/PhysRevD.101.055004}.

\begin{figure*}[t]
	\includegraphics[width=0.43\textwidth]{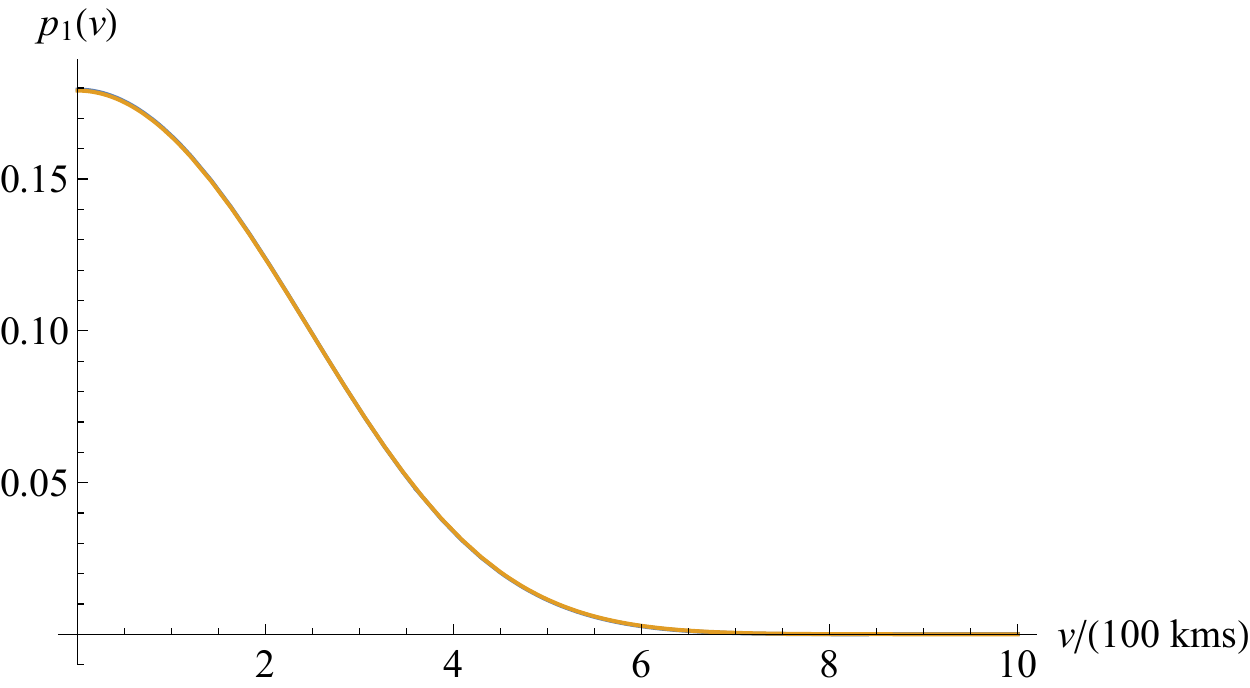}
	\includegraphics[width=0.43\textwidth]{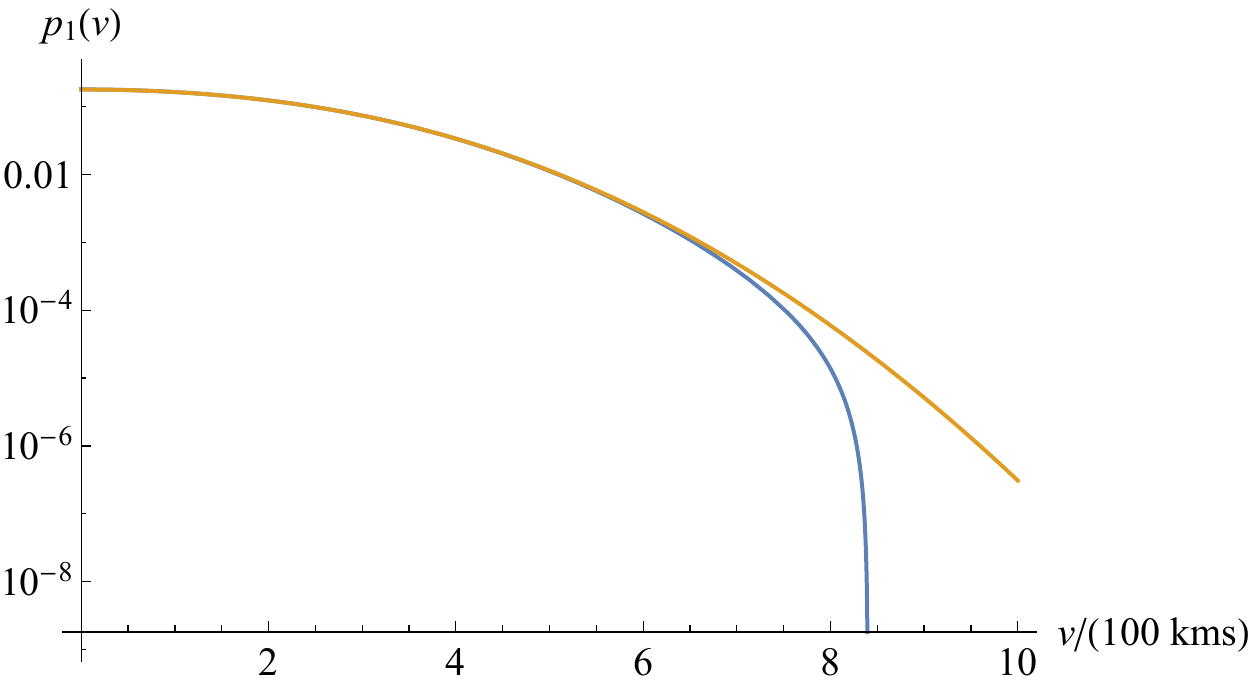}
	\caption{Direction-averaged velocity distribution
		at Earth
		for the truncated Maxwell-Boltzmann
		DM velocity distribution from
		Appendix~\ref{appdmvel}.
		The blue curve shows the distribution
		for $v_{\rm esc} = 600 \kms$,
		and the orange curve for
		$v_{\rm esc} \rightarrow \infty$.
		The difference between this curves
		is not visible in the linear-scale
		plot in the left panel, only in the log-scale
		plot in the right panel.}
	\label{figp1}
\end{figure*}

In the $v_{\rm esc} \rightarrow \infty$
limit (corresponding to a simple Maxwell-Boltzmann distribution
in the Galactic frame, which can be useful
for seeing the basic form of expressions),
we have
 $N_0 = \pi^{3/2} v_0^3$.
From Section~\ref{secsrate}, a useful quantity 
for computing DM scattering rates is
\begin{equation}
	p_1(v_z) = 
	\int dv_x dv_y \frac{d\Omega_e}{4\pi} f(v)
\end{equation}
For a Maxwell-Boltzmann distribution, this is 
\begin{equation}
	p_1(v_z) = \frac{1}{4 v_e}\left({\rm erf}\left(\frac{v_e - v_z}{v_0}\right)
	+ {\rm erf}\left(\frac{v_e + v_z}{v_0}\right)\right)
\end{equation}
\begin{widetext}
	For the truncated Maxwell-Boltzmann distribution, we have
\begin{equation}
	p_1(v_z) = \begin{cases}
		\frac{\pi^{3/2} v_0^3}{4 v_e N_0}	\left({\rm erf}\left(\frac{v_e - v_z}{v_0}\right)
	+ {\rm erf}\left(\frac{v_e + v_z}{v_0}\right)\right)
		- \frac{\pi v_0^2}{N_0} e^{-v_{\rm esc}^2/v_0^2} &
		v_z < v_{\rm esc} - v_e \\
		\frac{\pi^{3/2} v_0^3}{4 v_e N_0}	
		\left({\rm erf}\left(\frac{v_{\rm esc}}{v_0}\right)
	+ {\rm erf}\left(\frac{v_e - v_z}{v_0}\right)\right)
		- \frac{\pi v_0^2}{2 N_0} \frac{v_e + v_{\rm esc} - v_z}{v_e} e^{-v_{\rm esc}^2/v_0^2} &
		v_z < v_{\rm esc} + v_e \\
		0 &  v_z > v_{\rm esc} + v_e
	\end{cases}
\end{equation}
	Figure~\ref{figp1} plots $p_1(v_z)$ for the standard
	parameter values, illustrating that the difference
	between the truncated and non-truncated Maxwell-Boltzmann
	distributions is only important at high velocities,
	and correspondingly small $p_1$ values.
\end{widetext}

\section{Other EM sum rules}
\label{appsum}

As well as the $\int \frac{d\omega}{\omega}
\Imag(-1/\epsilon_L(\omega))$
sum rule,
there are also other constraints that the longitudinal
dielectric function should satisfy~\cite{10.1007/978-1-4757-5714-9,Dressel2002}.
For example, suppose that $\epsilon(\omega) \simeq 1 -
\omega_p^2 / \omega^2$ for large enough $|\omega|$.
Then, from the Kramers-Kronig relations,
\begin{equation}
	\int_0^\infty d\omega \, \omega \,
	\Imag \left(\frac{-1}{\epsilon_L(\omega,k)}\right)
	 = \frac{\pi}{2} \omega_p^2
	 \label{eqsrule2}
\end{equation}
The $d\omega/\omega$ sum rule (Eq.~\eqref{eqsr1})
can be viewed as corresponding to energy absorption
from a delta-function pulse in time.
In contrast, the $d\omega \ \omega$
sum rule from Eq.~\eqref{eqsrule2}
corresponds to energy absorption from a 
second-derivative-of-delta-function pulse
(a rapid down-up-down sequence). This emphasises
absorption at higher frequencies.
Physically, since we are interested in low-mass
DM absorption, and correspondingly, in smaller energy
transfers, the $d\omega/\omega$
sum rule will be more useful for our purposes.


\bibliography{dmscattering}

\begin{thebibliography}{79}%
\makeatletter
\providecommand \@ifxundefined [1]{%
 \@ifx{#1\undefined}
}%
\providecommand \@ifnum [1]{%
 \ifnum #1\expandafter \@firstoftwo
 \else \expandafter \@secondoftwo
 \fi
}%
\providecommand \@ifx [1]{%
 \ifx #1\expandafter \@firstoftwo
 \else \expandafter \@secondoftwo
 \fi
}%
\providecommand \natexlab [1]{#1}%
\providecommand \enquote  [1]{``#1''}%
\providecommand \bibnamefont  [1]{#1}%
\providecommand \bibfnamefont [1]{#1}%
\providecommand \citenamefont [1]{#1}%
\providecommand \href@noop [0]{\@secondoftwo}%
\providecommand \href [0]{\begingroup \@sanitize@url \@href}%
\providecommand \@href[1]{\@@startlink{#1}\@@href}%
\providecommand \@@href[1]{\endgroup#1\@@endlink}%
\providecommand \@sanitize@url [0]{\catcode `\\12\catcode `\$12\catcode
  `\&12\catcode `\#12\catcode `\^12\catcode `\_12\catcode `\%12\relax}%
\providecommand \@@startlink[1]{}%
\providecommand \@@endlink[0]{}%
\providecommand \url  [0]{\begingroup\@sanitize@url \@url }%
\providecommand \@url [1]{\endgroup\@href {#1}{\urlprefix }}%
\providecommand \urlprefix  [0]{URL }%
\providecommand \Eprint [0]{\href }%
\providecommand \doibase [0]{http://dx.doi.org/}%
\providecommand \selectlanguage [0]{\@gobble}%
\providecommand \bibinfo  [0]{\@secondoftwo}%
\providecommand \bibfield  [0]{\@secondoftwo}%
\providecommand \translation [1]{[#1]}%
\providecommand \BibitemOpen [0]{}%
\providecommand \bibitemStop [0]{}%
\providecommand \bibitemNoStop [0]{.\EOS\space}%
\providecommand \EOS [0]{\spacefactor3000\relax}%
\providecommand \BibitemShut  [1]{\csname bibitem#1\endcsname}%
\let\auto@bib@innerbib\@empty
\bibitem [{\citenamefont {Schumann}(2019)}]{1903.03026}%
  \BibitemOpen
  \bibfield  {author} {\bibinfo {author} {\bibfnamefont {M.}~\bibnamefont
  {Schumann}},\ }\href {\doibase 10.1088/1361-6471/ab2ea5} {\bibfield
  {journal} {\bibinfo  {journal} {J. Phys. G}\ }\textbf {\bibinfo {volume}
  {46}},\ \bibinfo {pages} {103003} (\bibinfo {year} {2019})},\ \Eprint
  {http://arxiv.org/abs/1903.03026} {arXiv:1903.03026 [astro-ph.CO]}
  \BibitemShut {NoStop}%
\bibitem [{\citenamefont {Lee}\ and\ \citenamefont
  {Weinberg}(1977)}]{10.1103/PhysRevLett.39.165}%
  \BibitemOpen
  \bibfield  {author} {\bibinfo {author} {\bibfnamefont {B.~W.}\ \bibnamefont
  {Lee}}\ and\ \bibinfo {author} {\bibfnamefont {S.}~\bibnamefont {Weinberg}},\
  }\href {\doibase 10.1103/physrevlett.39.165} {\bibfield  {journal} {\bibinfo
  {journal} {Physical Review Letters}\ }\textbf {\bibinfo {volume} {39}},\
  \bibinfo {pages} {165} (\bibinfo {year} {1977})}\BibitemShut {NoStop}%
\bibitem [{\citenamefont {B{\oe}hm}\ and\ \citenamefont
  {Fayet}(2004)}]{10.1016/j.nuclphysb.2004.01.015}%
  \BibitemOpen
  \bibfield  {author} {\bibinfo {author} {\bibfnamefont {C.}~\bibnamefont
  {B{\oe}hm}}\ and\ \bibinfo {author} {\bibfnamefont {P.}~\bibnamefont
  {Fayet}},\ }\href {\doibase 10.1016/j.nuclphysb.2004.01.015} {\bibfield
  {journal} {\bibinfo  {journal} {Nuclear Physics B}\ }\textbf {\bibinfo
  {volume} {683}},\ \bibinfo {pages} {219} (\bibinfo {year}
  {2004})}\BibitemShut {NoStop}%
\bibitem [{\citenamefont {Chang}\ \emph {et~al.}(2021)\citenamefont {Chang},
  \citenamefont {Essig},\ and\ \citenamefont {Reinert}}]{1911.03389}%
  \BibitemOpen
  \bibfield  {author} {\bibinfo {author} {\bibfnamefont {J.~H.}\ \bibnamefont
  {Chang}}, \bibinfo {author} {\bibfnamefont {R.}~\bibnamefont {Essig}}, \ and\
  \bibinfo {author} {\bibfnamefont {A.}~\bibnamefont {Reinert}},\ }\href
  {\doibase 10.1007/jhep03(2021)141} {\bibfield  {journal} {\bibinfo  {journal}
  {Journal of High Energy Physics}\ }\textbf {\bibinfo {volume} {2021}}
  (\bibinfo {year} {2021}),\ 10.1007/jhep03(2021)141}\BibitemShut {NoStop}%
\bibitem [{\citenamefont {Dvorkin}\ \emph {et~al.}(2019)\citenamefont
  {Dvorkin}, \citenamefont {Lin},\ and\ \citenamefont
  {Schutz}}]{10.1103/PhysRevD.99.115009}%
  \BibitemOpen
  \bibfield  {author} {\bibinfo {author} {\bibfnamefont {C.}~\bibnamefont
  {Dvorkin}}, \bibinfo {author} {\bibfnamefont {T.}~\bibnamefont {Lin}}, \ and\
  \bibinfo {author} {\bibfnamefont {K.}~\bibnamefont {Schutz}},\ }\href
  {\doibase 10.1103/physrevd.99.115009} {\bibfield  {journal} {\bibinfo
  {journal} {Physical Review D}\ }\textbf {\bibinfo {volume} {99}} (\bibinfo
  {year} {2019}),\ 10.1103/physrevd.99.115009}\BibitemShut {NoStop}%
\bibitem [{\citenamefont {Hall}\ \emph {et~al.}(2010)\citenamefont {Hall},
  \citenamefont {Jedamzik}, \citenamefont {March-Russell},\ and\ \citenamefont
  {West}}]{10.1007/JHEP03(2010)080}%
  \BibitemOpen
  \bibfield  {author} {\bibinfo {author} {\bibfnamefont {L.~J.}\ \bibnamefont
  {Hall}}, \bibinfo {author} {\bibfnamefont {K.}~\bibnamefont {Jedamzik}},
  \bibinfo {author} {\bibfnamefont {J.}~\bibnamefont {March-Russell}}, \ and\
  \bibinfo {author} {\bibfnamefont {S.~M.}\ \bibnamefont {West}},\ }\href
  {\doibase 10.1007/jhep03(2010)080} {\bibfield  {journal} {\bibinfo  {journal}
  {Journal of High Energy Physics}\ }\textbf {\bibinfo {volume} {2010}}
  (\bibinfo {year} {2010}),\ 10.1007/jhep03(2010)080}\BibitemShut {NoStop}%
\bibitem [{\citenamefont {Essig}\ \emph
  {et~al.}(2012{\natexlab{a}})\citenamefont {Essig}, \citenamefont {Mardon},\
  and\ \citenamefont {Volansky}}]{Essig:2011nj}%
  \BibitemOpen
  \bibfield  {author} {\bibinfo {author} {\bibfnamefont {R.}~\bibnamefont
  {Essig}}, \bibinfo {author} {\bibfnamefont {J.}~\bibnamefont {Mardon}}, \
  and\ \bibinfo {author} {\bibfnamefont {T.}~\bibnamefont {Volansky}},\ }\href
  {\doibase 10.1103/PhysRevD.85.076007} {\bibfield  {journal} {\bibinfo
  {journal} {Phys. Rev. D}\ }\textbf {\bibinfo {volume} {85}},\ \bibinfo
  {pages} {076007} (\bibinfo {year} {2012}{\natexlab{a}})},\ \Eprint
  {http://arxiv.org/abs/1108.5383} {arXiv:1108.5383 [hep-ph]} \BibitemShut
  {NoStop}%
\bibitem [{\citenamefont {Essig}\ \emph
  {et~al.}(2012{\natexlab{b}})\citenamefont {Essig}, \citenamefont
  {Manalaysay}, \citenamefont {Mardon}, \citenamefont {Sorensen},\ and\
  \citenamefont {Volansky}}]{Essig:2012yx}%
  \BibitemOpen
  \bibfield  {author} {\bibinfo {author} {\bibfnamefont {R.}~\bibnamefont
  {Essig}}, \bibinfo {author} {\bibfnamefont {A.}~\bibnamefont {Manalaysay}},
  \bibinfo {author} {\bibfnamefont {J.}~\bibnamefont {Mardon}}, \bibinfo
  {author} {\bibfnamefont {P.}~\bibnamefont {Sorensen}}, \ and\ \bibinfo
  {author} {\bibfnamefont {T.}~\bibnamefont {Volansky}},\ }\href {\doibase
  10.1103/PhysRevLett.109.021301} {\bibfield  {journal} {\bibinfo  {journal}
  {Phys. Rev. Lett.}\ }\textbf {\bibinfo {volume} {109}},\ \bibinfo {pages}
  {021301} (\bibinfo {year} {2012}{\natexlab{b}})},\ \Eprint
  {http://arxiv.org/abs/1206.2644} {arXiv:1206.2644 [astro-ph.CO]} \BibitemShut
  {NoStop}%
\bibitem [{\citenamefont {Essig}\ \emph {et~al.}(2016)\citenamefont {Essig},
  \citenamefont {Fernandez-Serra}, \citenamefont {Mardon}, \citenamefont
  {Soto}, \citenamefont {Volansky},\ and\ \citenamefont {Yu}}]{Essig:2015cda}%
  \BibitemOpen
  \bibfield  {author} {\bibinfo {author} {\bibfnamefont {R.}~\bibnamefont
  {Essig}}, \bibinfo {author} {\bibfnamefont {M.}~\bibnamefont
  {Fernandez-Serra}}, \bibinfo {author} {\bibfnamefont {J.}~\bibnamefont
  {Mardon}}, \bibinfo {author} {\bibfnamefont {A.}~\bibnamefont {Soto}},
  \bibinfo {author} {\bibfnamefont {T.}~\bibnamefont {Volansky}}, \ and\
  \bibinfo {author} {\bibfnamefont {T.-T.}\ \bibnamefont {Yu}},\ }\href
  {\doibase 10.1007/JHEP05(2016)046} {\bibfield  {journal} {\bibinfo  {journal}
  {JHEP}\ }\textbf {\bibinfo {volume} {05}},\ \bibinfo {pages} {046} (\bibinfo
  {year} {2016})},\ \Eprint {http://arxiv.org/abs/1509.01598} {arXiv:1509.01598
  [hep-ph]} \BibitemShut {NoStop}%
\bibitem [{\citenamefont {Hochberg}\ \emph
  {et~al.}(2016{\natexlab{a}})\citenamefont {Hochberg}, \citenamefont {Zhao},\
  and\ \citenamefont {Zurek}}]{Hochberg:2015pha}%
  \BibitemOpen
  \bibfield  {author} {\bibinfo {author} {\bibfnamefont {Y.}~\bibnamefont
  {Hochberg}}, \bibinfo {author} {\bibfnamefont {Y.}~\bibnamefont {Zhao}}, \
  and\ \bibinfo {author} {\bibfnamefont {K.~M.}\ \bibnamefont {Zurek}},\ }\href
  {\doibase 10.1103/PhysRevLett.116.011301} {\bibfield  {journal} {\bibinfo
  {journal} {Phys. Rev. Lett.}\ }\textbf {\bibinfo {volume} {116}},\ \bibinfo
  {pages} {011301} (\bibinfo {year} {2016}{\natexlab{a}})},\ \Eprint
  {http://arxiv.org/abs/1504.07237} {arXiv:1504.07237 [hep-ph]} \BibitemShut
  {NoStop}%
\bibitem [{\citenamefont {Hochberg}\ \emph
  {et~al.}(2016{\natexlab{b}})\citenamefont {Hochberg}, \citenamefont {Pyle},
  \citenamefont {Zhao},\ and\ \citenamefont {Zurek}}]{Hochberg:2015fth}%
  \BibitemOpen
  \bibfield  {author} {\bibinfo {author} {\bibfnamefont {Y.}~\bibnamefont
  {Hochberg}}, \bibinfo {author} {\bibfnamefont {M.}~\bibnamefont {Pyle}},
  \bibinfo {author} {\bibfnamefont {Y.}~\bibnamefont {Zhao}}, \ and\ \bibinfo
  {author} {\bibfnamefont {K.~M.}\ \bibnamefont {Zurek}},\ }\href {\doibase
  10.1007/JHEP08(2016)057} {\bibfield  {journal} {\bibinfo  {journal} {JHEP}\
  }\textbf {\bibinfo {volume} {08}},\ \bibinfo {pages} {057} (\bibinfo {year}
  {2016}{\natexlab{b}})},\ \Eprint {http://arxiv.org/abs/1512.04533}
  {arXiv:1512.04533 [hep-ph]} \BibitemShut {NoStop}%
\bibitem [{\citenamefont {Hochberg}\ \emph {et~al.}(2017)\citenamefont
  {Hochberg}, \citenamefont {Kahn}, \citenamefont {Lisanti}, \citenamefont
  {Tully},\ and\ \citenamefont {Zurek}}]{Hochberg:2016ntt}%
  \BibitemOpen
  \bibfield  {author} {\bibinfo {author} {\bibfnamefont {Y.}~\bibnamefont
  {Hochberg}}, \bibinfo {author} {\bibfnamefont {Y.}~\bibnamefont {Kahn}},
  \bibinfo {author} {\bibfnamefont {M.}~\bibnamefont {Lisanti}}, \bibinfo
  {author} {\bibfnamefont {C.~G.}\ \bibnamefont {Tully}}, \ and\ \bibinfo
  {author} {\bibfnamefont {K.~M.}\ \bibnamefont {Zurek}},\ }\href {\doibase
  10.1016/j.physletb.2017.06.051} {\bibfield  {journal} {\bibinfo  {journal}
  {Phys. Lett. B}\ }\textbf {\bibinfo {volume} {772}},\ \bibinfo {pages} {239}
  (\bibinfo {year} {2017})},\ \Eprint {http://arxiv.org/abs/1606.08849}
  {arXiv:1606.08849 [hep-ph]} \BibitemShut {NoStop}%
\bibitem [{\citenamefont {Hochberg}\ \emph {et~al.}(2018)\citenamefont
  {Hochberg}, \citenamefont {Kahn}, \citenamefont {Lisanti}, \citenamefont
  {Zurek}, \citenamefont {Grushin}, \citenamefont {Ilan}, \citenamefont
  {Griffin}, \citenamefont {Liu}, \citenamefont {Weber},\ and\ \citenamefont
  {Neaton}}]{1708.08929}%
  \BibitemOpen
  \bibfield  {author} {\bibinfo {author} {\bibfnamefont {Y.}~\bibnamefont
  {Hochberg}}, \bibinfo {author} {\bibfnamefont {Y.}~\bibnamefont {Kahn}},
  \bibinfo {author} {\bibfnamefont {M.}~\bibnamefont {Lisanti}}, \bibinfo
  {author} {\bibfnamefont {K.~M.}\ \bibnamefont {Zurek}}, \bibinfo {author}
  {\bibfnamefont {A.~G.}\ \bibnamefont {Grushin}}, \bibinfo {author}
  {\bibfnamefont {R.}~\bibnamefont {Ilan}}, \bibinfo {author} {\bibfnamefont
  {S.~M.}\ \bibnamefont {Griffin}}, \bibinfo {author} {\bibfnamefont {Z.-F.}\
  \bibnamefont {Liu}}, \bibinfo {author} {\bibfnamefont {S.~F.}\ \bibnamefont
  {Weber}}, \ and\ \bibinfo {author} {\bibfnamefont {J.~B.}\ \bibnamefont
  {Neaton}},\ }\href {\doibase 10.1103/PhysRevD.97.015004} {\bibfield
  {journal} {\bibinfo  {journal} {Phys. Rev. D}\ }\textbf {\bibinfo {volume}
  {97}},\ \bibinfo {pages} {015004} (\bibinfo {year} {2018})},\ \Eprint
  {http://arxiv.org/abs/1708.08929} {arXiv:1708.08929 [hep-ph]} \BibitemShut
  {NoStop}%
\bibitem [{\citenamefont {Derenzo}\ \emph {et~al.}(2017)\citenamefont
  {Derenzo}, \citenamefont {Essig}, \citenamefont {Massari}, \citenamefont
  {Soto},\ and\ \citenamefont {Yu}}]{Derenzo:2016fse}%
  \BibitemOpen
  \bibfield  {author} {\bibinfo {author} {\bibfnamefont {S.}~\bibnamefont
  {Derenzo}}, \bibinfo {author} {\bibfnamefont {R.}~\bibnamefont {Essig}},
  \bibinfo {author} {\bibfnamefont {A.}~\bibnamefont {Massari}}, \bibinfo
  {author} {\bibfnamefont {A.}~\bibnamefont {Soto}}, \ and\ \bibinfo {author}
  {\bibfnamefont {T.-T.}\ \bibnamefont {Yu}},\ }\href {\doibase
  10.1103/PhysRevD.96.016026} {\bibfield  {journal} {\bibinfo  {journal} {Phys.
  Rev. D}\ }\textbf {\bibinfo {volume} {96}},\ \bibinfo {pages} {016026}
  (\bibinfo {year} {2017})},\ \Eprint {http://arxiv.org/abs/1607.01009}
  {arXiv:1607.01009 [hep-ph]} \BibitemShut {NoStop}%
\bibitem [{\citenamefont {Kurinsky}\ \emph {et~al.}(2019)\citenamefont
  {Kurinsky}, \citenamefont {Yu}, \citenamefont {Hochberg},\ and\ \citenamefont
  {Cabrera}}]{Kurinsky:2019pgb}%
  \BibitemOpen
  \bibfield  {author} {\bibinfo {author} {\bibfnamefont {N.~A.}\ \bibnamefont
  {Kurinsky}}, \bibinfo {author} {\bibfnamefont {T.~C.}\ \bibnamefont {Yu}},
  \bibinfo {author} {\bibfnamefont {Y.}~\bibnamefont {Hochberg}}, \ and\
  \bibinfo {author} {\bibfnamefont {B.}~\bibnamefont {Cabrera}},\ }\href
  {\doibase 10.1103/PhysRevD.99.123005} {\bibfield  {journal} {\bibinfo
  {journal} {Phys. Rev. D}\ }\textbf {\bibinfo {volume} {99}},\ \bibinfo
  {pages} {123005} (\bibinfo {year} {2019})},\ \Eprint
  {http://arxiv.org/abs/1901.07569} {arXiv:1901.07569 [hep-ex]} \BibitemShut
  {NoStop}%
\bibitem [{\citenamefont {Griffin}\ \emph {et~al.}(2020)\citenamefont
  {Griffin}, \citenamefont {Inzani}, \citenamefont {Trickle}, \citenamefont
  {Zhang},\ and\ \citenamefont {Zurek}}]{10.1103/PhysRevD.101.055004}%
  \BibitemOpen
  \bibfield  {author} {\bibinfo {author} {\bibfnamefont {S.~M.}\ \bibnamefont
  {Griffin}}, \bibinfo {author} {\bibfnamefont {K.}~\bibnamefont {Inzani}},
  \bibinfo {author} {\bibfnamefont {T.}~\bibnamefont {Trickle}}, \bibinfo
  {author} {\bibfnamefont {Z.}~\bibnamefont {Zhang}}, \ and\ \bibinfo {author}
  {\bibfnamefont {K.~M.}\ \bibnamefont {Zurek}},\ }\href {\doibase
  10.1103/PhysRevD.101.055004} {\bibfield  {journal} {\bibinfo  {journal}
  {Phys. Rev. D}\ }\textbf {\bibinfo {volume} {101}},\ \bibinfo {pages}
  {055004} (\bibinfo {year} {2020})}\BibitemShut {NoStop}%
\bibitem [{\citenamefont {Blanco}\ \emph {et~al.}(2020)\citenamefont {Blanco},
  \citenamefont {Collar}, \citenamefont {Kahn},\ and\ \citenamefont
  {Lillard}}]{Blanco:2019lrf}%
  \BibitemOpen
  \bibfield  {author} {\bibinfo {author} {\bibfnamefont {C.}~\bibnamefont
  {Blanco}}, \bibinfo {author} {\bibfnamefont {J.~I.}\ \bibnamefont {Collar}},
  \bibinfo {author} {\bibfnamefont {Y.}~\bibnamefont {Kahn}}, \ and\ \bibinfo
  {author} {\bibfnamefont {B.}~\bibnamefont {Lillard}},\ }\href {\doibase
  10.1103/PhysRevD.101.056001} {\bibfield  {journal} {\bibinfo  {journal}
  {Phys. Rev. D}\ }\textbf {\bibinfo {volume} {101}},\ \bibinfo {pages}
  {056001} (\bibinfo {year} {2020})},\ \Eprint
  {http://arxiv.org/abs/1912.02822} {arXiv:1912.02822 [hep-ph]} \BibitemShut
  {NoStop}%
\bibitem [{\citenamefont {Trickle}\ \emph {et~al.}(2020)\citenamefont
  {Trickle}, \citenamefont {Zhang}, \citenamefont {Zurek}, \citenamefont
  {Inzani},\ and\ \citenamefont {Griffin}}]{Trickle:2019nya}%
  \BibitemOpen
  \bibfield  {author} {\bibinfo {author} {\bibfnamefont {T.}~\bibnamefont
  {Trickle}}, \bibinfo {author} {\bibfnamefont {Z.}~\bibnamefont {Zhang}},
  \bibinfo {author} {\bibfnamefont {K.~M.}\ \bibnamefont {Zurek}}, \bibinfo
  {author} {\bibfnamefont {K.}~\bibnamefont {Inzani}}, \ and\ \bibinfo {author}
  {\bibfnamefont {S.}~\bibnamefont {Griffin}},\ }\href {\doibase
  10.1007/JHEP03(2020)036} {\bibfield  {journal} {\bibinfo  {journal} {JHEP}\
  }\textbf {\bibinfo {volume} {03}},\ \bibinfo {pages} {036} (\bibinfo {year}
  {2020})},\ \Eprint {http://arxiv.org/abs/1910.08092} {arXiv:1910.08092
  [hep-ph]} \BibitemShut {NoStop}%
\bibitem [{\citenamefont {Geilhufe}\ \emph {et~al.}(2020)\citenamefont
  {Geilhufe}, \citenamefont {Kahlhoefer},\ and\ \citenamefont
  {Winkler}}]{1910.02091}%
  \BibitemOpen
  \bibfield  {author} {\bibinfo {author} {\bibfnamefont {R.~M.}\ \bibnamefont
  {Geilhufe}}, \bibinfo {author} {\bibfnamefont {F.}~\bibnamefont
  {Kahlhoefer}}, \ and\ \bibinfo {author} {\bibfnamefont {M.~W.}\ \bibnamefont
  {Winkler}},\ }\href {\doibase 10.1103/physrevd.101.055005} {\bibfield
  {journal} {\bibinfo  {journal} {Physical Review D}\ }\textbf {\bibinfo
  {volume} {101}} (\bibinfo {year} {2020}),\
  10.1103/physrevd.101.055005}\BibitemShut {NoStop}%
\bibitem [{\citenamefont {Hochberg}\ \emph {et~al.}(2019)\citenamefont
  {Hochberg}, \citenamefont {Charaev}, \citenamefont {Nam}, \citenamefont
  {Verma}, \citenamefont {Colangelo},\ and\ \citenamefont
  {Berggren}}]{10.1103/PhysRevLett.123.151802}%
  \BibitemOpen
  \bibfield  {author} {\bibinfo {author} {\bibfnamefont {Y.}~\bibnamefont
  {Hochberg}}, \bibinfo {author} {\bibfnamefont {I.}~\bibnamefont {Charaev}},
  \bibinfo {author} {\bibfnamefont {S.-W.}\ \bibnamefont {Nam}}, \bibinfo
  {author} {\bibfnamefont {V.}~\bibnamefont {Verma}}, \bibinfo {author}
  {\bibfnamefont {M.}~\bibnamefont {Colangelo}}, \ and\ \bibinfo {author}
  {\bibfnamefont {K.~K.}\ \bibnamefont {Berggren}},\ }\href {\doibase
  10.1103/PhysRevLett.123.151802} {\bibfield  {journal} {\bibinfo  {journal}
  {Phys. Rev. Lett.}\ }\textbf {\bibinfo {volume} {123}},\ \bibinfo {pages}
  {151802} (\bibinfo {year} {2019})},\ \Eprint
  {http://arxiv.org/abs/1903.05101} {arXiv:1903.05101 [hep-ph]} \BibitemShut
  {NoStop}%
\bibitem [{\citenamefont {Coskuner}\ \emph
  {et~al.}(2021{\natexlab{a}})\citenamefont {Coskuner}, \citenamefont
  {Mitridate}, \citenamefont {Olivares},\ and\ \citenamefont
  {Zurek}}]{1909.09170}%
  \BibitemOpen
  \bibfield  {author} {\bibinfo {author} {\bibfnamefont {A.}~\bibnamefont
  {Coskuner}}, \bibinfo {author} {\bibfnamefont {A.}~\bibnamefont {Mitridate}},
  \bibinfo {author} {\bibfnamefont {A.}~\bibnamefont {Olivares}}, \ and\
  \bibinfo {author} {\bibfnamefont {K.~M.}\ \bibnamefont {Zurek}},\ }\href
  {\doibase 10.1103/physrevd.103.016006} {\bibfield  {journal} {\bibinfo
  {journal} {Physical Review D}\ }\textbf {\bibinfo {volume} {103}} (\bibinfo
  {year} {2021}{\natexlab{a}}),\ 10.1103/physrevd.103.016006}\BibitemShut
  {NoStop}%
\bibitem [{\citenamefont {Griffin}\ \emph {et~al.}(2021)\citenamefont
  {Griffin}, \citenamefont {Hochberg}, \citenamefont {Inzani}, \citenamefont
  {Kurinsky}, \citenamefont {Lin},\ and\ \citenamefont {Yu}}]{Griffin:2020lgd}%
  \BibitemOpen
  \bibfield  {author} {\bibinfo {author} {\bibfnamefont {S.~M.}\ \bibnamefont
  {Griffin}}, \bibinfo {author} {\bibfnamefont {Y.}~\bibnamefont {Hochberg}},
  \bibinfo {author} {\bibfnamefont {K.}~\bibnamefont {Inzani}}, \bibinfo
  {author} {\bibfnamefont {N.}~\bibnamefont {Kurinsky}}, \bibinfo {author}
  {\bibfnamefont {T.}~\bibnamefont {Lin}}, \ and\ \bibinfo {author}
  {\bibfnamefont {T.~C.}\ \bibnamefont {Yu}},\ }\href {\doibase
  10.1103/PhysRevD.103.075002} {\bibfield  {journal} {\bibinfo  {journal}
  {Phys. Rev. D}\ }\textbf {\bibinfo {volume} {103}},\ \bibinfo {pages}
  {075002} (\bibinfo {year} {2021})},\ \Eprint
  {http://arxiv.org/abs/2008.08560} {arXiv:2008.08560 [hep-ph]} \BibitemShut
  {NoStop}%
\bibitem [{\citenamefont {Knapen}\ \emph {et~al.}(2017)\citenamefont {Knapen},
  \citenamefont {Lin},\ and\ \citenamefont
  {Zurek}}]{10.1103/PhysRevD.96.115021}%
  \BibitemOpen
  \bibfield  {author} {\bibinfo {author} {\bibfnamefont {S.}~\bibnamefont
  {Knapen}}, \bibinfo {author} {\bibfnamefont {T.}~\bibnamefont {Lin}}, \ and\
  \bibinfo {author} {\bibfnamefont {K.~M.}\ \bibnamefont {Zurek}},\ }\href
  {\doibase 10.1103/PhysRevD.96.115021} {\bibfield  {journal} {\bibinfo
  {journal} {Phys. Rev. D}\ }\textbf {\bibinfo {volume} {96}},\ \bibinfo
  {pages} {115021} (\bibinfo {year} {2017})},\ \Eprint
  {http://arxiv.org/abs/1709.07882} {arXiv:1709.07882 [hep-ph]} \BibitemShut
  {NoStop}%
\bibitem [{\citenamefont {Hochberg}\ \emph
  {et~al.}(2021{\natexlab{a}})\citenamefont {Hochberg}, \citenamefont {Kahn},
  \citenamefont {Kurinsky}, \citenamefont {Lehmann}, \citenamefont {Yu},\ and\
  \citenamefont {Berggren}}]{2101.08263}%
  \BibitemOpen
  \bibfield  {author} {\bibinfo {author} {\bibfnamefont {Y.}~\bibnamefont
  {Hochberg}}, \bibinfo {author} {\bibfnamefont {Y.}~\bibnamefont {Kahn}},
  \bibinfo {author} {\bibfnamefont {N.}~\bibnamefont {Kurinsky}}, \bibinfo
  {author} {\bibfnamefont {B.~V.}\ \bibnamefont {Lehmann}}, \bibinfo {author}
  {\bibfnamefont {T.~C.}\ \bibnamefont {Yu}}, \ and\ \bibinfo {author}
  {\bibfnamefont {K.~K.}\ \bibnamefont {Berggren}},\ }\href {\doibase
  10.1103/PhysRevLett.127.151802} {\bibfield  {journal} {\bibinfo  {journal}
  {Phys. Rev. Lett.}\ }\textbf {\bibinfo {volume} {127}},\ \bibinfo {pages}
  {151802} (\bibinfo {year} {2021}{\natexlab{a}})},\ \Eprint
  {http://arxiv.org/abs/2101.08263} {arXiv:2101.08263 [hep-ph]} \BibitemShut
  {NoStop}%
\bibitem [{\citenamefont {Knapen}\ \emph
  {et~al.}(2021{\natexlab{a}})\citenamefont {Knapen}, \citenamefont
  {Kozaczuk},\ and\ \citenamefont {Lin}}]{2101.08275}%
  \BibitemOpen
  \bibfield  {author} {\bibinfo {author} {\bibfnamefont {S.}~\bibnamefont
  {Knapen}}, \bibinfo {author} {\bibfnamefont {J.}~\bibnamefont {Kozaczuk}}, \
  and\ \bibinfo {author} {\bibfnamefont {T.}~\bibnamefont {Lin}},\ }\href
  {\doibase 10.1103/physrevd.104.015031} {\bibfield  {journal} {\bibinfo
  {journal} {Physical Review D}\ }\textbf {\bibinfo {volume} {104}} (\bibinfo
  {year} {2021}{\natexlab{a}}),\ 10.1103/physrevd.104.015031}\BibitemShut
  {NoStop}%
\bibitem [{\citenamefont {Nozi\`eres}\ and\ \citenamefont
  {Pines}(1959)}]{10.1103/PhysRev.113.1254}%
  \BibitemOpen
  \bibfield  {author} {\bibinfo {author} {\bibfnamefont {P.}~\bibnamefont
  {Nozi\`eres}}\ and\ \bibinfo {author} {\bibfnamefont {D.}~\bibnamefont
  {Pines}},\ }\href {\doibase 10.1103/PhysRev.113.1254} {\bibfield  {journal}
  {\bibinfo  {journal} {Phys. Rev.}\ }\textbf {\bibinfo {volume} {113}},\
  \bibinfo {pages} {1254} (\bibinfo {year} {1959})}\BibitemShut {NoStop}%
\bibitem [{\citenamefont {Mahan}(2000)}]{10.1007/978-1-4757-5714-9}%
  \BibitemOpen
  \bibfield  {author} {\bibinfo {author} {\bibfnamefont {G.~D.}\ \bibnamefont
  {Mahan}},\ }\href@noop {} {\emph {\bibinfo {title} {Many-Particle Physics}}}\
  (\bibinfo  {publisher} {Springer},\ \bibinfo {address} {Boston, MA},\
  \bibinfo {year} {2000})\BibitemShut {NoStop}%
\bibitem [{\citenamefont {Dressel}\ and\ \citenamefont
  {Gr\"{u}ner}(2002)}]{Dressel2002}%
  \BibitemOpen
  \bibfield  {author} {\bibinfo {author} {\bibfnamefont {M.}~\bibnamefont
  {Dressel}}\ and\ \bibinfo {author} {\bibfnamefont {G.}~\bibnamefont
  {Gr\"{u}ner}},\ }\href {\doibase 10.1017/cbo9780511606168} {\emph {\bibinfo
  {title} {Electrodynamics of Solids}}}\ (\bibinfo  {publisher} {Cambridge
  University Press},\ \bibinfo {year} {2002})\BibitemShut {NoStop}%
\bibitem [{\citenamefont
  {Dreyling-Eschweiler}(2014)}]{Dreyling-Eschweiler:2014mxa}%
  \BibitemOpen
  \bibfield  {author} {\bibinfo {author} {\bibfnamefont {J.}~\bibnamefont
  {Dreyling-Eschweiler}} (\bibinfo {collaboration} {ALPS-II}),\ }in\ \href
  {\doibase 10.3204/DESY-PROC-2014-03/dreyling-eschweiler_jan} {\emph {\bibinfo
  {booktitle} {{Proceedings, 10th Patras Workshop on Axions, WIMPs and WISPs
  (AXION-WIMP 2014): Geneva, Switzerland, June 29-July 4, 2014}}}}\ (\bibinfo
  {year} {2014})\ pp.\ \bibinfo {pages} {63--66},\ \Eprint
  {http://arxiv.org/abs/1409.6992} {arXiv:1409.6992 [physics.ins-det]}
  \BibitemShut {NoStop}%
\bibitem [{\citenamefont {Dreyling-Eschweiler}\ \emph
  {et~al.}(2015)\citenamefont {Dreyling-Eschweiler}, \citenamefont {Bastidon},
  \citenamefont {Döbrich}, \citenamefont {Horns}, \citenamefont {Januschek},\
  and\ \citenamefont {Lindner}}]{Dreyling-Eschweiler:2015pja}%
  \BibitemOpen
  \bibfield  {author} {\bibinfo {author} {\bibfnamefont {J.}~\bibnamefont
  {Dreyling-Eschweiler}}, \bibinfo {author} {\bibfnamefont {N.}~\bibnamefont
  {Bastidon}}, \bibinfo {author} {\bibfnamefont {B.}~\bibnamefont {Döbrich}},
  \bibinfo {author} {\bibfnamefont {D.}~\bibnamefont {Horns}}, \bibinfo
  {author} {\bibfnamefont {F.}~\bibnamefont {Januschek}}, \ and\ \bibinfo
  {author} {\bibfnamefont {A.}~\bibnamefont {Lindner}},\ }\href {\doibase
  10.1080/09500340.2015.1021723} {\bibfield  {journal} {\bibinfo  {journal} {J.
  Mod. Opt.}\ }\textbf {\bibinfo {volume} {62}},\ \bibinfo {pages} {1132}
  (\bibinfo {year} {2015})},\ \Eprint {http://arxiv.org/abs/1502.07878}
  {arXiv:1502.07878 [physics.ins-det]} \BibitemShut {NoStop}%
\bibitem [{\citenamefont {Cabrera}\ \emph {et~al.}(1998)\citenamefont
  {Cabrera}, \citenamefont {Clarke}, \citenamefont {Colling}, \citenamefont
  {Miller}, \citenamefont {Nam},\ and\ \citenamefont {Romani}}]{Cabrera1998}%
  \BibitemOpen
  \bibfield  {author} {\bibinfo {author} {\bibfnamefont {B.}~\bibnamefont
  {Cabrera}}, \bibinfo {author} {\bibfnamefont {R.~M.}\ \bibnamefont {Clarke}},
  \bibinfo {author} {\bibfnamefont {P.}~\bibnamefont {Colling}}, \bibinfo
  {author} {\bibfnamefont {A.~J.}\ \bibnamefont {Miller}}, \bibinfo {author}
  {\bibfnamefont {S.}~\bibnamefont {Nam}}, \ and\ \bibinfo {author}
  {\bibfnamefont {R.~W.}\ \bibnamefont {Romani}},\ }\href {\doibase
  10.1063/1.121984} {\bibfield  {journal} {\bibinfo  {journal} {Applied Physics
  Letters}\ }\textbf {\bibinfo {volume} {73}},\ \bibinfo {pages} {735}
  (\bibinfo {year} {1998})},\ \Eprint
  {http://arxiv.org/abs/https://doi.org/10.1063/1.121984}
  {https://doi.org/10.1063/1.121984} \BibitemShut {NoStop}%
\bibitem [{\citenamefont {Karasik}\ \emph {et~al.}(2012)\citenamefont
  {Karasik}, \citenamefont {Pereverzev}, \citenamefont {Soibel}, \citenamefont
  {Santavicca}, \citenamefont {Prober}, \citenamefont {Olaya},\ and\
  \citenamefont {Gershenson}}]{Karasik:2012rb}%
  \BibitemOpen
  \bibfield  {author} {\bibinfo {author} {\bibfnamefont {B.~S.}\ \bibnamefont
  {Karasik}}, \bibinfo {author} {\bibfnamefont {S.~V.}\ \bibnamefont
  {Pereverzev}}, \bibinfo {author} {\bibfnamefont {A.}~\bibnamefont {Soibel}},
  \bibinfo {author} {\bibfnamefont {D.~F.}\ \bibnamefont {Santavicca}},
  \bibinfo {author} {\bibfnamefont {D.~E.}\ \bibnamefont {Prober}}, \bibinfo
  {author} {\bibfnamefont {D.}~\bibnamefont {Olaya}}, \ and\ \bibinfo {author}
  {\bibfnamefont {M.~E.}\ \bibnamefont {Gershenson}},\ }\href {\doibase
  10.1063/1.4739839} {\bibfield  {journal} {\bibinfo  {journal} {Appl. Phys.
  Lett.}\ }\textbf {\bibinfo {volume} {101}},\ \bibinfo {pages} {052601}
  (\bibinfo {year} {2012})},\ \Eprint {http://arxiv.org/abs/1207.2164}
  {arXiv:1207.2164 [physics.ins-det]} \BibitemShut {NoStop}%
\bibitem [{\citenamefont {Lita}\ \emph {et~al.}(2008)\citenamefont {Lita},
  \citenamefont {Miller},\ and\ \citenamefont {Nam}}]{Lita:08}%
  \BibitemOpen
  \bibfield  {author} {\bibinfo {author} {\bibfnamefont {A.~E.}\ \bibnamefont
  {Lita}}, \bibinfo {author} {\bibfnamefont {A.~J.}\ \bibnamefont {Miller}}, \
  and\ \bibinfo {author} {\bibfnamefont {S.~W.}\ \bibnamefont {Nam}},\ }\href
  {\doibase 10.1364/OE.16.003032} {\bibfield  {journal} {\bibinfo  {journal}
  {Opt. Express}\ }\textbf {\bibinfo {volume} {16}},\ \bibinfo {pages} {3032}
  (\bibinfo {year} {2008})}\BibitemShut {NoStop}%
\bibitem [{\citenamefont {Bastidon}\ \emph {et~al.}(2015)\citenamefont
  {Bastidon}, \citenamefont {Horns},\ and\ \citenamefont
  {Lindner}}]{Bastidon:2015aha}%
  \BibitemOpen
  \bibfield  {author} {\bibinfo {author} {\bibfnamefont {N.}~\bibnamefont
  {Bastidon}}, \bibinfo {author} {\bibfnamefont {D.}~\bibnamefont {Horns}}, \
  and\ \bibinfo {author} {\bibfnamefont {A.}~\bibnamefont {Lindner}}\
  }(\bibinfo {year} {2015})\ \Eprint {http://arxiv.org/abs/1509.02064}
  {arXiv:1509.02064 [physics.ins-det]} \BibitemShut {NoStop}%
\bibitem [{\citenamefont {{Mazin}}(2009)}]{Mazin}%
  \BibitemOpen
  \bibfield  {author} {\bibinfo {author} {\bibfnamefont {B.~A.}\ \bibnamefont
  {{Mazin}}},\ }in\ \href {\doibase 10.1063/1.3292300} {\emph {\bibinfo
  {booktitle} {American Institute of Physics Conference Series}}},\ \bibinfo
  {series} {American Institute of Physics Conference Series}, Vol.\ \bibinfo
  {volume} {1185},\ \bibinfo {editor} {edited by\ \bibinfo {editor}
  {\bibfnamefont {B.}~\bibnamefont {{Young}}}, \bibinfo {editor} {\bibfnamefont
  {B.}~\bibnamefont {{Cabrera}}}, \ and\ \bibinfo {editor} {\bibfnamefont
  {A.}~\bibnamefont {{Miller}}}}\ (\bibinfo {year} {2009})\ pp.\ \bibinfo
  {pages} {135--142}\BibitemShut {NoStop}%
\bibitem [{\citenamefont {Day}\ \emph {et~al.}(2003)\citenamefont {Day},
  \citenamefont {Leduc}, \citenamefont {A~Mazin}, \citenamefont {Vayonakis},\
  and\ \citenamefont {Zmuidzinas}}]{DayLeduc}%
  \BibitemOpen
  \bibfield  {author} {\bibinfo {author} {\bibfnamefont {P.}~\bibnamefont
  {Day}}, \bibinfo {author} {\bibfnamefont {H.}~\bibnamefont {Leduc}}, \bibinfo
  {author} {\bibfnamefont {B.}~\bibnamefont {A~Mazin}}, \bibinfo {author}
  {\bibfnamefont {A.}~\bibnamefont {Vayonakis}}, \ and\ \bibinfo {author}
  {\bibfnamefont {J.}~\bibnamefont {Zmuidzinas}},\ }\bibfield  {booktitle}
  {\emph {\bibinfo {booktitle} {Nature}},\ }\href@noop {} {\ \textbf {\bibinfo
  {volume} {425}},\ \bibinfo {pages} {817} (\bibinfo {year}
  {2003})}\BibitemShut {NoStop}%
\bibitem [{\citenamefont {Gao}\ \emph {et~al.}(2012)\citenamefont {Gao},
  \citenamefont {Vissers}, \citenamefont {Sandberg}, \citenamefont {da~Silva},
  \citenamefont {Nam}, \citenamefont {Pappas}, \citenamefont {Wisbey},
  \citenamefont {Langman}, \citenamefont {Meeker}, \citenamefont {Mazin},
  \citenamefont {Leduc}, \citenamefont {Zmuidzinas},\ and\ \citenamefont
  {Irwin}}]{GaoMazin}%
  \BibitemOpen
  \bibfield  {author} {\bibinfo {author} {\bibfnamefont {J.}~\bibnamefont
  {Gao}}, \bibinfo {author} {\bibfnamefont {M.~R.}\ \bibnamefont {Vissers}},
  \bibinfo {author} {\bibfnamefont {M.~O.}\ \bibnamefont {Sandberg}}, \bibinfo
  {author} {\bibfnamefont {F.~C.~S.}\ \bibnamefont {da~Silva}}, \bibinfo
  {author} {\bibfnamefont {S.~W.}\ \bibnamefont {Nam}}, \bibinfo {author}
  {\bibfnamefont {D.~P.}\ \bibnamefont {Pappas}}, \bibinfo {author}
  {\bibfnamefont {D.~S.}\ \bibnamefont {Wisbey}}, \bibinfo {author}
  {\bibfnamefont {E.~C.}\ \bibnamefont {Langman}}, \bibinfo {author}
  {\bibfnamefont {S.~R.}\ \bibnamefont {Meeker}}, \bibinfo {author}
  {\bibfnamefont {B.~A.}\ \bibnamefont {Mazin}}, \bibinfo {author}
  {\bibfnamefont {H.~G.}\ \bibnamefont {Leduc}}, \bibinfo {author}
  {\bibfnamefont {J.}~\bibnamefont {Zmuidzinas}}, \ and\ \bibinfo {author}
  {\bibfnamefont {K.~D.}\ \bibnamefont {Irwin}},\ }\href {\doibase
  10.1063/1.4756916} {\bibfield  {journal} {\bibinfo  {journal} {Applied
  Physics Letters}\ }\textbf {\bibinfo {volume} {101}},\ \bibinfo {pages}
  {142602} (\bibinfo {year} {2012})},\ \Eprint
  {http://arxiv.org/abs/https://doi.org/10.1063/1.4756916}
  {https://doi.org/10.1063/1.4756916} \BibitemShut {NoStop}%
\bibitem [{\citenamefont {Rosfjord}\ \emph
  {et~al.}(2006{\natexlab{a}})\citenamefont {Rosfjord}, \citenamefont {Yang},
  \citenamefont {Dauler}, \citenamefont {Kerman}, \citenamefont {Anant},
  \citenamefont {Voronov}, \citenamefont {Gol'tsman},\ and\ \citenamefont
  {Berggren}}]{Rosfjord:06}%
  \BibitemOpen
  \bibfield  {author} {\bibinfo {author} {\bibfnamefont {K.~M.}\ \bibnamefont
  {Rosfjord}}, \bibinfo {author} {\bibfnamefont {J.~K.~W.}\ \bibnamefont
  {Yang}}, \bibinfo {author} {\bibfnamefont {E.~A.}\ \bibnamefont {Dauler}},
  \bibinfo {author} {\bibfnamefont {A.~J.}\ \bibnamefont {Kerman}}, \bibinfo
  {author} {\bibfnamefont {V.}~\bibnamefont {Anant}}, \bibinfo {author}
  {\bibfnamefont {B.~M.}\ \bibnamefont {Voronov}}, \bibinfo {author}
  {\bibfnamefont {G.~N.}\ \bibnamefont {Gol'tsman}}, \ and\ \bibinfo {author}
  {\bibfnamefont {K.~K.}\ \bibnamefont {Berggren}},\ }\href {\doibase
  10.1364/OPEX.14.000527} {\bibfield  {journal} {\bibinfo  {journal} {Opt.
  Express}\ }\textbf {\bibinfo {volume} {14}},\ \bibinfo {pages} {527}
  (\bibinfo {year} {2006}{\natexlab{a}})}\BibitemShut {NoStop}%
\bibitem [{\citenamefont {Reddy}\ \emph {et~al.}(2020)\citenamefont {Reddy},
  \citenamefont {Nerem}, \citenamefont {Nam}, \citenamefont {Mirin},\ and\
  \citenamefont {Verma}}]{Reddy20}%
  \BibitemOpen
  \bibfield  {author} {\bibinfo {author} {\bibfnamefont {D.~V.}\ \bibnamefont
  {Reddy}}, \bibinfo {author} {\bibfnamefont {R.~R.}\ \bibnamefont {Nerem}},
  \bibinfo {author} {\bibfnamefont {S.~W.}\ \bibnamefont {Nam}}, \bibinfo
  {author} {\bibfnamefont {R.~P.}\ \bibnamefont {Mirin}}, \ and\ \bibinfo
  {author} {\bibfnamefont {V.~B.}\ \bibnamefont {Verma}},\ }\href {\doibase
  10.1364/OPTICA.400751} {\bibfield  {journal} {\bibinfo  {journal} {Optica}\
  }\textbf {\bibinfo {volume} {7}},\ \bibinfo {pages} {1649} (\bibinfo {year}
  {2020})}\BibitemShut {NoStop}%
\bibitem [{\citenamefont {Verma}\ \emph
  {et~al.}(2020{\natexlab{a}})\citenamefont {Verma} \emph
  {et~al.}}]{verma2020singlephoton}%
  \BibitemOpen
  \bibfield  {author} {\bibinfo {author} {\bibfnamefont {V.~B.}\ \bibnamefont
  {Verma}} \emph {et~al.},\ }\href@noop {} {\enquote {\bibinfo {title}
  {Single-photon detection in the mid-infrared up to 10 micron wavelength using
  tungsten silicide superconducting nanowire detectors},}\ } (\bibinfo {year}
  {2020}{\natexlab{a}}),\ \Eprint {http://arxiv.org/abs/2012.09979}
  {arXiv:2012.09979 [physics.ins-det]} \BibitemShut {NoStop}%
\bibitem [{\citenamefont {Wollman}\ \emph {et~al.}(2017)\citenamefont {Wollman}
  \emph {et~al.}}]{Wollman17}%
  \BibitemOpen
  \bibfield  {author} {\bibinfo {author} {\bibfnamefont {E.~E.}\ \bibnamefont
  {Wollman}} \emph {et~al.},\ }\href {\doibase 10.1364/OE.25.026792} {\bibfield
   {journal} {\bibinfo  {journal} {Opt. Express}\ }\textbf {\bibinfo {volume}
  {25}},\ \bibinfo {pages} {26792} (\bibinfo {year} {2017})}\BibitemShut
  {NoStop}%
\bibitem [{\citenamefont {Hochberg}\ \emph
  {et~al.}(2021{\natexlab{b}})\citenamefont {Hochberg}, \citenamefont
  {Lehmann}, \citenamefont {Charaev}, \citenamefont {Chiles}, \citenamefont
  {Nam},\ and\ \citenamefont {Berggren}}]{yonit}%
  \BibitemOpen
  \bibfield  {author} {\bibinfo {author} {\bibfnamefont {Y.}~\bibnamefont
  {Hochberg}}, \bibinfo {author} {\bibfnamefont {B.}~\bibnamefont {Lehmann}},
  \bibinfo {author} {\bibfnamefont {I.}~\bibnamefont {Charaev}}, \bibinfo
  {author} {\bibfnamefont {J.}~\bibnamefont {Chiles}}, \bibinfo {author}
  {\bibfnamefont {S.-W.}\ \bibnamefont {Nam}}, \ and\ \bibinfo {author}
  {\bibfnamefont {K.~K.}\ \bibnamefont {Berggren}},\ }\href@noop {} {\
  (\bibinfo {year} {2021}{\natexlab{b}})},\ \Eprint
  {http://arxiv.org/abs/2110.01586} {arXiv:2110.01586 [hep-ph]} \BibitemShut
  {NoStop}%
\bibitem [{\citenamefont {Bunting}\ \emph {et~al.}(2017)\citenamefont
  {Bunting}, \citenamefont {Gratta}, \citenamefont {Melia},\ and\ \citenamefont
  {Rajendran}}]{1701.06566}%
  \BibitemOpen
  \bibfield  {author} {\bibinfo {author} {\bibfnamefont {P.~C.}\ \bibnamefont
  {Bunting}}, \bibinfo {author} {\bibfnamefont {G.}~\bibnamefont {Gratta}},
  \bibinfo {author} {\bibfnamefont {T.}~\bibnamefont {Melia}}, \ and\ \bibinfo
  {author} {\bibfnamefont {S.}~\bibnamefont {Rajendran}},\ }\href {\doibase
  10.1103/PhysRevD.95.095001} {\bibfield  {journal} {\bibinfo  {journal} {Phys.
  Rev. D}\ }\textbf {\bibinfo {volume} {95}},\ \bibinfo {pages} {095001}
  (\bibinfo {year} {2017})},\ \Eprint {http://arxiv.org/abs/1701.06566}
  {arXiv:1701.06566 [hep-ph]} \BibitemShut {NoStop}%
\bibitem [{\citenamefont {Dolgov}\ \emph {et~al.}(1981)\citenamefont {Dolgov},
  \citenamefont {Kirzhnits},\ and\ \citenamefont
  {Maksimov}}]{10.1103/RevModPhys.53.81}%
  \BibitemOpen
  \bibfield  {author} {\bibinfo {author} {\bibfnamefont {O.~V.}\ \bibnamefont
  {Dolgov}}, \bibinfo {author} {\bibfnamefont {D.~A.}\ \bibnamefont
  {Kirzhnits}}, \ and\ \bibinfo {author} {\bibfnamefont {E.~G.}\ \bibnamefont
  {Maksimov}},\ }\href {\doibase 10.1103/revmodphys.53.81} {\bibfield
  {journal} {\bibinfo  {journal} {Reviews of Modern Physics}\ }\textbf
  {\bibinfo {volume} {53}},\ \bibinfo {pages} {81} (\bibinfo {year}
  {1981})}\BibitemShut {NoStop}%
\bibitem [{\citenamefont {DeRocco}\ \emph {et~al.}(2022)\citenamefont
  {DeRocco}, \citenamefont {Galanis},\ and\ \citenamefont {Lasenby}}]{wdr}%
  \BibitemOpen
  \bibfield  {author} {\bibinfo {author} {\bibfnamefont {W.}~\bibnamefont
  {DeRocco}}, \bibinfo {author} {\bibfnamefont {M.}~\bibnamefont {Galanis}}, \
  and\ \bibinfo {author} {\bibfnamefont {R.}~\bibnamefont {Lasenby}},\
  }\href@noop {} {\  (\bibinfo {year} {2022})},\ \Eprint
  {http://arxiv.org/abs/2201.05167} {arXiv:2201.05167 [hep-ph]} \BibitemShut
  {NoStop}%
\bibitem [{\citenamefont {Carney}\ \emph {et~al.}(2021)\citenamefont {Carney},
  \citenamefont {H\"affner}, \citenamefont {Moore},\ and\ \citenamefont
  {Taylor}}]{2104.05737}%
  \BibitemOpen
  \bibfield  {author} {\bibinfo {author} {\bibfnamefont {D.}~\bibnamefont
  {Carney}}, \bibinfo {author} {\bibfnamefont {H.}~\bibnamefont {H\"affner}},
  \bibinfo {author} {\bibfnamefont {D.~C.}\ \bibnamefont {Moore}}, \ and\
  \bibinfo {author} {\bibfnamefont {J.~M.}\ \bibnamefont {Taylor}},\
  }\href@noop {} {\  (\bibinfo {year} {2021})},\ \Eprint
  {http://arxiv.org/abs/2104.05737} {arXiv:2104.05737 [quant-ph]} \BibitemShut
  {NoStop}%
\bibitem [{\citenamefont {Budker}\ \emph {et~al.}(2021)\citenamefont {Budker},
  \citenamefont {Graham}, \citenamefont {Ramani}, \citenamefont
  {Schmidt-Kaler}, \citenamefont {Smorra},\ and\ \citenamefont
  {Ulmer}}]{2108.05283}%
  \BibitemOpen
  \bibfield  {author} {\bibinfo {author} {\bibfnamefont {D.}~\bibnamefont
  {Budker}}, \bibinfo {author} {\bibfnamefont {P.~W.}\ \bibnamefont {Graham}},
  \bibinfo {author} {\bibfnamefont {H.}~\bibnamefont {Ramani}}, \bibinfo
  {author} {\bibfnamefont {F.}~\bibnamefont {Schmidt-Kaler}}, \bibinfo {author}
  {\bibfnamefont {C.}~\bibnamefont {Smorra}}, \ and\ \bibinfo {author}
  {\bibfnamefont {S.}~\bibnamefont {Ulmer}},\ }\href@noop {} {\  (\bibinfo
  {year} {2021})},\ \Eprint {http://arxiv.org/abs/2108.05283} {arXiv:2108.05283
  [hep-ph]} \BibitemShut {NoStop}%
\bibitem [{\citenamefont {Sun}\ \emph {et~al.}(2016)\citenamefont {Sun},
  \citenamefont {Xu}, \citenamefont {Da}, \citenamefont {Mao},\ and\
  \citenamefont {Ding}}]{10.1063/1674-0068/29/cjcp1605110}%
  \BibitemOpen
  \bibfield  {author} {\bibinfo {author} {\bibfnamefont {Y.}~\bibnamefont
  {Sun}}, \bibinfo {author} {\bibfnamefont {H.}~\bibnamefont {Xu}}, \bibinfo
  {author} {\bibfnamefont {B.}~\bibnamefont {Da}}, \bibinfo {author}
  {\bibfnamefont {S.-f.}\ \bibnamefont {Mao}}, \ and\ \bibinfo {author}
  {\bibfnamefont {Z.-j.}\ \bibnamefont {Ding}},\ }\href {\doibase
  10.1063/1674-0068/29/cjcp1605110} {\bibfield  {journal} {\bibinfo  {journal}
  {Chinese Journal of Chemical Physics}\ }\textbf {\bibinfo {volume} {29}},\
  \bibinfo {pages} {663} (\bibinfo {year} {2016})},\ \Eprint
  {http://arxiv.org/abs/https://doi.org/10.1063/1674-0068/29/cjcp1605110}
  {https://doi.org/10.1063/1674-0068/29/cjcp1605110} \BibitemShut {NoStop}%
\bibitem [{\citenamefont {Knapen}\ \emph
  {et~al.}(2021{\natexlab{b}})\citenamefont {Knapen}, \citenamefont
  {Kozaczuk},\ and\ \citenamefont {Lin}}]{2104.12786}%
  \BibitemOpen
  \bibfield  {author} {\bibinfo {author} {\bibfnamefont {S.}~\bibnamefont
  {Knapen}}, \bibinfo {author} {\bibfnamefont {J.}~\bibnamefont {Kozaczuk}}, \
  and\ \bibinfo {author} {\bibfnamefont {T.}~\bibnamefont {Lin}},\ }\href@noop
  {} {\  (\bibinfo {year} {2021}{\natexlab{b}})},\ \Eprint
  {http://arxiv.org/abs/2104.12786} {arXiv:2104.12786 [hep-ph]} \BibitemShut
  {NoStop}%
\bibitem [{\citenamefont {Coskuner}\ \emph
  {et~al.}(2021{\natexlab{b}})\citenamefont {Coskuner}, \citenamefont
  {Trickle}, \citenamefont {Zhang},\ and\ \citenamefont {Zurek}}]{2102.09567}%
  \BibitemOpen
  \bibfield  {author} {\bibinfo {author} {\bibfnamefont {A.}~\bibnamefont
  {Coskuner}}, \bibinfo {author} {\bibfnamefont {T.}~\bibnamefont {Trickle}},
  \bibinfo {author} {\bibfnamefont {Z.}~\bibnamefont {Zhang}}, \ and\ \bibinfo
  {author} {\bibfnamefont {K.~M.}\ \bibnamefont {Zurek}},\ }\href@noop {} {\
  (\bibinfo {year} {2021}{\natexlab{b}})},\ \Eprint
  {http://arxiv.org/abs/2102.09567} {arXiv:2102.09567 [hep-ph]} \BibitemShut
  {NoStop}%
\bibitem [{\citenamefont {Knapen}\ \emph
  {et~al.}(2018{\natexlab{a}})\citenamefont {Knapen}, \citenamefont {Lin},
  \citenamefont {Pyle},\ and\ \citenamefont {Zurek}}]{1712.06598}%
  \BibitemOpen
  \bibfield  {author} {\bibinfo {author} {\bibfnamefont {S.}~\bibnamefont
  {Knapen}}, \bibinfo {author} {\bibfnamefont {T.}~\bibnamefont {Lin}},
  \bibinfo {author} {\bibfnamefont {M.}~\bibnamefont {Pyle}}, \ and\ \bibinfo
  {author} {\bibfnamefont {K.~M.}\ \bibnamefont {Zurek}},\ }\href {\doibase
  10.1016/j.physletb.2018.08.064} {\bibfield  {journal} {\bibinfo  {journal}
  {Phys. Lett. B}\ }\textbf {\bibinfo {volume} {785}},\ \bibinfo {pages} {386}
  (\bibinfo {year} {2018}{\natexlab{a}})},\ \Eprint
  {http://arxiv.org/abs/1712.06598} {arXiv:1712.06598 [hep-ph]} \BibitemShut
  {NoStop}%
\bibitem [{\citenamefont {Griffin}\ \emph
  {et~al.}(2018{\natexlab{a}})\citenamefont {Griffin}, \citenamefont {Knapen},
  \citenamefont {Lin},\ and\ \citenamefont {Zurek}}]{1807.10291}%
  \BibitemOpen
  \bibfield  {author} {\bibinfo {author} {\bibfnamefont {S.}~\bibnamefont
  {Griffin}}, \bibinfo {author} {\bibfnamefont {S.}~\bibnamefont {Knapen}},
  \bibinfo {author} {\bibfnamefont {T.}~\bibnamefont {Lin}}, \ and\ \bibinfo
  {author} {\bibfnamefont {K.~M.}\ \bibnamefont {Zurek}},\ }\href {\doibase
  10.1103/PhysRevD.98.115034} {\bibfield  {journal} {\bibinfo  {journal} {Phys.
  Rev. D}\ }\textbf {\bibinfo {volume} {98}},\ \bibinfo {pages} {115034}
  (\bibinfo {year} {2018}{\natexlab{a}})},\ \Eprint
  {http://arxiv.org/abs/1807.10291} {arXiv:1807.10291 [hep-ph]} \BibitemShut
  {NoStop}%
\bibitem [{kzh()}]{kzhang}%
  \BibitemOpen
  \href@noop {} {}\bibinfo {howpublished} {T. Trickle and Z.\ Zhang, personal
  communication}\BibitemShut {NoStop}%
\bibitem [{\citenamefont {Gelmini}\ \emph {et~al.}(2020)\citenamefont
  {Gelmini}, \citenamefont {Takhistov},\ and\ \citenamefont
  {Vitagliano}}]{2006.13909}%
  \BibitemOpen
  \bibfield  {author} {\bibinfo {author} {\bibfnamefont {G.~B.}\ \bibnamefont
  {Gelmini}}, \bibinfo {author} {\bibfnamefont {V.}~\bibnamefont {Takhistov}},
  \ and\ \bibinfo {author} {\bibfnamefont {E.}~\bibnamefont {Vitagliano}},\
  }\href {\doibase 10.1016/j.physletb.2020.135779} {\bibfield  {journal}
  {\bibinfo  {journal} {Phys. Lett. B}\ }\textbf {\bibinfo {volume} {809}},\
  \bibinfo {pages} {135779} (\bibinfo {year} {2020})},\ \Eprint
  {http://arxiv.org/abs/2006.13909} {arXiv:2006.13909 [hep-ph]} \BibitemShut
  {NoStop}%
\bibitem [{\citenamefont {Fasolino}\ \emph {et~al.}(1978)\citenamefont
  {Fasolino}, \citenamefont {Parrinello},\ and\ \citenamefont
  {Tosi}}]{10.1016/0375-9601(78)90013-0}%
  \BibitemOpen
  \bibfield  {author} {\bibinfo {author} {\bibfnamefont {A.}~\bibnamefont
  {Fasolino}}, \bibinfo {author} {\bibfnamefont {M.}~\bibnamefont
  {Parrinello}}, \ and\ \bibinfo {author} {\bibfnamefont {M.}~\bibnamefont
  {Tosi}},\ }\href {\doibase 10.1016/0375-9601(78)90013-0} {\bibfield
  {journal} {\bibinfo  {journal} {Physics Letters A}\ }\textbf {\bibinfo
  {volume} {66}},\ \bibinfo {pages} {119} (\bibinfo {year} {1978})}\BibitemShut
  {NoStop}%
\bibitem [{\citenamefont {Geilhufe}\ \emph {et~al.}(2018)\citenamefont
  {Geilhufe}, \citenamefont {Olsthoorn}, \citenamefont {Ferella}, \citenamefont
  {Koski}, \citenamefont {Kahlhoefer}, \citenamefont {Conrad},\ and\
  \citenamefont {Balatsky}}]{1806.06040}%
  \BibitemOpen
  \bibfield  {author} {\bibinfo {author} {\bibfnamefont {R.~M.}\ \bibnamefont
  {Geilhufe}}, \bibinfo {author} {\bibfnamefont {B.}~\bibnamefont {Olsthoorn}},
  \bibinfo {author} {\bibfnamefont {A.~D.}\ \bibnamefont {Ferella}}, \bibinfo
  {author} {\bibfnamefont {T.}~\bibnamefont {Koski}}, \bibinfo {author}
  {\bibfnamefont {F.}~\bibnamefont {Kahlhoefer}}, \bibinfo {author}
  {\bibfnamefont {J.}~\bibnamefont {Conrad}}, \ and\ \bibinfo {author}
  {\bibfnamefont {A.~V.}\ \bibnamefont {Balatsky}},\ }\href {\doibase
  10.1002/pssr.201800293} {\bibfield  {journal} {\bibinfo  {journal} {physica
  status solidi ({RRL}) - Rapid Research Letters}\ }\textbf {\bibinfo {volume}
  {12}},\ \bibinfo {pages} {1800293} (\bibinfo {year} {2018})}\BibitemShut
  {NoStop}%
\bibitem [{\citenamefont {Knickerbocker}\ and\ \citenamefont
  {Kulkarni}(1996)}]{10.1116/1.580384}%
  \BibitemOpen
  \bibfield  {author} {\bibinfo {author} {\bibfnamefont {S.~A.}\ \bibnamefont
  {Knickerbocker}}\ and\ \bibinfo {author} {\bibfnamefont {A.~K.}\ \bibnamefont
  {Kulkarni}},\ }\href {\doibase 10.1116/1.580384} {\bibfield  {journal}
  {\bibinfo  {journal} {Journal of Vacuum Science {\&} Technology A: Vacuum,
  Surfaces, and Films}\ }\textbf {\bibinfo {volume} {14}},\ \bibinfo {pages}
  {757} (\bibinfo {year} {1996})}\BibitemShut {NoStop}%
\bibitem [{\citenamefont {Bozovic}(1990)}]{10.1103/PhysRevB.42.1969}%
  \BibitemOpen
  \bibfield  {author} {\bibinfo {author} {\bibfnamefont {I.}~\bibnamefont
  {Bozovic}},\ }\href {\doibase 10.1103/physrevb.42.1969} {\bibfield  {journal}
  {\bibinfo  {journal} {Physical Review B}\ }\textbf {\bibinfo {volume} {42}},\
  \bibinfo {pages} {1969} (\bibinfo {year} {1990})}\BibitemShut {NoStop}%
\bibitem [{\citenamefont {Kuz'michev}\ and\ \citenamefont
  {Motulevich}(1983)}]{nbn}%
  \BibitemOpen
  \bibfield  {author} {\bibinfo {author} {\bibfnamefont {N.~D.}\ \bibnamefont
  {Kuz'michev}}\ and\ \bibinfo {author} {\bibfnamefont {G.~P.}\ \bibnamefont
  {Motulevich}},\ }\href@noop {} {\enquote {\bibinfo {title} {Determination of
  the electron characteristics of niobium nitride by an optical method},}\ }
  (\bibinfo {year} {1983})\BibitemShut {NoStop}%
\bibitem [{\citenamefont {Cai}\ and\ \citenamefont {Shalaev}(2010)}]{Cai2010}%
  \BibitemOpen
  \bibfield  {author} {\bibinfo {author} {\bibfnamefont {W.}~\bibnamefont
  {Cai}}\ and\ \bibinfo {author} {\bibfnamefont {V.}~\bibnamefont {Shalaev}},\
  }\href {\doibase 10.1007/978-1-4419-1151-3} {\emph {\bibinfo {title} {Optical
  Metamaterials}}}\ (\bibinfo  {publisher} {Springer New York},\ \bibinfo
  {year} {2010})\BibitemShut {NoStop}%
\bibitem [{\citenamefont {Maleeva}\ \emph {et~al.}(2018)\citenamefont
  {Maleeva}, \citenamefont {Gr\"{u}nhaupt}, \citenamefont {Klein},
  \citenamefont {Levy-Bertrand}, \citenamefont {Dupre}, \citenamefont {Calvo},
  \citenamefont {Valenti}, \citenamefont {Winkel}, \citenamefont {Friedrich},
  \citenamefont {Wernsdorfer}, \citenamefont {Ustinov}, \citenamefont
  {Rotzinger}, \citenamefont {Monfardini}, \citenamefont {Fistul},\ and\
  \citenamefont {Pop}}]{Maleeva2018}%
  \BibitemOpen
  \bibfield  {author} {\bibinfo {author} {\bibfnamefont {N.}~\bibnamefont
  {Maleeva}}, \bibinfo {author} {\bibfnamefont {L.}~\bibnamefont
  {Gr\"{u}nhaupt}}, \bibinfo {author} {\bibfnamefont {T.}~\bibnamefont
  {Klein}}, \bibinfo {author} {\bibfnamefont {F.}~\bibnamefont
  {Levy-Bertrand}}, \bibinfo {author} {\bibfnamefont {O.}~\bibnamefont
  {Dupre}}, \bibinfo {author} {\bibfnamefont {M.}~\bibnamefont {Calvo}},
  \bibinfo {author} {\bibfnamefont {F.}~\bibnamefont {Valenti}}, \bibinfo
  {author} {\bibfnamefont {P.}~\bibnamefont {Winkel}}, \bibinfo {author}
  {\bibfnamefont {F.}~\bibnamefont {Friedrich}}, \bibinfo {author}
  {\bibfnamefont {W.}~\bibnamefont {Wernsdorfer}}, \bibinfo {author}
  {\bibfnamefont {A.~V.}\ \bibnamefont {Ustinov}}, \bibinfo {author}
  {\bibfnamefont {H.}~\bibnamefont {Rotzinger}}, \bibinfo {author}
  {\bibfnamefont {A.}~\bibnamefont {Monfardini}}, \bibinfo {author}
  {\bibfnamefont {M.~V.}\ \bibnamefont {Fistul}}, \ and\ \bibinfo {author}
  {\bibfnamefont {I.~M.}\ \bibnamefont {Pop}},\ }\href {\doibase
  10.1038/s41467-018-06386-9} {\bibfield  {journal} {\bibinfo  {journal}
  {Nature Communications}\ }\textbf {\bibinfo {volume} {9}} (\bibinfo {year}
  {2018}),\ 10.1038/s41467-018-06386-9}\BibitemShut {NoStop}%
\bibitem [{\citenamefont {Vogel}\ and\ \citenamefont
  {Redondo}(2014)}]{10.1088/1475-7516/2014/02/029}%
  \BibitemOpen
  \bibfield  {author} {\bibinfo {author} {\bibfnamefont {H.}~\bibnamefont
  {Vogel}}\ and\ \bibinfo {author} {\bibfnamefont {J.}~\bibnamefont
  {Redondo}},\ }\href {\doibase 10.1088/1475-7516/2014/02/029} {\bibfield
  {journal} {\bibinfo  {journal} {Journal of Cosmology and Astroparticle
  Physics}\ }\textbf {\bibinfo {volume} {2014}},\ \bibinfo {pages} {029}
  (\bibinfo {year} {2014})}\BibitemShut {NoStop}%
\bibitem [{\citenamefont {Chang}\ \emph {et~al.}(2018)\citenamefont {Chang},
  \citenamefont {Essig},\ and\ \citenamefont
  {McDermott}}]{10.1007/JHEP09(2018)051}%
  \BibitemOpen
  \bibfield  {author} {\bibinfo {author} {\bibfnamefont {J.~H.}\ \bibnamefont
  {Chang}}, \bibinfo {author} {\bibfnamefont {R.}~\bibnamefont {Essig}}, \ and\
  \bibinfo {author} {\bibfnamefont {S.~D.}\ \bibnamefont {McDermott}},\ }\href
  {\doibase 10.1007/jhep09(2018)051} {\bibfield  {journal} {\bibinfo  {journal}
  {Journal of High Energy Physics}\ }\textbf {\bibinfo {volume} {2018}}
  (\bibinfo {year} {2018}),\ 10.1007/jhep09(2018)051}\BibitemShut {NoStop}%
\bibitem [{\citenamefont {Essig}\ \emph
  {et~al.}(2012{\natexlab{c}})\citenamefont {Essig}, \citenamefont
  {Manalaysay}, \citenamefont {Mardon}, \citenamefont {Sorensen},\ and\
  \citenamefont {Volansky}}]{10.1103/PhysRevLett.109.021301}%
  \BibitemOpen
  \bibfield  {author} {\bibinfo {author} {\bibfnamefont {R.}~\bibnamefont
  {Essig}}, \bibinfo {author} {\bibfnamefont {A.}~\bibnamefont {Manalaysay}},
  \bibinfo {author} {\bibfnamefont {J.}~\bibnamefont {Mardon}}, \bibinfo
  {author} {\bibfnamefont {P.}~\bibnamefont {Sorensen}}, \ and\ \bibinfo
  {author} {\bibfnamefont {T.}~\bibnamefont {Volansky}},\ }\href {\doibase
  10.1103/physrevlett.109.021301} {\bibfield  {journal} {\bibinfo  {journal}
  {Physical Review Letters}\ }\textbf {\bibinfo {volume} {109}} (\bibinfo
  {year} {2012}{\natexlab{c}}),\ 10.1103/physrevlett.109.021301}\BibitemShut
  {NoStop}%
\bibitem [{\citenamefont {Essig}\ \emph {et~al.}(2017)\citenamefont {Essig},
  \citenamefont {Volansky},\ and\ \citenamefont
  {Yu}}]{10.1103/PhysRevD.96.043017}%
  \BibitemOpen
  \bibfield  {author} {\bibinfo {author} {\bibfnamefont {R.}~\bibnamefont
  {Essig}}, \bibinfo {author} {\bibfnamefont {T.}~\bibnamefont {Volansky}}, \
  and\ \bibinfo {author} {\bibfnamefont {T.-T.}\ \bibnamefont {Yu}},\ }\href
  {\doibase 10.1103/physrevd.96.043017} {\bibfield  {journal} {\bibinfo
  {journal} {Physical Review D}\ }\textbf {\bibinfo {volume} {96}} (\bibinfo
  {year} {2017}),\ 10.1103/physrevd.96.043017}\BibitemShut {NoStop}%
\bibitem [{\citenamefont {Yakubovsky}\ \emph {et~al.}(2017)\citenamefont
  {Yakubovsky}, \citenamefont {Arsenin}, \citenamefont {Stebunov},
  \citenamefont {Fedyanin},\ and\ \citenamefont
  {Volkov}}]{10.1364/OE.25.025574}%
  \BibitemOpen
  \bibfield  {author} {\bibinfo {author} {\bibfnamefont {D.~I.}\ \bibnamefont
  {Yakubovsky}}, \bibinfo {author} {\bibfnamefont {A.~V.}\ \bibnamefont
  {Arsenin}}, \bibinfo {author} {\bibfnamefont {Y.~V.}\ \bibnamefont
  {Stebunov}}, \bibinfo {author} {\bibfnamefont {D.~Y.}\ \bibnamefont
  {Fedyanin}}, \ and\ \bibinfo {author} {\bibfnamefont {V.~S.}\ \bibnamefont
  {Volkov}},\ }\href {\doibase 10.1364/oe.25.025574} {\bibfield  {journal}
  {\bibinfo  {journal} {Optics Express}\ }\textbf {\bibinfo {volume} {25}},\
  \bibinfo {pages} {25574} (\bibinfo {year} {2017})}\BibitemShut {NoStop}%
\bibitem [{\citenamefont {Chiles}\ \emph {et~al.}(2021)\citenamefont {Chiles},
  \citenamefont {Charaev}, \citenamefont {Lasenby}, \citenamefont {Baryakhtar},
  \citenamefont {Huang}, \citenamefont {Roshko}, \citenamefont {Burton},
  \citenamefont {Colangelo}, \citenamefont {Van~Tilburg}, \citenamefont
  {Arvanitaki}, \citenamefont {Nam},\ and\ \citenamefont {Berggren}}]{lampost}%
  \BibitemOpen
  \bibfield  {author} {\bibinfo {author} {\bibfnamefont {J.}~\bibnamefont
  {Chiles}}, \bibinfo {author} {\bibfnamefont {I.}~\bibnamefont {Charaev}},
  \bibinfo {author} {\bibfnamefont {R.}~\bibnamefont {Lasenby}}, \bibinfo
  {author} {\bibfnamefont {M.}~\bibnamefont {Baryakhtar}}, \bibinfo {author}
  {\bibfnamefont {J.}~\bibnamefont {Huang}}, \bibinfo {author} {\bibfnamefont
  {A.}~\bibnamefont {Roshko}}, \bibinfo {author} {\bibfnamefont
  {G.}~\bibnamefont {Burton}}, \bibinfo {author} {\bibfnamefont
  {M.}~\bibnamefont {Colangelo}}, \bibinfo {author} {\bibfnamefont
  {K.}~\bibnamefont {Van~Tilburg}}, \bibinfo {author} {\bibfnamefont
  {A.}~\bibnamefont {Arvanitaki}}, \bibinfo {author} {\bibfnamefont {S.-W.}\
  \bibnamefont {Nam}}, \ and\ \bibinfo {author} {\bibfnamefont {K.~K.}\
  \bibnamefont {Berggren}},\ }\href@noop {} {\  (\bibinfo {year} {2021})},\
  \Eprint {http://arxiv.org/abs/2110.01582} {arXiv:2110.01582 [hep-ex]}
  \BibitemShut {NoStop}%
\bibitem [{\citenamefont {Polakovic}\ \emph {et~al.}(2020)\citenamefont
  {Polakovic}, \citenamefont {Armstrong}, \citenamefont {Karapetrov},
  \citenamefont {Meziani},\ and\ \citenamefont
  {Novosad}}]{10.3390/nano10061198}%
  \BibitemOpen
  \bibfield  {author} {\bibinfo {author} {\bibfnamefont {T.}~\bibnamefont
  {Polakovic}}, \bibinfo {author} {\bibfnamefont {W.}~\bibnamefont
  {Armstrong}}, \bibinfo {author} {\bibfnamefont {G.}~\bibnamefont
  {Karapetrov}}, \bibinfo {author} {\bibfnamefont {Z.-E.}\ \bibnamefont
  {Meziani}}, \ and\ \bibinfo {author} {\bibfnamefont {V.}~\bibnamefont
  {Novosad}},\ }\href {\doibase 10.3390/nano10061198} {\bibfield  {journal}
  {\bibinfo  {journal} {Nanomaterials}\ }\textbf {\bibinfo {volume} {10}},\
  \bibinfo {pages} {1198} (\bibinfo {year} {2020})}\BibitemShut {NoStop}%
\bibitem [{\citenamefont {Knapen}\ \emph
  {et~al.}(2018{\natexlab{b}})\citenamefont {Knapen}, \citenamefont {Lin},
  \citenamefont {Pyle},\ and\ \citenamefont {Zurek}}]{Knapen_2018}%
  \BibitemOpen
  \bibfield  {author} {\bibinfo {author} {\bibfnamefont {S.}~\bibnamefont
  {Knapen}}, \bibinfo {author} {\bibfnamefont {T.}~\bibnamefont {Lin}},
  \bibinfo {author} {\bibfnamefont {M.}~\bibnamefont {Pyle}}, \ and\ \bibinfo
  {author} {\bibfnamefont {K.~M.}\ \bibnamefont {Zurek}},\ }\href {\doibase
  10.1016/j.physletb.2018.08.064} {\bibfield  {journal} {\bibinfo  {journal}
  {Physics Letters B}\ }\textbf {\bibinfo {volume} {785}},\ \bibinfo {pages}
  {386–390} (\bibinfo {year} {2018}{\natexlab{b}})}\BibitemShut {NoStop}%
\bibitem [{\citenamefont {Griffin}\ \emph
  {et~al.}(2018{\natexlab{b}})\citenamefont {Griffin}, \citenamefont {Knapen},
  \citenamefont {Lin},\ and\ \citenamefont {Zurek}}]{Griffin2018}%
  \BibitemOpen
  \bibfield  {author} {\bibinfo {author} {\bibfnamefont {S.}~\bibnamefont
  {Griffin}}, \bibinfo {author} {\bibfnamefont {S.}~\bibnamefont {Knapen}},
  \bibinfo {author} {\bibfnamefont {T.}~\bibnamefont {Lin}}, \ and\ \bibinfo
  {author} {\bibfnamefont {K.~M.}\ \bibnamefont {Zurek}},\ }\href {\doibase
  10.1103/PhysRevD.98.115034} {\bibfield  {journal} {\bibinfo  {journal} {Phys.
  Rev. D}\ }\textbf {\bibinfo {volume} {98}},\ \bibinfo {pages} {115034}
  (\bibinfo {year} {2018}{\natexlab{b}})}\BibitemShut {NoStop}%
\bibitem [{\citenamefont {Rosfjord}\ \emph
  {et~al.}(2006{\natexlab{b}})\citenamefont {Rosfjord}, \citenamefont {Yang},
  \citenamefont {Dauler}, \citenamefont {Kerman}, \citenamefont {Anant},
  \citenamefont {Voronov}, \citenamefont {Gol'tsman},\ and\ \citenamefont
  {Berggren}}]{Rosfjord06}%
  \BibitemOpen
  \bibfield  {author} {\bibinfo {author} {\bibfnamefont {K.~M.}\ \bibnamefont
  {Rosfjord}}, \bibinfo {author} {\bibfnamefont {J.~K.~W.}\ \bibnamefont
  {Yang}}, \bibinfo {author} {\bibfnamefont {E.~A.}\ \bibnamefont {Dauler}},
  \bibinfo {author} {\bibfnamefont {A.~J.}\ \bibnamefont {Kerman}}, \bibinfo
  {author} {\bibfnamefont {V.}~\bibnamefont {Anant}}, \bibinfo {author}
  {\bibfnamefont {B.~M.}\ \bibnamefont {Voronov}}, \bibinfo {author}
  {\bibfnamefont {G.~N.}\ \bibnamefont {Gol'tsman}}, \ and\ \bibinfo {author}
  {\bibfnamefont {K.~K.}\ \bibnamefont {Berggren}},\ }\href {\doibase
  10.1364/OPEX.14.000527} {\bibfield  {journal} {\bibinfo  {journal} {Opt.
  Express}\ }\textbf {\bibinfo {volume} {14}},\ \bibinfo {pages} {527}
  (\bibinfo {year} {2006}{\natexlab{b}})}\BibitemShut {NoStop}%
\bibitem [{\citenamefont {Verma}\ \emph
  {et~al.}(2020{\natexlab{b}})\citenamefont {Verma}, \citenamefont {Korzh},
  \citenamefont {Walter}, \citenamefont {Lita}, \citenamefont {Briggs},
  \citenamefont {Colangelo}, \citenamefont {Zhai}, \citenamefont {Wollman},
  \citenamefont {Beyer}, \citenamefont {Allmaras}, \citenamefont {Bumble},
  \citenamefont {Vora}, \citenamefont {Zhu}, \citenamefont {Schmidt},
  \citenamefont {Berggren}, \citenamefont {Mirin}, \citenamefont {Nam},\ and\
  \citenamefont {Shaw}}]{verma2020}%
  \BibitemOpen
  \bibfield  {author} {\bibinfo {author} {\bibfnamefont {V.~B.}\ \bibnamefont
  {Verma}}, \bibinfo {author} {\bibfnamefont {B.}~\bibnamefont {Korzh}},
  \bibinfo {author} {\bibfnamefont {A.~B.}\ \bibnamefont {Walter}}, \bibinfo
  {author} {\bibfnamefont {A.~E.}\ \bibnamefont {Lita}}, \bibinfo {author}
  {\bibfnamefont {R.~M.}\ \bibnamefont {Briggs}}, \bibinfo {author}
  {\bibfnamefont {M.}~\bibnamefont {Colangelo}}, \bibinfo {author}
  {\bibfnamefont {Y.}~\bibnamefont {Zhai}}, \bibinfo {author} {\bibfnamefont
  {E.~E.}\ \bibnamefont {Wollman}}, \bibinfo {author} {\bibfnamefont {A.~D.}\
  \bibnamefont {Beyer}}, \bibinfo {author} {\bibfnamefont {J.~P.}\ \bibnamefont
  {Allmaras}}, \bibinfo {author} {\bibfnamefont {B.}~\bibnamefont {Bumble}},
  \bibinfo {author} {\bibfnamefont {H.}~\bibnamefont {Vora}}, \bibinfo {author}
  {\bibfnamefont {D.}~\bibnamefont {Zhu}}, \bibinfo {author} {\bibfnamefont
  {E.}~\bibnamefont {Schmidt}}, \bibinfo {author} {\bibfnamefont {K.~K.}\
  \bibnamefont {Berggren}}, \bibinfo {author} {\bibfnamefont {R.~P.}\
  \bibnamefont {Mirin}}, \bibinfo {author} {\bibfnamefont {S.~W.}\ \bibnamefont
  {Nam}}, \ and\ \bibinfo {author} {\bibfnamefont {M.~D.}\ \bibnamefont
  {Shaw}},\ }\href@noop {} {\enquote {\bibinfo {title} {Single-photon detection
  in the mid-infrared up to 10 micron wavelength using tungsten silicide
  superconducting nanowire detectors},}\ } (\bibinfo {year}
  {2020}{\natexlab{b}}),\ \Eprint {http://arxiv.org/abs/2012.09979}
  {arXiv:2012.09979 [physics.ins-det]} \BibitemShut {NoStop}%
\bibitem [{\citenamefont {Bell}\ \emph {et~al.}(2020)\citenamefont {Bell},
  \citenamefont {Busoni}, \citenamefont {Robles},\ and\ \citenamefont
  {Virgato}}]{2004.14888}%
  \BibitemOpen
  \bibfield  {author} {\bibinfo {author} {\bibfnamefont {N.~F.}\ \bibnamefont
  {Bell}}, \bibinfo {author} {\bibfnamefont {G.}~\bibnamefont {Busoni}},
  \bibinfo {author} {\bibfnamefont {S.}~\bibnamefont {Robles}}, \ and\ \bibinfo
  {author} {\bibfnamefont {M.}~\bibnamefont {Virgato}},\ }\href {\doibase
  10.1088/1475-7516/2020/09/028} {\bibfield  {journal} {\bibinfo  {journal}
  {JCAP}\ }\textbf {\bibinfo {volume} {09}},\ \bibinfo {pages} {028} (\bibinfo
  {year} {2020})},\ \Eprint {http://arxiv.org/abs/2004.14888} {arXiv:2004.14888
  [hep-ph]} \BibitemShut {NoStop}%
\bibitem [{\citenamefont {Bell}\ \emph
  {et~al.}(2021{\natexlab{a}})\citenamefont {Bell}, \citenamefont {Busoni},
  \citenamefont {Robles},\ and\ \citenamefont {Virgato}}]{2010.13257}%
  \BibitemOpen
  \bibfield  {author} {\bibinfo {author} {\bibfnamefont {N.~F.}\ \bibnamefont
  {Bell}}, \bibinfo {author} {\bibfnamefont {G.}~\bibnamefont {Busoni}},
  \bibinfo {author} {\bibfnamefont {S.}~\bibnamefont {Robles}}, \ and\ \bibinfo
  {author} {\bibfnamefont {M.}~\bibnamefont {Virgato}},\ }\href {\doibase
  10.1088/1475-7516/2021/03/086} {\bibfield  {journal} {\bibinfo  {journal}
  {JCAP}\ }\textbf {\bibinfo {volume} {03}},\ \bibinfo {pages} {086} (\bibinfo
  {year} {2021}{\natexlab{a}})},\ \Eprint {http://arxiv.org/abs/2010.13257}
  {arXiv:2010.13257 [hep-ph]} \BibitemShut {NoStop}%
\bibitem [{\citenamefont {Bell}\ \emph
  {et~al.}(2021{\natexlab{b}})\citenamefont {Bell}, \citenamefont {Busoni},
  \citenamefont {Ramirez-Quezada}, \citenamefont {Robles},\ and\ \citenamefont
  {Virgato}}]{2104.14367}%
  \BibitemOpen
  \bibfield  {author} {\bibinfo {author} {\bibfnamefont {N.~F.}\ \bibnamefont
  {Bell}}, \bibinfo {author} {\bibfnamefont {G.}~\bibnamefont {Busoni}},
  \bibinfo {author} {\bibfnamefont {M.~E.}\ \bibnamefont {Ramirez-Quezada}},
  \bibinfo {author} {\bibfnamefont {S.}~\bibnamefont {Robles}}, \ and\ \bibinfo
  {author} {\bibfnamefont {M.}~\bibnamefont {Virgato}},\ }\href@noop {} {\
  (\bibinfo {year} {2021}{\natexlab{b}})},\ \Eprint
  {http://arxiv.org/abs/2104.14367} {arXiv:2104.14367 [hep-ph]} \BibitemShut
  {NoStop}%
\bibitem [{\citenamefont {Bellac}(2011)}]{Bellac}%
  \BibitemOpen
  \bibfield  {author} {\bibinfo {author} {\bibfnamefont {M.~L.}\ \bibnamefont
  {Bellac}},\ }\href
  {http://www.cambridge.org/mw/academic/subjects/physics/theoretical-physics-and-mathematical-physics/thermal-field-theory?format=AR}
  {\emph {\bibinfo {title} {{Thermal Field Theory}}}}\ (\bibinfo  {publisher}
  {Cambridge University Press},\ \bibinfo {year} {2011})\BibitemShut {NoStop}%
\bibitem [{\citenamefont {Altherr}\ and\ \citenamefont
  {Kraemmer}(1992)}]{10.1016/0927-6505(92)90014-Q}%
  \BibitemOpen
  \bibfield  {author} {\bibinfo {author} {\bibfnamefont {T.}~\bibnamefont
  {Altherr}}\ and\ \bibinfo {author} {\bibfnamefont {U.}~\bibnamefont
  {Kraemmer}},\ }\href {\doibase 10.1016/0927-6505(92)90014-q} {\bibfield
  {journal} {\bibinfo  {journal} {Astroparticle Physics}\ }\textbf {\bibinfo
  {volume} {1}},\ \bibinfo {pages} {133} (\bibinfo {year} {1992})}\BibitemShut
  {NoStop}%
\bibitem [{\citenamefont {McCabe}(2010)}]{10.1103/PhysRevD.82.023530}%
  \BibitemOpen
  \bibfield  {author} {\bibinfo {author} {\bibfnamefont {C.}~\bibnamefont
  {McCabe}},\ }\href {\doibase 10.1103/physrevd.82.023530} {\bibfield
  {journal} {\bibinfo  {journal} {Physical Review D}\ }\textbf {\bibinfo
  {volume} {82}} (\bibinfo {year} {2010}),\
  10.1103/physrevd.82.023530}\BibitemShut {NoStop}%
\bibitem [{\citenamefont {Lisanti}(2016)}]{1603.03797}%
  \BibitemOpen
  \bibfield  {author} {\bibinfo {author} {\bibfnamefont {M.}~\bibnamefont
  {Lisanti}},\ }in\ \href {\doibase 10.1142/9789813149441_0007} {\emph
  {\bibinfo {booktitle} {New Frontiers in Fields and Strings}}}\ (\bibinfo
  {publisher} {{WORLD} {SCIENTIFIC}},\ \bibinfo {year} {2016})\BibitemShut
  {NoStop}%
\end{thebibliography}%

\end{document}